\documentclass[a4paper,fleqn,usenatbib]{mnras}

\usepackage{newtxtext,newtxmath}

\usepackage[T1]{fontenc}
\usepackage{ae,aecompl}

\usepackage{graphicx}	
\usepackage{amssymb}	

\newcommand{\profound}{{\sc ProFound}}	
\newcommand{\profit}{{\sc ProFit}}
\newcommand{\sex}{{\sc SExtractor}}
\newcommand{\ebimage}{{\sc EBImage}}
\newcommand{\lambdar}{{\sc LAMBDAR}}
\newcommand{\uv}{UltraVISTA}
\newcommand{\R}{{\sc R}}
\newcommand{\sersic}{S\'{e}rsic}
\newcommand{\change}{}
\newcommand{\changetwo}{}
\newcommand{\changethree}{}

\title[ProFound: Source Extraction]{ProFound: Source Extraction and Application to Modern Survey Data}

\author[A.~S.~G. Robotham et al.]{
A.~S.~G. Robotham,$^{1}$\thanks{E-mail: aaron.robotham@uwa.edu.au}
L.~J.~M. Davies,$^{1}$
S.~P. Driver,$^{1}$
S. Koushan,$^{1}$
\newauthor
D.~S. Taranu,$^{1,2}$
S. Casura,$^{3}$
J. Liske$^{3}$
\\\\
$^{1}$ICRAR, M468, University of Western Australia, Crawley, WA 6009, Australia\\
$^{2}$ARC Centre of Excellence for All-sky Astrophysics (CAASTRO)\\
$^{3}$Hamburger Sternwarte, Universit{\"a}t Hamburg, Gojenbergsweg 112, 21029 Hamburg, Germany\\
}

\date{Accepted XXX. Received YYY; in original form ZZZ}

\pubyear{2018}

\begin{document}
\label{firstpage}
\pagerange{\pageref{firstpage}--\pageref{lastpage}}
\maketitle

\begin{abstract}
We introduce \profound{}, a source finding and image analysis package. \profound{} provides methods to detect sources in noisy images, generate segmentation maps identifying the pixels belonging to each source, and measure statistics like flux, size and ellipticity. These inputs are key requirements of \profit{}, our recently released galaxy profiling package, where the design aim is that these two software packages will be used in unison to semi-automatically profile large samples of galaxies. The key novel feature introduced in \profound{} is that all photometry is executed on dilated segmentation maps that fully contain the identifiable flux, rather than using more traditional circular or ellipse based photometry. Also, to be less sensitive to pathological segmentation issues, the de-blending is made across saddle points in flux. We apply \profound{} in a number of simulated and real world cases, and demonstrate that it behaves reasonably given its stated design goals. In particular, it offers good initial parameter estimation for \profit{}, and also segmentation maps that follow the sometimes complex geometry of resolved sources, whilst capturing nearly all of the flux. A number of bulge-disc decomposition projects are already making use of the \profound{} and \profit{} pipeline, and adoption is being encouraged by publicly releasing the software for the open source \R{} data analysis platform under an LGPL-3 license on GitHub (github.com/asgr/ProFound).
\end{abstract}




\section{Introduction}

Consistent and reliable source detection and photometric extraction has been a rich vein of research in astronomy. Clearly it is preferable to have a quantitative and reproducible means to analyse images, and over the years a number of fully automatic tools have been developed to achieve such outcomes \citep[e.g.][]{bert96}.

Our group recently developed the \profit{} 2D galaxy profiling tool \citep{robo17}, which requires a number of reasonable inputs that require tools outside of the package. Critically important for achieving a good fit are: a pixel matched sigma map (reflecting the local uncertainty in the image provided); a segmentation map that flags the pixels to use when computing the fit likelihoods; a careful sky subtraction; and reasonable initial guesses for the profile parameters.

A mixture of tools written in a number of languages cover most of the input requirements for \profit{}, however in practice how these tools are combined when scripting \profit{} for a large automatic analysis of galaxy profiles has a critical impact on how successful the fitting procedure is. With a particular focus on sky subtraction, object segmentation and initial parameter estimates (the three most difficult aspects of galaxy profiling outside of the optimization problem itself) we developed the \profound{} photometry package using the \R{} data language \citep{rcor16}. Ostensibly this package is used to create good quality automatic inputs for further 2D decompositions with \profit{}, however it also serves as an extensively featured source detection and photometric extraction package in its own right. {\change \profound{} is designed to work well with relatively deep large-area images where at least a significant minority (25+\%) of the pixels belong to the sky, i.e. of the type that you might use for galaxy profiling. To combat image artefacts it supports the use of per pixel masks, but in general it works best of smoothly varying well calibrated images, i.e.\ images without serious pedestal mosaicking discontinuities.}

Blind source finding, as it is often known, has a long history in astronomy \citep[see the recent detailed review in][]{masi12, masi13}. In the earliest days it was a necessarily visual and heuristic process, where astronomers would identify sources in photographic images essentially by eye. As technology moved towards the era of digital detectors and large arrays of imaging pixels, computer techniques advanced to automate these results in a more deterministic manner. Early techniques included simple sigma thresholding of the data in reference to the root mean square (RMS) fluctuations measured in the sky. This approach works well when the sources of interest are well above the sky noise. When sources move closer the surface brightness limit of the data, this technique can become increasingly ineffective, and lead to a higher than ideal false-positive rate. That is the number of new real sources can become subdominant compared to statistical fluctuations in the sky \citep{davi16}.

To combat this effect many improvements have been identified in the literature, e.g. simple schemes that smooth the data with an appropriate kernel and require a certain number of pixels to be above the RMS threshold and within a certain spatial separation on sky \citep[see][for a discussion on such techniques for uncovering marginally detected sources]{saba03}. {\changethree In practice a matched filter is often the optimal smoothing kernel, where for convolution this is the transpose of the point spread function of the image \citep[for the one dimensional matched filtering argument see][]{vanv46}.} Whilst these approaches are often applied in an ad-hoc manner (e.g. the exact matched filter is often not chosen as the convolution kernel), they work well to qualitatively reduce the false-positive rate by essentially requiring a spatial correlation in image fluctuations, which reduces the chance sky noise fluctuations far below that implied by the threshold applied.

Another area that is heuristic in nature but has been seen to work quite well in practice is source de-blending. This is a complex problem that is only satisfactorily resolved using a full generative model, e.g. the 2D galaxy profiling code \profit{} offers a mechanism for doing such an extraction. However, in many applications this approach is prohibitively computationally expensive. A pragmatic option has been to process the image pixels with a source de-blending algorithm. These usually work on a variant of the so-called `watershed' de-blending. How these operate can differ in detail, but a generic feature is they separate the image into regions of distinct flux by approximating the image flux as belonging to different topographic structures, i.e. if the image was inverted these would approximately be seen as valleys (the positive flux sources) and flat noisy regions (the sky and the sky noise). If this topographic structure was steadily filled with water it is easy to see that structures that begin as distinct bodies of water will start to merge together as the image becomes entirely flooded. There are various methods to use this insight to define genuinely distinct sources, but the basic approach is the same.

Mixed in with the above issues, there are a large number of subtle effects that must be handled carefully. These include the sky estimation, the sky RMS estimation, and the growth of apertures to fully contain the flux. Each of these have long histories in astronomy literature, but they largely all share a heuristic approach. This is usually for pragmatic reasons of computational complexity rather than aiming to be the ideal solution in a demonstrative sense. The calculation of the sky and sky RMS are simpler problems to tackle for the most part (exceptions include very crowded fields and confusion limited data), and a large part of the Methods (Section \ref{sec:methods}) discusses the main approaches in detail. Choosing an appropriate method to fully capture the flux present is a more difficult problem to solve, and many approaches have been advocated in the literature.

The earliest attempt at systematically identifying the flux for extended sources can be found in \citet{petr76}. The basic idea behind the Petrosian magnitude is to determine the radius to be scaled based on surface brightness properties of the galaxy, namely the ratio of the integrated surface brightness within some radius compared to the instantaneous surface brightness at the same radius. By incorporating the surface brightness in the numerator and the denominator when calculating the Petrosian radius the results of many observational effects are naturally removed, e.g. cosmological surface brightness dimming, variable imaging depth of the data under consideration and different observing conditions which can produce variable seeing amongst other effects.

In theory the Petrosian magnitude is an elegant route to extract flux measurements, since in principle extracted extended source fluxes are not highly sensitive to the observing conditions. In practice things are not as simple as we would like, and galaxies are not well represented as having a shared fundamental profile, which whilst not immediately clear is implicitly assumed in the Petrosian magnitude system \citep{grah05}.  Since galaxies can have a very broad range of profiles the magnitude extracted is in fact highly sensitive to the profile \sersic{} index \citep{sers63, grah05}. Since galaxies are also convolved with the atmospheric seeing in ground based data there is an additional dependence beyond the intrinsic profile, namely that the same galaxy shifted to higher redshift will return a different fraction of the true flux because the profile will have evolved away from its intrinsic value towards the atmospheric value, which for a mixture of reasons is usually very close to a canonical \sersic{} index of 0.5, i.e. a Normal distribution.

A popular and computationally simpler alternative was presented in \citet{kron80}. In this approach the inner bright moments of light are used to estimate the Kron radius. It is then up to the user to identify a sensible Kron multiplier to scale this radius by in order to capture a certain quantity of the flux. For stars this factor is not especially important since the majority of the object flux is captured within the inner part of the profile. For more extended objects the number chosen for this multiplier can have a significant impact since much of the flux of extended objects comes from the outer parts of the profile. The `AUTO' magnitude returned by \sex{} \citep{bert96} is closely related to this version of the Kron magnitude, and can be run in such a mode that this multiplying factor is chosen intelligently for each source rather than using a single fixed value.

A key feature of both of the above methods for extracting photometry is that they use either circular or elliptical apertures. The earliest applications predominantly used circular apertures, with the elliptical variants being more modern and widely popular today. This suggests an obvious limitation of either approach: galaxies are not simple ellipses, and decisions have to be made regarding how flux is distributed between potentially overlapping apertures.

The above motivated the most novel aspect of the \profound{} code presented in this paper: a move away from simple elliptical apertures and towards apertures that properly identify the parts of galaxies containing the significant proportion of the flux. A few methods were looked at when first addressing this issue, with the result that dilated segments that follow the surface brightness distribution of the galaxies act as a better method to identify the true flux belonging to a given galaxy.

In the regime of bright compact elliptical sources, which are fairly isolated from other sources, the dilated segments follow the extent of traditional apertures fairly closely. However, for very extended sources the differences can be quite pronounced, with complex source geometry not being accurately captured by simple circular or even elliptical apertures. The dilation approach also offers a few other advantages when it comes to source de-blending, namely that segments are never allowed to overlap on the sky in the way the expanded apertures can. In fact it is non-trivial to determine how best to split flux between adjacent and overlapping apertures, and a number of different ad-hoc and heuristic schemes are usually applied to account for these effects.

The hard flux boundaries created by using segmented apertures also have a number of advantageous side-effects when it comes to determining fluxes in regions that have pathological issues such as very bright halos around bright stars that have saturated the detector. These often create biased photometry over an extended region, where the Kron or Petrosian aperture is often compromised and the expanded aperture overlaps with multiple fainter sources. We highlight some specific examples of this later in the paper, but it is a common feature of survey data \citep{wrig16}. The method of iterative dilation used in \profound{} naturally prevents extreme expansion artefacts since segments are not allowed to grow into each other. This is not to say the photometry extracted will not be compromised at all, but the segmentation map and the approximate sources properties are much nearer to the intrinsic values and serve as better inputs for \profit{}, which was the initial design goal of the new software.

In Section \ref{sec:methods} we describe the methodology behind the most critical aspects of the package design.
In Section \ref{sec:sims} we look at the application of \profound{} to fully simulated wide-field images, with a focus on the completeness and purity of the detection, and the accuracy of the photometric properties. In Section \ref{sec:UV} we apply \profound{} to the \uv{} multi-band imaging data, with a detailed comparison of some of the output properties compared to the public catalogues.

\section{Methods}
\label{sec:methods}

\begin{figure}
	\includegraphics[width=\columnwidth]{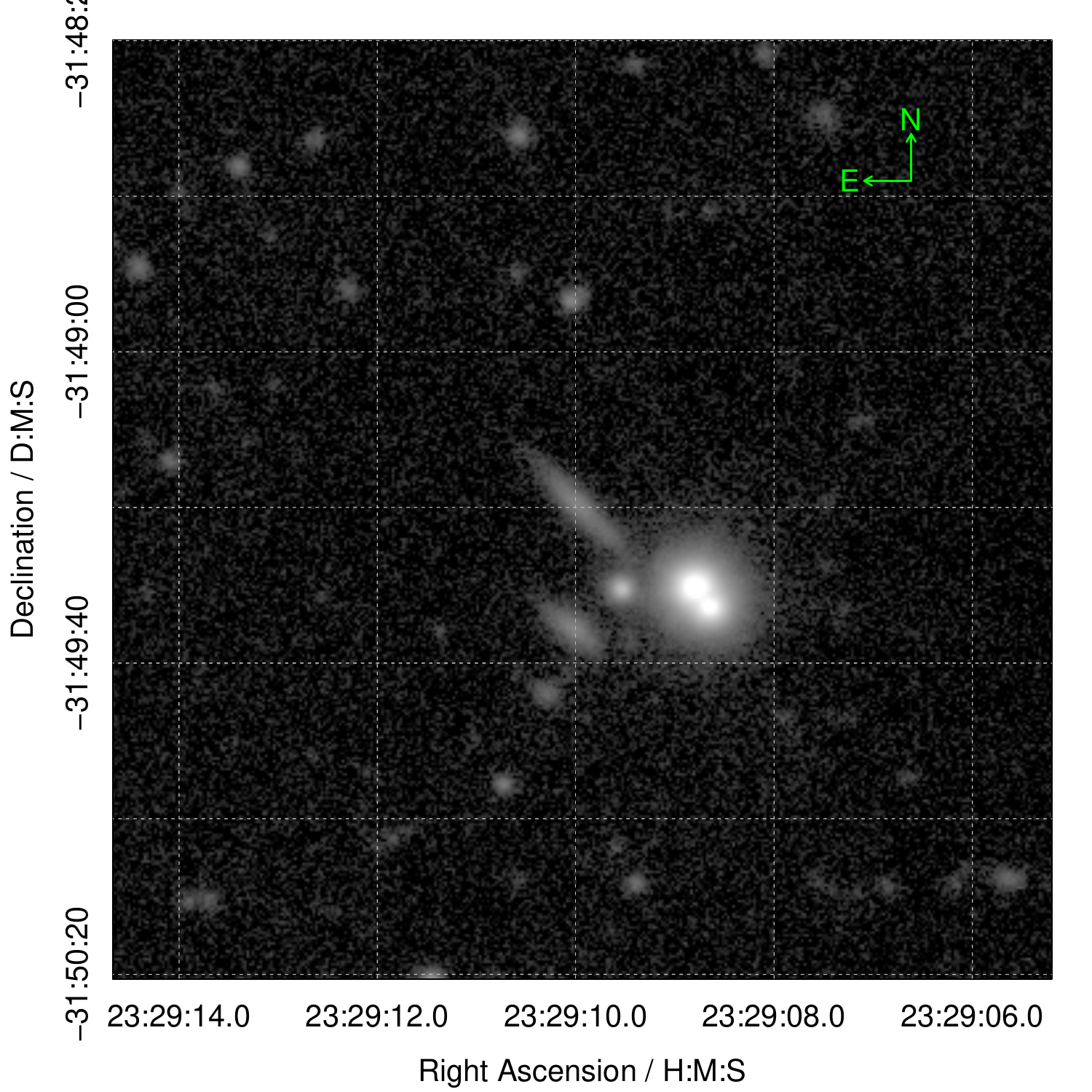}
    	\caption{Example VISTA Z-band data taken from the VIKING survey included with the \profound{} package and used in various parts of this paper (GAMA galaxy ID G5458748). In this Figure we stretch the z-scale to make the double star nature of the two bright central sources clear. In latter Figures we use a different mapping that enhances the contrast of fainter sources and visually merges these two stars together.}
    	\label{fig:example_image}
\end{figure}

In the following description of the main methods behind \profound{} source extraction we use the same test Z-band data shown in Figure \ref{fig:example_image}. This was taken from the public VISTA (Visible and Infrared Survey Telescope for Astronomy) Kilo-degree INfrared Galaxy \citep[VIKING;][]{edge13} survey that used the VIRCAM instrument on ESO's 4m VISTA facility. The galaxy at the centre of the image was a main survey target (G5458748) for the Galaxy And Mass Assembly survey \cite[GAMA;][]{driv11, lisk15}. This image has a a number of properties that make it ideal as a small case study: it contains a mixture of bright and faint galaxies and stars; it contains a mixture of compact and extended galaxies, the central region contains a number of reasonably confused sources; the background root mean square (RMS) in the sky varies distinctly in the frame due to its stacked origin; and it has objects contained entirely within the image and near to the edge.

The example data is included with the \profound{} package so it is easy for a new user to recreate the plots in this paper using the many worked examples and vignettes\footnote{http://rpubs.com/asgr/}. The thorough package documentation (the embedded PDF manual is 61 pages, with every function, variable and output described) and long-form vignettes have been influenced by the clear utility of the `\sex{} for Dummies' guide which has been hugely beneficial to the community who regularly use \sex{} \citep{holw05}.

This paper is not intended as a user manual, so we will not discuss the technicalities of the detailed settings and parameters here. Except where mentioned explicitly the code has been run in close to default mode, with the notable difference being the setting of the magnitude zero point (which has to be set explicitly since there is no standard format to specify this in FITS headers). Otherwise meta-data is largely extracted from the FITS header and extraction properties are estimated dynamically using the data itself.

Functions in the \profound{} package are all named with a leading lowercase `profound'. This is to remove the potential for clashing function names since \R{} does not trivially support package aliasing in the way that some high level languages do (e.g. {\sc Python}). The \profound{} package includes a few different hierarchies of functions. The expectation is that some of these will be used routinely (e.g.\ the highest level \profound{} object extraction and photometric measurement function, also called \profound{}), and some will rarely be used by a typical end user (e.g.\ the linear interpolation function {\sc Interp2D}).

Between these two extremes there are a large number of mid-level functions that more advanced users might want to use directly in order to manipulate the data in a specific manner. The highest level \profound{} function effectively links a large number of these mid-level functions together in a manner that achieves good quality source extraction and photometric analysis for a range of typical two dimensional astronomy data (particularly imaging and radio continuum data, but not limited to such applications). During development the focus has been on optical and NIR survey data, but it has also been used successfully on ultra-violet (UV) data and far-infrared (FIR) data.

\subsection{\profound{} Source Extraction}
\label{sec:methods_profound} 

\begin{figure*}
	\centering{Code Flow of \profound}
	\includegraphics[width=17cm]{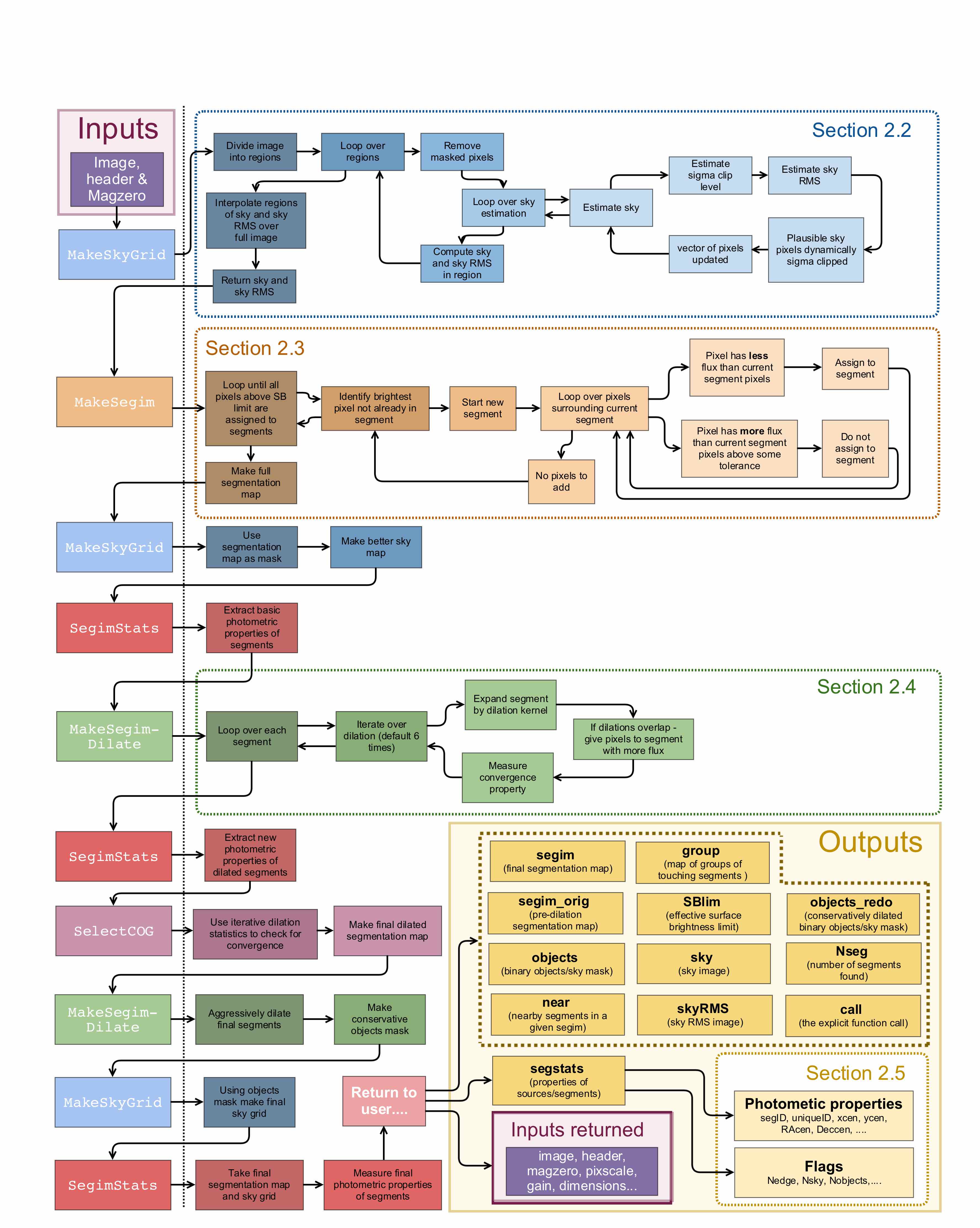}
    	\caption{Detailed code diagram for the highest level \profound{} function. This calls a number of mid-level functions that we discuss in detail in the relevant labelled Section.}
    	\label{fig:code-flow}
\end{figure*}

The highest level \profound{} function is ultimately a structured calling of a range of mid-level functions. The full code diagram is presented in Figure \ref{fig:code-flow} with the various parts we discuss in more detail labelled by Section. The simplified version of Figure \ref{fig:code-flow} is provided below, where the mid-level function executed is named at each relevant stage.

\begin{enumerate}
\item Make a rough sky map (see Section \ref{sec:methods_skysub}) --- {\sc MakeSkyGrid}
\item Using this rough sky map, make an initial segmentation map (see Section \ref{sec:methods_segmap}) --- {\sc MakeSegim}
\item Using this segmentation map, make a better sky map --- {\sc MakeSkyGrid}
\item Using this better sky map, extract basic photometric properties (see Section \ref{sec:methods_photometry}) --- {\sc SegimStats}
\item Using the current segmentation map, dilate segments and re-measure photometric properties for the new image segments, by default it iterates six times (see Section \ref{sec:methods_dilation}) --- {\sc MakeSegimDilate} / {\sc SegimStats}
\item Using the iterative dilation statistics, every object is checked for convergence, by default convergence of flux is used (see Section \ref{sec:methods_dilation}) --- {\sc selectCoG}
\item Make final segmentation map by combining the segments when each source has converged in flux
\item Make a conservative object mask by aggressively dilating the final segmentation map --- {\sc MakeSegimDilate}
\item Using this conservative objects mask make a final sky map --- {\sc MakeSkyGrid}
\item Using the final segmentation map and the final sky map compute the final comprehensive photometric properties --- {\sc SegimStats}
\item Return a list containing the input image pixel-matched final segmentation map (called {\bf segim}), the pre-dilation segmentation map ({\bf segim\_orig}) , the binary object/sky mask ({\bf objects}), the conservatively dilated binary object/sky mask ({\bf objects\_redo}), the sky image ({\bf sky}), the sky-RMS image ({\bf skyRMS}) and the effective surface brightness limit image ({\bf SBlim})
\item Return the data-frame of photometric properties for every detected source ({\bf segstats})
\end{enumerate}

Some other simple properties are passed through and included in the output of ProFound: the original image, the image header ({\bf header}) if it is attached to the input image, the magnitude zero point specified ({\bf magzero}), the gain in electrons per astronomical data unit ({\bf gain}), the pixel scale in arc-seconds per pixel ({\bf pixscale}) and finally the full function call ({\bf call}).

The above detection and extraction sequence was developed on a range of test data, where the desire was that good quality results should be achievable when running on default parameters. The latter was deemed important since parameter tuning can be non-obvious and complicated for novice users. Even in examples where qualitatively better extraction could be achieved by changing the parameters away from defaults, the changes were usually small and the impact marginal (i.e.\ pathological failure is very rare).

The emphasis during development was on robustness rather than speed. E.g.\ as written the sky determination is a relatively expensive operation, and by default this is done three times with increasingly aggressive object masks in order to be robust against biases due to extended objects and crowded fields. Even with the iterative object dilation and sky subtraction routines turned off, \profound{} is still notably slower and more memory intensive than \sex{} \citep{bert96}, typically a factor of a few slower for the same data when achieving a similar number of source extractions. Since it is largely written in \R{} in a highly functional manner, there is a lot of data copying between different levels of functions, although efforts are made to minimise this whilst still preserving the safe and functional nature of R code.

Typically the processing time scales with the number of pixels, i.e\ a 2k$\times$2k image will take a similar time to process as four 1k$\times$1k images. However, this scaling only continues until the point the various objects produced within \profound{} during processing can still be held in random access memory (RAM) without having to use disk based virtual memory. Since various internal objects are necessarily created that have the same dimensions as the input image, the maximum practical image size without subsetting the image is of order a tenth the available RAM (before potentially compromising the processing speed). On a modern computer with $\sim$8GB of RAM this is in the region of a 10k$\times$10k image. On a MacBook Pro with 16GB of RAM we were able to process 20k$\times$20k images without resorting to disk based virtual memory and the associated slow down. There is also a low memory mode which radically cuts down the RAM requirements to be nearer to double the image size, but removes many of the useful outputs to save RAM (e.g.\ the sky and sky-RMS maps).

A limitation compared to \sex{} is that the main \profound{} function cannot natively handle matrix-like images much larger than 46k$\times$46k pixels (strictly a hard limit of $2^{31}-1$ pixels in total), even on machines with much more memory available. A higher-level function ({\sc ProFoundLarge}) is included to process very large FITS files, which extracts overlapping subregions directly from a target FITS file, and recombines them in an unambiguous manner. The \uv{} survey data discussed later in this paper was processed using the {\sc ProFoundLarge} function since it is marginally too large to be extracted in a single pass and served as useful test data, although the sub-region of interest for the Deep Extragalactic VIsible Legacy Survey (DEVILS; Davies et~al. in prep) is ultimately smaller than the $2^{31}-1$ pixel limit. {\change A reason to take this approach is that {\sc ProFoundLarge} can compute the data in an embarrassingly parallel manner, whereas most of the routines within \profound{} are inherently single-threaded in nature. Routes exist to compile \R{} with support for multi-threading, but this requires reasonably advanced knowledge, and our assumption is most users of \profound{} will be using the standard single-threaded version of \R{}.}

For a typical survey image run with default parameters the various stages take fairly predictable proportions of the total computing time: calculating the full watershed de-blend for the segmentation map dominates the total time ($\sim$50\%), followed by calculating the sky map (done three times by default, $\sim$20\%), iteratively dilating the image segments (done six times by default, $\sim$20\%) and calculating the photometric properties of the segments (done once per dilation step and again at the end, so seven times in total, $\sim$10\%). Since the watershed stage dominates the time and uses an efficient external function written in C, the potential optimization gains are quite moderate. Reductions in processing time are possible if the number of sky calculations and/or the number of dilations are reduced from the defaults. When run in matched segment mode the processing time is significantly reduced since the watershed de-blend stage is no longer required. If the sky subtraction and dilation steps are also turned off then the total processing time can be reduced by a factor $\sim$10, i.e. the minimum requirement is that photometric properties of the provided segments are computed.

\subsection{Sky Subtraction}
\label{sec:methods_skysub}

An important step in almost any approach to object extraction from astronomical imaging data is the sky subtraction. Depending on the origin of the data the image might arrive to the user fairly flat and featureless in the background (e.g.\ optical drift scan data with the Sloan telescope; \citealt{ahn14}) or complex and variable at a number of different scales (e.g.\ near-infrared data with variable fringing artefacts; \citealt{andr14}).

Given the potential complexity of the sky background it is rarely sensible to attempt to construct a formal statistical model of the sky since it is very difficult to meaningfully parameterise the range of behaviour observed \citep[see][]{bija80, irwi85}. Instead most popular astronomy applications have taken the route of a heuristic but visually appealing and pragmatically achievable scheme: coarsely sampled sky measurements combined with a polynomial interpolation \citep{bert96}. Key parts of this process are that {\it true} sky pixels need to be identifiable in the target image, and some manner of estimating their variance and absolute level is possible.

The above is achieved in a practical manner by clipping likely objects out of the data and using a sliding box car filter on a grid to measure image properties. With a meaningful sampling of the sky and sky-variance a traditional scheme to interpolate between grid points (e.g.\ bilinear or bicubic) can then be used to construct a per-pixel estimate of the sky. The caveats to this process are that objects need to be well masked (else they will systematically bias the estimators) and the box car scale has to be well chosen so as to remove the real sky variations and not structure that belongs to objects in the image. The sky is extrapolated at the edges, which can cause artefacts if it is changing rapidly and bicubic interpolation is used. In this scenario bilinear interpolation is the safer option since it does not use higher order polynomial terms that cause this effect.

The main high-level sky estimation routine in \profound{} ({\sc MakeSkyGrid}) allows users to pass an image, masks (both for flagged pixels and for identified objects), the box car size, the grid sampling and the type of sky interpolation to use (bilinear of bicubic). This fairly simple functional interface allows for a large amount of flexibility in usage. The default code flow is as follows (names of the \R{} functions used are displayed in small caps):

\begin{enumerate}
\item Divide the image into regions based on the requested box car and grid sampling. By default the grid sampling inherits the box car size, meaning pixels are evaluated once
\item Each sub region is analysed separately in a large loop:
	\begin{enumerate}
	\item The masked pixels are removed from analysis, leaving a vector of fiducial sky pixels ($sky_{fid-pix}$)
	\item Then the following is computed iteratively (either until the clipping is converged, or after 5 iterations):
		\begin{enumerate}
		\item The sky value is estimated as $sky=\textrm{\sc median}(sky_{fid-pix})$
		\item The dynamic sigma clip level is estimated to be $\sigma_{clip}=\textrm{\sc qnorm}(1-2/N_{sky})$
		\item The standard deviation of the sky pixels is estimated as $sky_{RMS}=\textrm{\sc quantile}(sky_{fid-pix},0.5) - \textrm{\sc quantile}(sky_{fid-pix}, 0.159)$
		\item The plausible sky pixels are dynamically sigma clipped such that pixels satisfying $sky_{fid-pix} > sky + sky_{RMS} \sigma_{clip} \lor sky_{fid-pix} < sky - sky_{RMS} \sigma_{clip}$ are removed
		\item The vector of $sky_{fid-pix}$ is updated and these new fiducial sky pixels are used for the next iteration
		\end{enumerate}
	\item Once convergence has been achieved the final computed $sky$ and $sky_{RMS}$ values are returned for the region under consideration
	\end{enumerate}
\item With all regions having a unique estimate of the $sky$ and $sky_{RMS}$ a bilinear or bicubic interpolation scheme is used to calculate plausible $sky$ and $sky_{RMS}$ values for all pixels
\item The $sky$ and $sky_{RMS}$ images are returned to the user or higher level function in a list
\end{enumerate}

As discussed above, the sky is measured a number of times during \profound{} source extraction. The basic design philosophy is that it is possible to measure more accurate values for the sky and sky RMS as the objects are extracted, leaving behind increasingly certain sky pixels, more accurate identification of true sources, and better photometric measurements of the segments.

\subsubsection{Different Sky Estimators}

In principle the clipping process works on positive and negative pixels, but unless there are serious artefacts in the data it will work predominantly on the positively valued pixels by removing undetected and un-extracted sources. The clipping process can be changed so as to not clip out possibly biased pixels, and the type of estimator for the sky can also be changed from median to mean or mode. Whether or not these options are used depends on the type of image being analysed, and what the aim of the source extraction is. {\change An implicit assumption in \profound{} (and indeed \sex{}) is that the sky fluctuations are symmetrically distributed around some intrinsic value. This will only be true in detail when the number of sky photons is fairly large (many dozens or more) and we can use the Normal $\sim$ Poisson approximation.}

The difference between the sky level estimators (most commonly the mean, median or mode) and sky RMS level estimator (e.g.\ quantile versus standard-deviation) is an interesting point to consider. In the toy situation described it is preferable to use the {\change mode (or possibly median if the mode is noisy and/or poorly sampled)} for the sky and the quantile for the sky RMS, since these are both systematically nearer to the specified `intrinsic' sky for these simple estimators \citep{irwi85}. However, what really matters is the origin of the positively biased flux that skews the distribution. There are a number of processes that generate the `sky' and contribute to the extended area signal in a typical digital detector image, in roughly descending order of importance:

\begin{enumerate}
\item The actual night sky caused by Earth's atmosphere glowing. Can be quite spatially and temporally variable (e.g.\ NIR images, distant artificial lights turning on and off)
\item Flux scattered around the image due to telescope optics or instruments producing scattered light and/or very broad $\sim$Lorentzian wings
\item Scattering of astronomical light by the Earth's atmosphere, usually at very small scales (close to Gaussian usually)
\item Intrinsically broad features caused by the Milky-Way's foreground cirrus
\item Intrinsically broad wings caused by extra-galactic sources (e.g.\ the low surface brightness wings of galaxies or intra halo/cluster light)
\item Undetectable faint compact sources, can be structured (Milky-Way stars) or effectively uniform in distribution (high redshift galaxies)
\end{enumerate}

The question is then: which of these do we wish to remove? The answer clearly depends on what sources we are trying to extract photometry for. If we are measuring stellar photometry of a bright star then we almost certainly want to remove 1/4/5/6, i.e. we want to model the light from the star that has been scattered by the atmosphere and the telescope (assuming it is the dominant contribution close to the star being modelled). If we want to profile a faint galaxy then you probably need to remove 1/2 (assuming it is mostly caused by other sources)/3 (assuming it is mostly caused by other sources)/4/6, i.e. we want to keep the faint wings of the target galaxy intact. There is not a trivially right answer, but \profound{} does offer a few routes to compute these different types of sky which are described at length with examples in some of the available vignettes\footnote{http://rpubs.com/asgr/}.

In summary, a \profound{} user might reasonably prefer a median, mode or mean type sky, depending on the use case. If the source sits on top of the sky (whatever makes it) then the user probably wants to use the more biased mean estimator. If the source is the dominant part of the observable background (e.g. when profiling the faint wings of galaxies) then the user might prefer the less biased {\change median or mode} estimators and more aggressive source clipping.

A final issue is whether you can extract better galaxy profile models by also using \profit{} to model the background for a given source, where \profit{} has the capacity to model a flat local floor for the sky. The three obvious options are:

\begin{enumerate}
\item Use \profound{} sky subtraction on the image and do not fit a sky background in \profit{}
\item Use \profound{} sky subtraction on the image and also fit a sky background in \profit{}
\item Do not use the \profound{} sky subtraction on the image and only fit a sky background in \profit{}
\end{enumerate}

For pragmatic reasons of needing to remove complex sky that cannot be fully generatively modelled with \profit{}, options (i) or (ii) are likely to be the best strategy for the majority of use cases.

\subsection{Segmentation Map}
\label{sec:methods_segmap}

Due to the emphasis on making segmentation maps that are useful inputs for \profit{} galaxy profiling, the method of making segmentation maps differs in important ways to the method used in \citet{bert96} (i.e.\ \sex{}, which in turn was inspired by \citet{bear90}). The most significant practical difference is that pixels that are flagged as being above a requested surface brightness threshold (be that stated in terms of sky RMS fluctuations or absolute surface brightness) are de-blended using a non-discretised watershed algorithm that creates flux de-blends through two-dimensional saddle-point cuts in image space, rather than one-dimensional cuts in flux space. The watershed approach outlined here is nearest in spirit to the `FOCUS' de-blend method of \citet{jarv81} and variants of the popular \citet{meye94} `priority flood' algorithm \citep[see][for recent astronomical applications]{zhan15, zhen15}. \profound{} uses the iterative process outlined below:

\begin{enumerate}
\item Identify the brightest pixel in the image above the specified surface brightness level which is not already assigned to a segment
\item Progressively search unassigned image pixels surrounding the current segment, for each pixel searched:
	\begin{enumerate}
	\item If a searched pixel has less flux than any neighbouring pixels already assigned to the segment, then assign to the current segment if no unassigned pixels neighbouring the pixel under consideration have more flux
	\item If a searched pixel has more flux than its neighbours already assigned to the segment above some tolerance level then do not assign it to the current segment
	\item If no more pixels can be assigned to the current segment then terminate the growing process
	\end{enumerate}
\item Select the next brightest unassigned pixel remaining in the image and assign it to a new segment, then repeat the above segment growing process
\item Once all pixels above the specified surface brightness level have been assigned to a segment terminate the watershed process
\end{enumerate}

There are a small number of parameters that have a significant effect on the watershed process, and in practice these need to be slightly altered to best segment the data under consideration. The most important is the `tolerance', which specifies to what degree pixel growth is allowed to traverse uphill within a segment. In practice this determines the level of de-blending between closely separated flux peaks, where a higher tolerance means less splitting up of extended regions of flux. This is always specified in terms of the RMS of the sky and is 4 by default, i.e. a fluctuation would need to be more than 4 deviations of the sky RMS above the neighbouring segment pixel in order to not be included as part of the current segment being grown. The next most important parameter is the smoothing applied to the image (sigma), where by default a Gaussian kernel with a standard deviation of 1 pixel is used to blur the image. The smoothing can be turned off entirely, but this is rarely a good idea since there is always a large degree of pixel-to-pixel noise in all but most correlated images (typically flux values only appear smooth on the scale of a few pixels). Instead it is sometimes justifiable to increase the smoothing size, with an upper limit of 3 or 4 sometimes more suitable for images of large well-resolved galaxies. Related to this is the `ext' parameter which is passed directly to the \ebimage{} {\sc watershed} function used in \profound{} \citep{greg10}. This determines the allowed search radius around each pixel (rather than restricting the search to immediately adjacent pixels), where the default is 2 pixels. The smoothing `sigma' and the search radius `ext' have a similar impact and small changes in at least one of them (over the range 1--4) are common when first applying \profound{} to a new dataset and tuning for optimal segmentation. Increasing the smoothing is an effective strategy for extracting extended low surface brightness sources.

\begin{figure*}
	\includegraphics[width=\columnwidth]{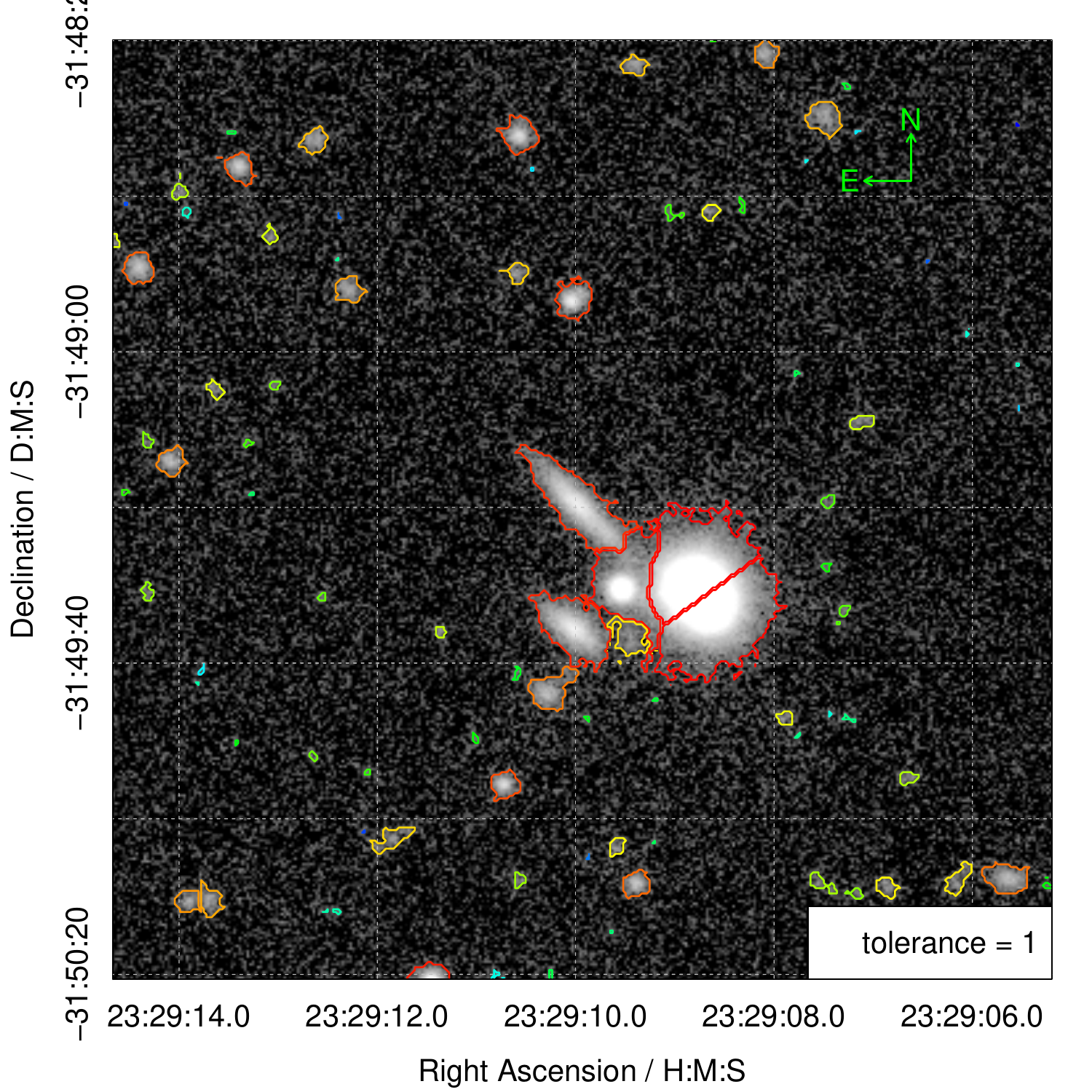}
	\includegraphics[width=\columnwidth]{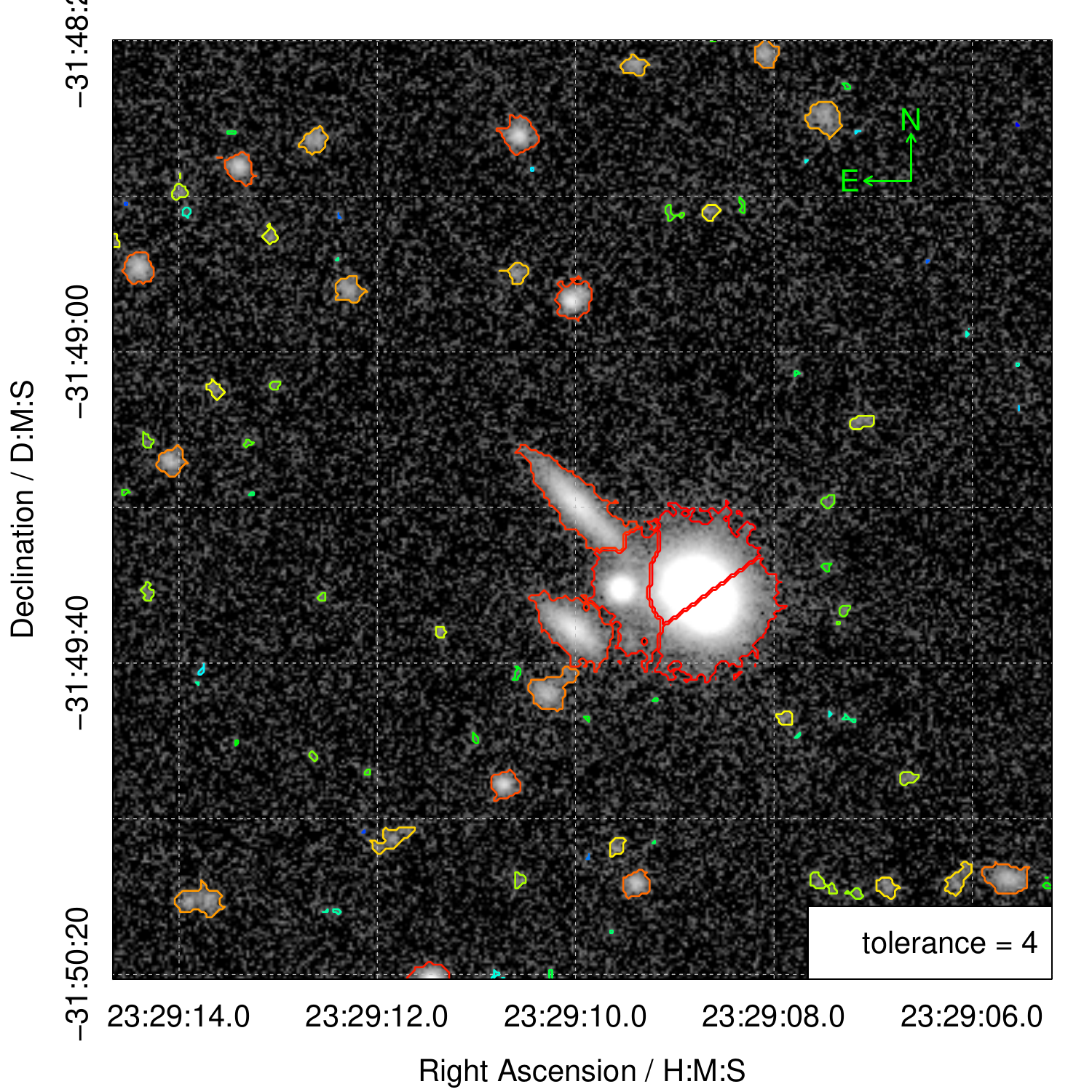}\\
	\includegraphics[width=\columnwidth]{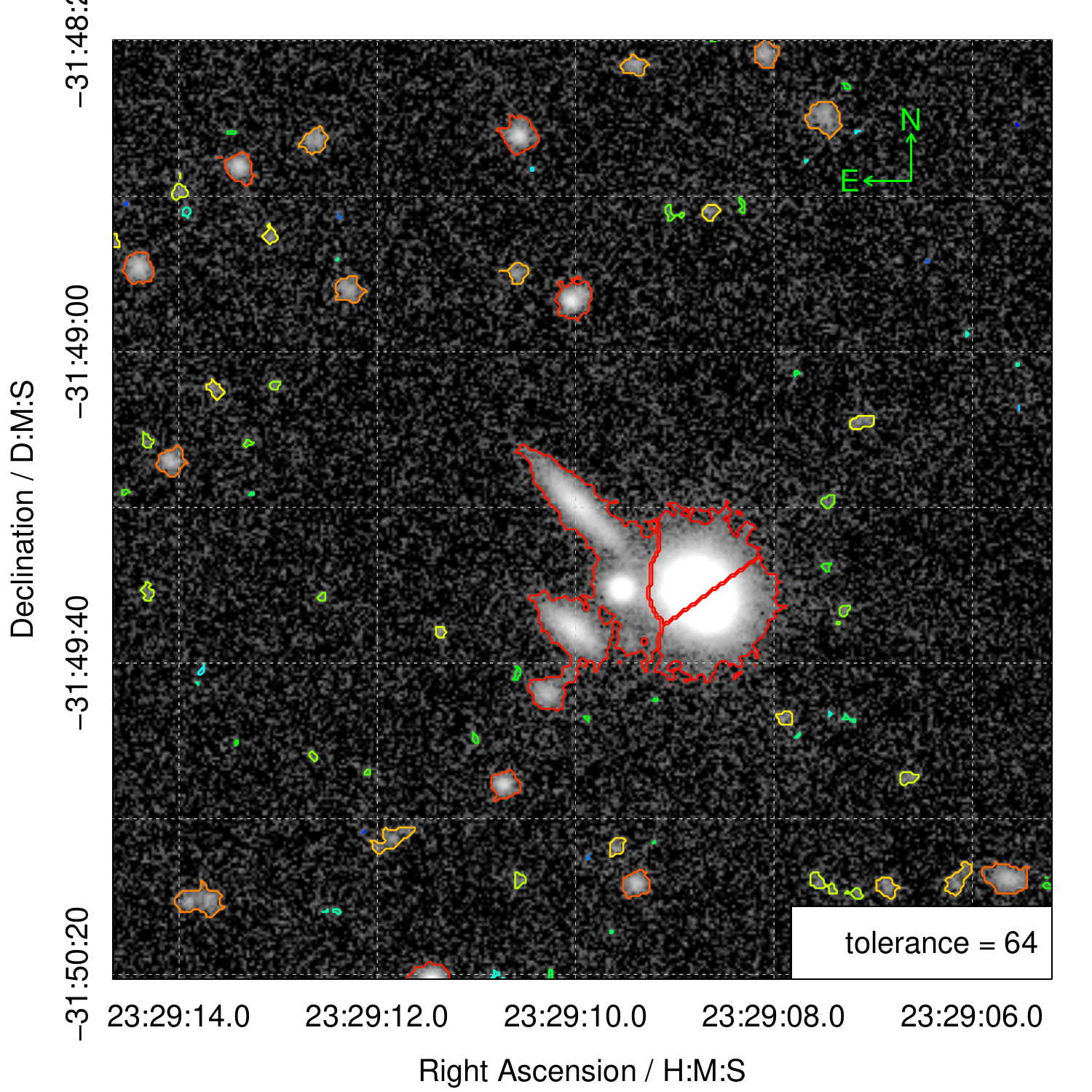}
	\includegraphics[width=\columnwidth]{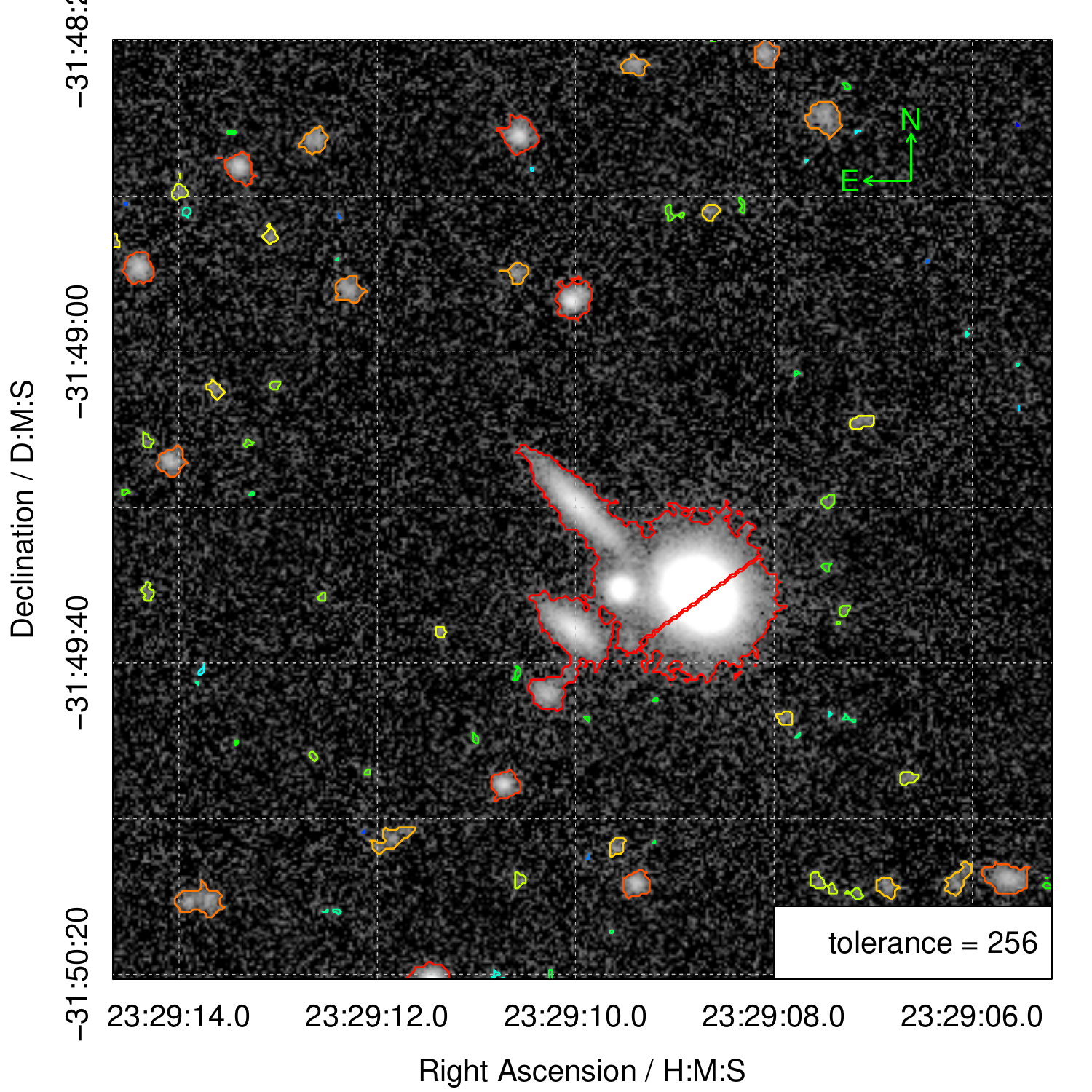}\\
    	\caption{Examples of using different watershed de-blend tolerance levels on the initial segmentation map, as per labelled on each panel. There is no truly objective approach to say which solution is preferred, but most professional astronomers would probably suggest the best answer lies in the regime of tolerance 1--4, where the central confused complex has been broken up into its plausible sub-components.}
    	\label{fig:watershed_tol}
\end{figure*}

Figure \ref{fig:watershed_tol} demonstrates how the de-blender used in \profound{} works in practice for different watershed tolerance levels. The main consequence of using the above approach is that islands of segments within larger segments can never be created. Instead the de-blended segments reflect the segment water would flow into if the flux map was turned upside down (ignoring dynamical effects like momentum, and just specifying the pixels that water would next flow into if the velocity was set to zero). There are other definitions of `watershed', but this is the most standardised in the field of geology and accurately reflects a true gravitationally influenced watershed map. An internal Sloan Digital Sky Survey memo discusses the relative merits of different de-blending approaches\footnote{http://www.astro.princeton.edu/~rhl/photomisc/deblender.pdf}, with the general remark that all approaches make compromises and assumptions. This is true even for full generative modelling, given the requirement to define the parameterization of the model.

For our stated aims of producing good segmentation maps for passing into \profit{}, the type of saddle point segmentation discussed above works well since it errs on the side of ignoring pixels compromised by nearby sources when modelling the profile of a target galaxy. Qualitatively we also see relatively few occasions of segments being grouped together in a common aperture erroneously, which we sometimes see to occur when running \sex{} \citep[see][and the \uv{} example below for examples of such situations]{wrig16}.

The full \profit{} model of the central complex shown in Figure \ref{fig:watershed_tol} yields photometric properties (most notably flux) not far removed from the \profound{} estimates (less than 0.5 mag differences). However, it is in general more important to get the correct number of segments and mode locations than having very good initial estimates for the fluxes and sizes. This is because estimating the number of components and mixtures is more difficult when galaxy modelling than optimising the parameters of a given model. Whilst \profound{} does include routines to improve object measurements based on symmetry and flux sharing, these are turned off by default since they are computationally costly and generally unnecessary (i.e.\ the raw measurements are good enough).

The underlying code that computes the watershed de-blend comes from the image processing package \ebimage{} that is already available in \R{} and widely used for low-level image processing \citep{greg10}. Its design and focus was for cellular biology (the EB in the name standing for European Bioinformatics) where the main task tackled was how to correctly segment images of cells taken by microscopes. As often noted anecdotally, there is much similarity between an image of biological systems and astronomy images, the former probably being the more complex to organise and segment in a systematic manner. For this reason it is not surprising that a tool developed for such an application works well for astronomical images. An important feature of the routine used is that it does not discretise the flux levels in the image (as \sex{} does) and uses the full flux resolution available. This fact means that it is quite slow (despite the underlying code being written in {\sc C}), and the watershed step usually dominates the computation time for larger images since it scales as $\mathcal{O}(n\log{}n)$ whereas nearly every other subroutine scales as $n$ or better (where $n$ is the number of pixels in the target image). The effect of this is that it can be faster to run \profound{} on sub-regions rather than one very large image. It is only beyond the size of 10k$\times$10k images (more than $10^8$ pixels) that the difference becomes worth considering.

For convenience when using the outputs of \profound{} the identities of neighbouring segments with respect to all other segments and also the friends-of-friends groups of de-blended regions can also be returned. The latter is particularly important when using \profound{} as an input to \profit{}, since you should minimally try to fit all the segments in a grouped friends-of-friends region when trying to profile blended objects. These two types of grouped structures are not trivially returned by \sex{}, so for creating profiling inputs this is a clear advantage of using \profound.

\subsection{Segment Dilation}
\label{sec:methods_dilation}

The next phase of the source extraction routine grows the segments using circular top-hat (by default) dilation operations until convergence has been achieved. As highlighted in the introduction, the segment dilation method is the most novel aspect of how \profound{} operates. At no stage are fluxes or object properties calculated using apertures (be they circular or elliptical). Instead all integrated properties related to the source are estimated using the dilated segments alone. The procedure is iterative and therefore relatively expensive compared to simply expanding the inner Kron or Petrosian radius by some factor that approximately contains a large fraction of the flux.

\begin{figure}
	\includegraphics[width=\columnwidth]{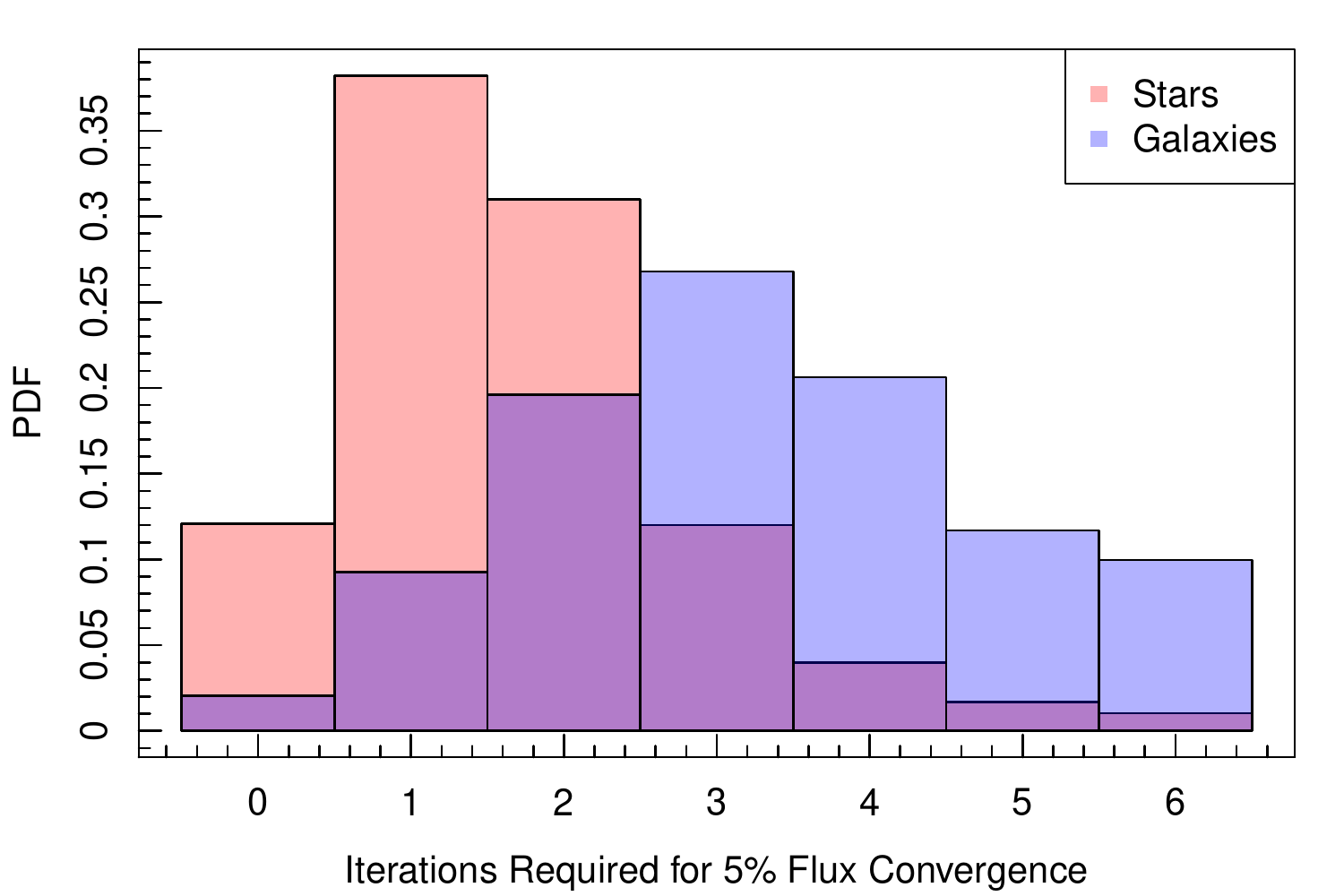}
    	\caption{Histograms of the number of iterations required to reach flux convergence for both stars and galaxies. The data is taken from the large suite of simulations that we discuss in Section \ref{sec:sims}.}
    	\label{fig:segim_iters}
\end{figure}

By default the code looks for convergence in flux at some tolerance level, but in theory any property of the catalogue generated can be used to ascertain whether the segment being grown has converged and the dilation should stop. Figure \ref{fig:segim_iters} shows the required number of dilation iterations before flux convergence is achieved for a large suite of simulated data containing 40k stars and galaxies that we discuss in detail later. It is notable that typically stars require fewer iterations than galaxies in order to achieve the default level of convergence (5\%). The minimum number of iterations is zero, but only a very small fraction of stars require so few dilations. More typically stars require two dilations and galaxies require four.

The dilation distribution has not fallen to zero for stars, and more clearly galaxies, even by iteration six. This suggests that galaxies in particular have such extended flux envelopes that they need even more dilations. It is possible to increase the maximum allowed number of dilations to greater than six, but this was considered to be a sensible compromise default value. The sources requiring six dilations are notable for having the lowest integrated surface brightness levels of all sources (i.e. these are the most marginal detections), so the danger of pushing to much more aggressive dilation levels is that a significant amount of noise is incorporated into the aperture, compromising the photometric properties of the object being measured.

With the main design considerations for the segment dilation process now justified, the basic flow is as follows:

\begin{enumerate}
\item Execute a number (the default is six) of dilation operations on all segments, for each dilation operation:
	\begin{enumerate}
	\item Expand every segment with a dilation kernel (by default this is a circular top-hat with a diameter of nine pixels)
	\item If segment dilations overlap then give all pixels to the segment containing more flux in the current iteration
	\item Measure the convergence property of interest for the new dilated segmentation map (by default this is the flux)
	\end{enumerate}
\item After all dilations have been made look through all segments and determine when each segment has converged within some tolerance (this is within a factor 1.05 in flux by default)
\item Put these converged segments together to make a final dilated segmentation map
\end{enumerate}

The dilation operation is executed by the {\sc dilate} function in the \ebimage{} package. This is optimised in design for detecting the full extent of pixels belonging to an already labelled biological cell. {\change A useful feature of \profound{} is that it can accept any segmentation map as long as it has the generic feature that the segments are non-zero integers and the sky is labelled as 0. This means it is perfectly possible to pass in a segmentation from another software package (e.g. \sex{}) and then use \profound{} to execute the dilation and flux convergence algorithms.}

There is trade off to be made between setting the initial detection threshold of the image higher and allowing a larger number of more aggressive dilation operations, but the key point is that at least three (by default) pixels must be identifiable as a segment above the detection threshold in order to even be dilated. The default parameters work well on a range of common survey imaging data (e.g.\ SDSS \citep{ahn14}, KiDS \citep{kuij15}, VIKING), so the need for large deviations from the defaults should be rare. In fact the settings related to the dilation operations are almost never altered in normal usage.

\begin{figure*}
	\includegraphics[width=\columnwidth]{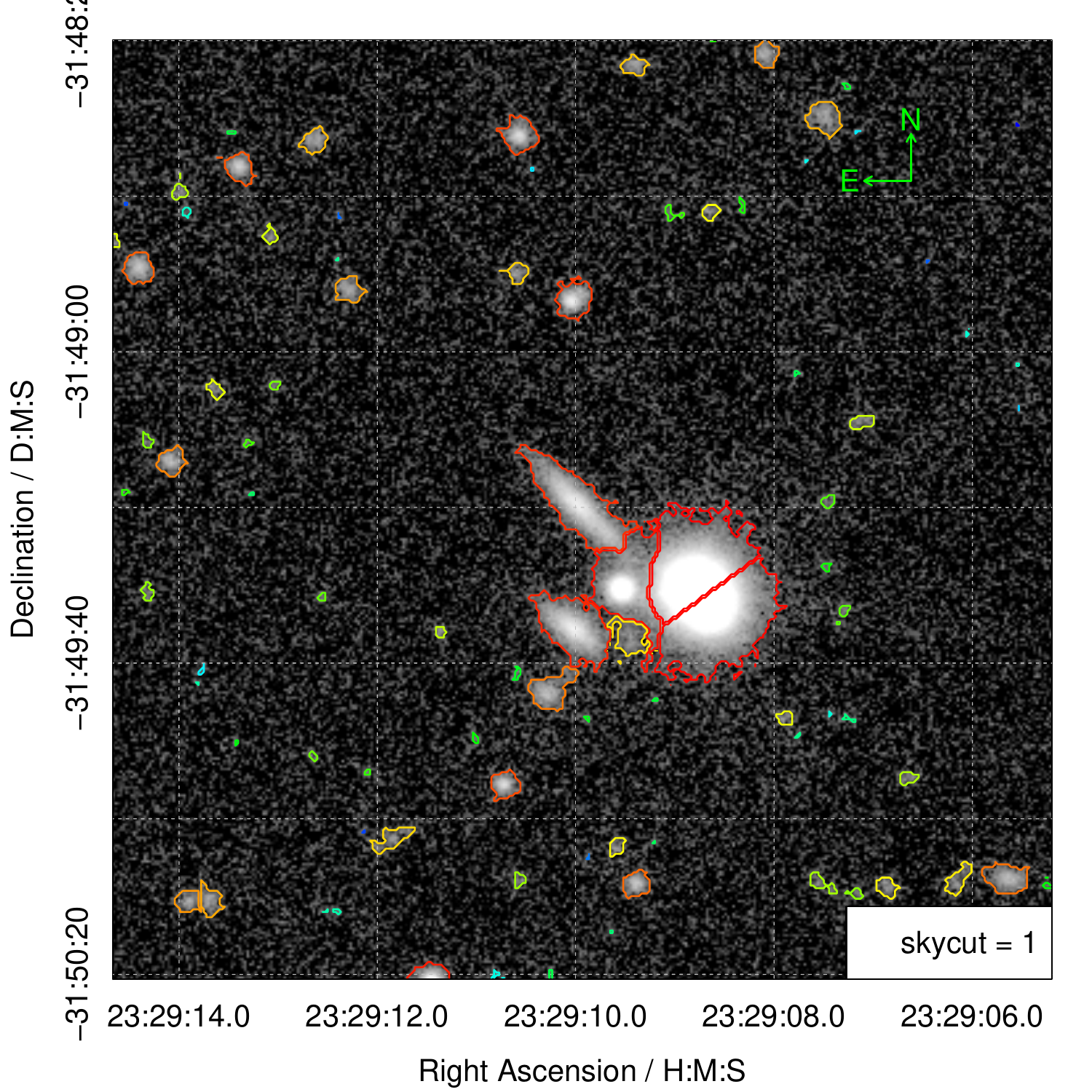}
	\includegraphics[width=\columnwidth]{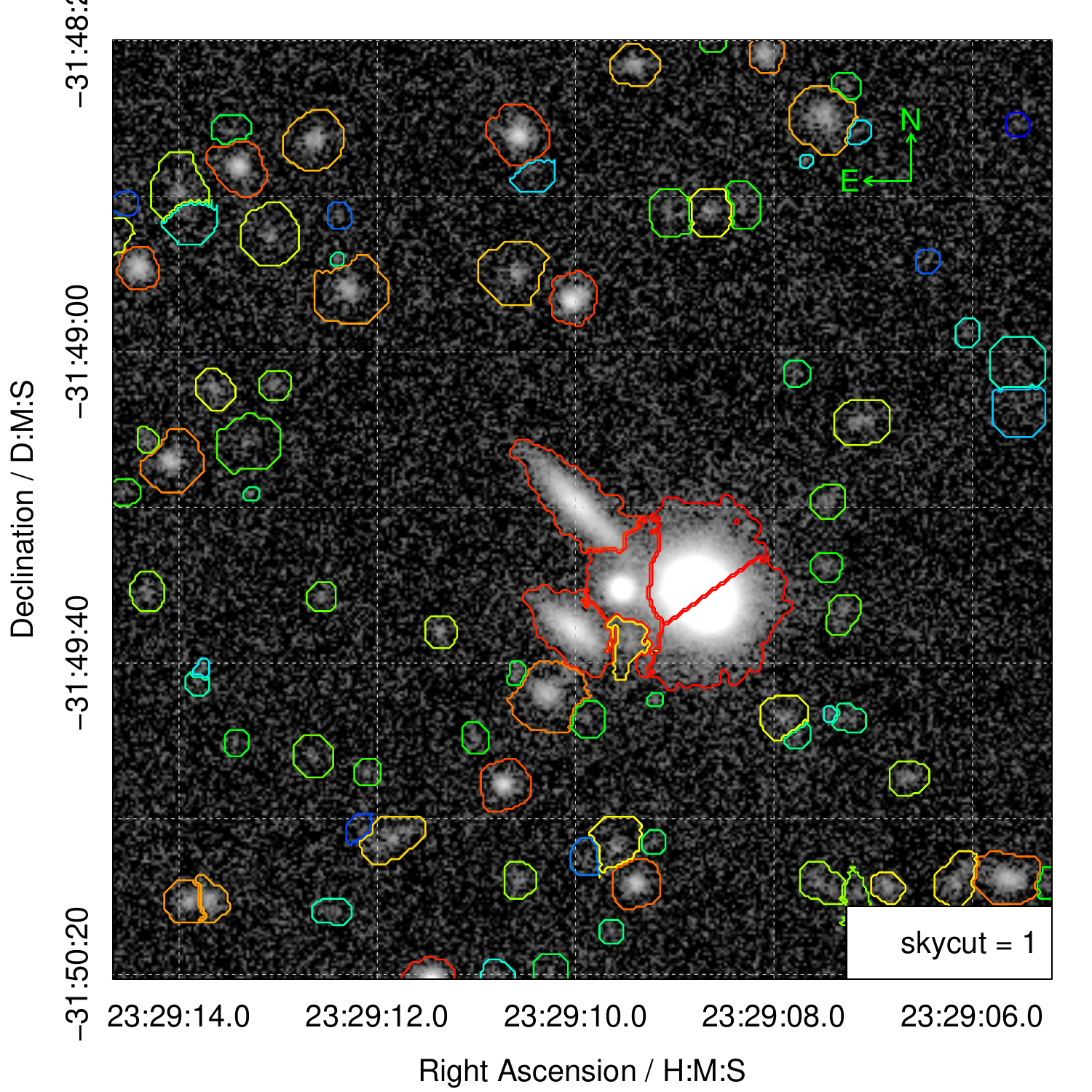}\\
    	\caption{An example of the default diagnostic output from the main \profound{} source extraction routine using the example VISTA Z-band data taken from the VIKING survey (as shown in Figure \ref{fig:example_image}). The left panel shows the identified flux (non-dilated) segments via multi-coloured contours (red showing the brightest sources, via green, through to blue showing the faintest). The right panel is similar, but the contours now represent the fully dilated and flux converged segments. The left panel is actually identical to the top-right panel in Figure \ref{fig:watershed_tol}, but it is repeated here to aid direct comparison of pre and post dilation.}
    	\label{fig:segim_orig}
\end{figure*}

Figure \ref{fig:segim_orig} shows the segments that define the bright initial components of the segmentation maps and the fully dilated segments for the same objects. Comparing the un-dilated segments (left panel) and the dilated segments (right panel) it is clear the bright stars do not dilate much before their flux converges, but fainter extended galaxies grow much larger in order to capture their converged share of flux. It is also notable that the geometries of the segments are kept largely intact during dilation.

\begin{figure}
	\includegraphics[width=\columnwidth]{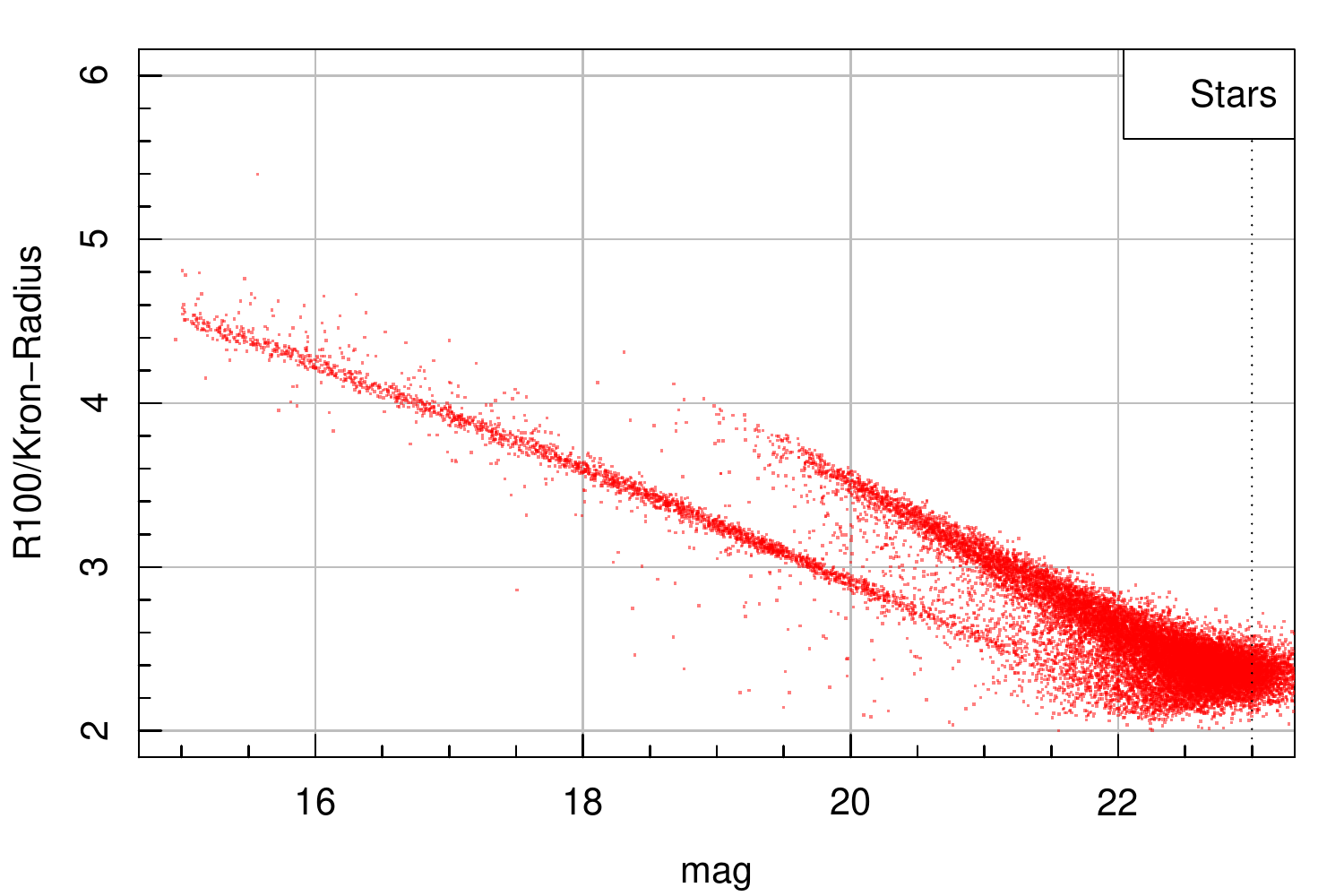}
	\includegraphics[width=\columnwidth]{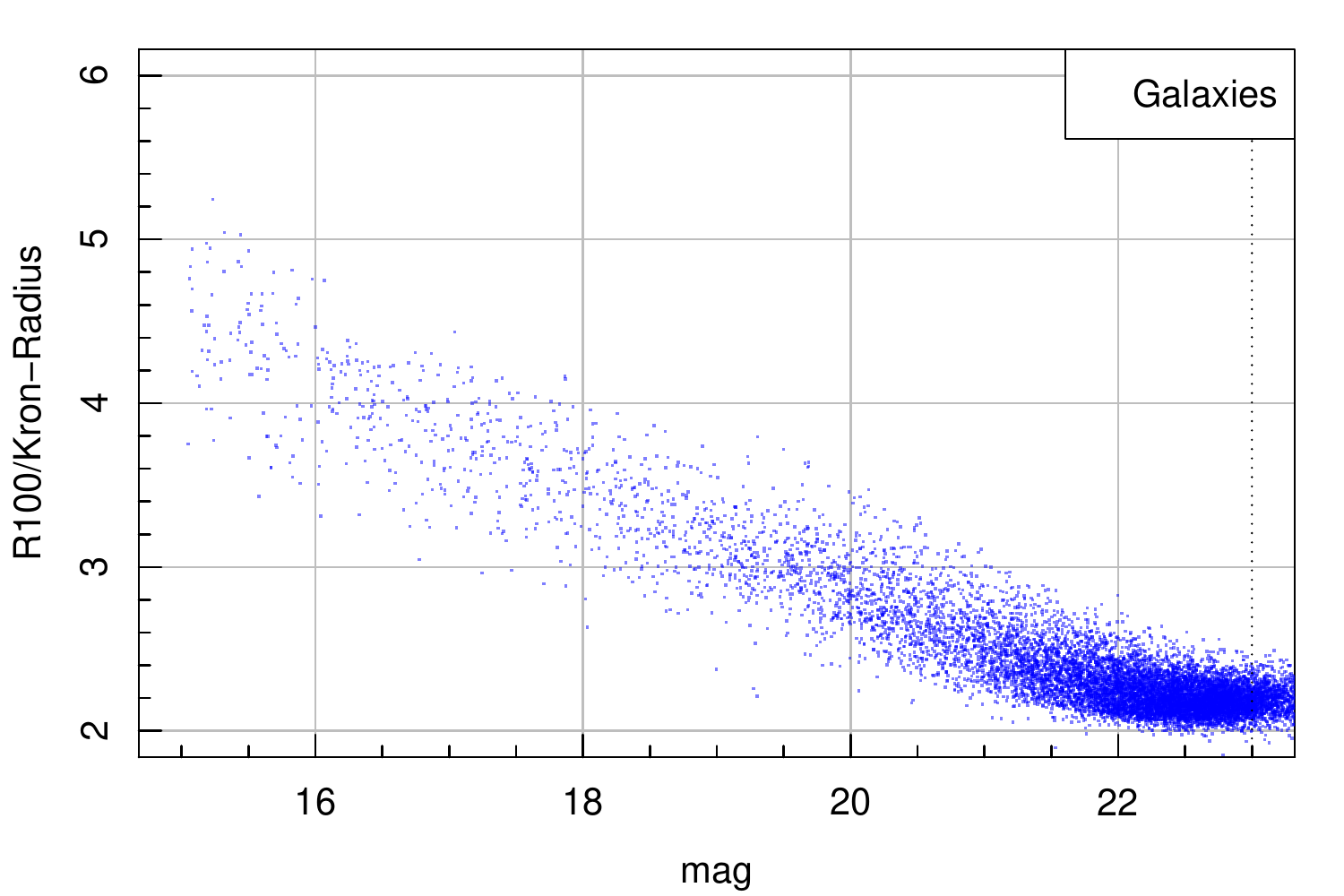}
    	\caption{The Kron to fully dilated \profound{} semi-major axis (R100) expansion factor for simulated stars and galaxies.}
    	\label{fig:R100oKron}
\end{figure}

Internally the dilated segments are used to compute a number of traditional geometric parameters such as effective size and ellipticity of the segment were it forced to be an ellipse. The approximate semi-major axis is output as `R100' in \profound{}, where the ratio between this and the standard Kron aperture (which is computed using the first order moments of the pixels in each segment) can be thought of as the expansion factor, which tends to be set to values between 2 and 4 when using aperture based photometric tools. Rather than being specified this is measured in \profound{}. Figure \ref{fig:R100oKron} shows the typical factors for a suite of simulated VIKING depth data (introduced and discussed in more detail later in this paper). It is clear there is an absolute lower limit near to two, and only the very brightest stars and galaxies require expansion factors beyond four. The discreteness seen for the stars is a consequence of the iteration process, where the thickest branch contains stars that require two dilations for convergence, as seen in Figure \ref{fig:segim_iters}.

\begin{figure}
	\includegraphics[width=\columnwidth]{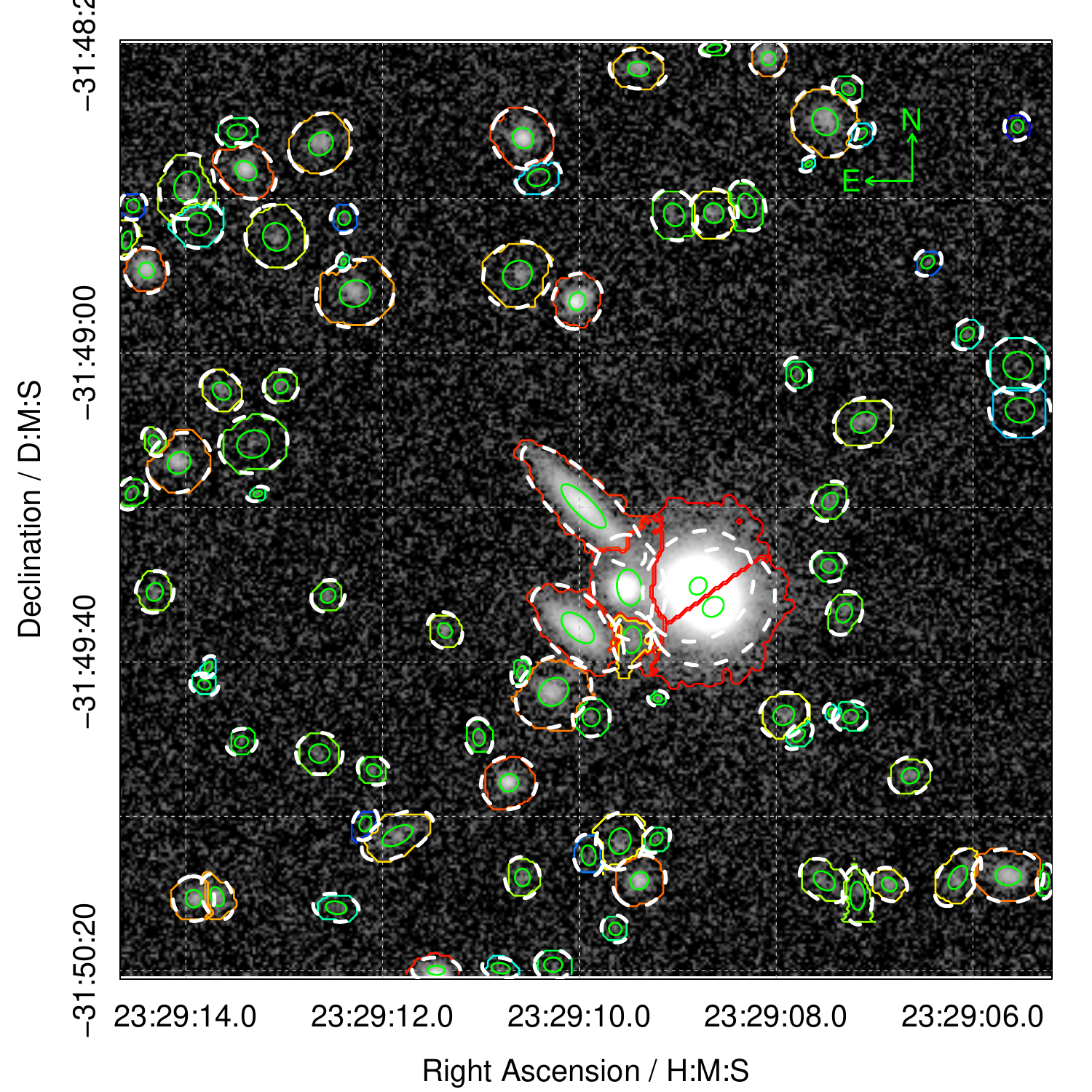}
    	\caption{Overlay of the intrinsic dilated segments (multi-coloured solid thin lines) and the inner Kron aperture (green ellipses) and approximated elliptical apertures that best describe the dilated segment geometries (thick white dashed ellipses).}
    	\label{fig:overlay}
\end{figure}

Figure \ref{fig:overlay} gives an idea of how similar the dilated segments and the approximated elliptical apertures are in practice. The thick dashed white ellipses would perfectly follow the multi-coloured segments if the relationship was perfect. In practice they trace fairly similar shapes and largely contain the same pixels, hence Figure \ref{fig:R100oKron} should give a broadly accurate impression of the true expansion factors required. Where this relationship breaks down is in the highly clustered (and therefore de-blended) regions, in particular in the example data the central cluster of objects that \profound{} has de-blended through multiple saddle points. Since these segments will have necessarily non-elliptical geometry there is no good elliptical approximation for the segment regions. Since the elliptical apertures are not used to compute photometry within \profound{} this is not a concern to people using the software in an isolated fashion, but it does mean that care has to be taken when attempting to apply these apertures using other programs \citep[e.g.\ \lambdar][]{wrig16}.

\subsection{Photometric Properties}
\label{sec:methods_photometry}

Once a segmentation map has been constructed a separate internal function is used to calculate a large suite of photometric properties and segment flags. Internally this is achieved by associating all pixels with their respective segments, looping through each segment, extracting the relevant pixels for the current segment, and computing photometric properties using just the pixels flagged as belonging to a particular segment. To achieve this process rapidly \profound{} uses the highly efficient {\sc data.table} package\footnote{https://CRAN.R-project.org/package=data.table} which is optimised for subsetting and processing on large datasets.

The main photometric properties returned are listed in Table \ref{tab:photo} (ignoring the various types of flags, alternative definitions of some quantities, and error columns). These outputs are sufficient to provide reasonable initial guesses for a single \sersic{} profile fit of a galaxy using \profit{}, which was the main initial design focus for \profound{}. The assumption is that a multi-component profile will be built in complexity iteratively, i.e.\ in order to achieve good inputs for a two component fit you would first start with a simple single component fit.

\begin{table*}
	\centering
	\caption{A selection of photometric properties computed in \profound{}}
	\label{tab:photo}
	\begin{tabular}{l | l} 
		\hline
		Name & Description\\
		\hline
		segID & Segmentation ID \\
  		uniqueID & Unique ID \\
  		xcen & Flux weighted x centre \\
  		ycen & Flux weighted y centre \\
		RAcen & Flux weighted Right Ascension centre \\
  		Deccen & Flux weighted Declination centre \\
		flux & Total flux in the segment in ADUs \\
		mag & The flux in the segment scaled to a magnitude \\
		flux\_reflect & Total flux in the segment in ADUs scaled by flux missing under a segment rotation\\
		mag\_reflect & The flux\_reflect in the segment scaled to a magnitude \\
		N50/90/100 & The number of pixels containing 50\% / 90\% / 100\% of the flux \\
		R50/90/100 & Approximate elliptical semi-major axis containing 50\% / 90\% / 100\% of the flux \\
		SB\_N50/90/100 & Mean surface brightness containing 50\% / 90\% / 100\% of the flux \\
		con & The concentration, defined here as R50/R90. \\
		axrat & Axial ratio of ellipse \\
		ang & Orientation angle of the ellipse \\
		\hline
	\end{tabular}
\end{table*}

\profound{} does not execute sophisticated algorithms to de-blend flux between neighbouring sources/segments. In comparison, methods presented in \citet{irwi85} (simultaneous maximum likelihood of sources) and \citet{bert96} (heuristic flux division via symmetry expectation) do. In order to extract truly optimal photometry the identified blended sources should be further run through generative modelling software such as \profit{} \citep[e.g.][see Section \ref{sec:combine} for an example]{kelv12}. However, \profound{} includes a number of schemes to flag and improve photometry in complex and confused regions. Improved flux reconstruction is possible using the `rotated' flux output for sources (`flux\_reflect' and `mag\_reflect' in Table \ref{tab:photo}), which assumes flux symmetry of sources. This option determines a plausible amount of missing flux by rotating each segment about the central pixel and determining how much flux does not fall onto a mirrored segmented pixel. Running on the VIKING example data, the median difference between the raw segment flux and the rotated version is $\sim$0.1 mag (i.e. the sources get brighter), so for most sources the difference is {\change fairly small. The scale of these differences is also in line with the kind of flux differences seen when attempting full profile modelling with \profit{} (see Section \ref{sec:combine}).}

\begin{table*}
	\centering
	\caption{Flags and diagnostics computed in \profound{}}
	\label{tab:flag}
	\begin{tabular}{l | l} 
		\hline
		Name & Description\\
		\hline
		Nedge & Number of edge segment pixels that make up the outer edge of the segment \\
		Nsky & Number of edge segment pixels that are touching sky \\
		Nobject & Number of edge segment pixels that are touching another object segment \\
		Nborder & Number of edge segment pixels that are touching the image border \\
		Nmask & Number of edge segment pixels that are touching a masked pixel \\
		edge\_frac & Fraction of edge segment pixels that are touching the sky \\
		edge\_excess & Ratio of the number of edge pixels to the expected number given elliptical geometry \\
		flag\_border & A binary flag telling the user which image borders the segment touches \\
		flag\_keep & A Boolean flag suggesting whether the object should be kept based on the flux growth and iterations\\
		\hline
	\end{tabular}
\end{table*}

\profound{} also returns a large number of flags that can help decide whether there are issues with the segmentation, or other potential issues with the photometry (like lying very close to a frame edge). Table \ref{tab:flag} is a summary of the major flags that are generated when the photometric properties of an image are computed. The edge\_frac flag is particularly useful for flagging well isolated objects, and when this drops much below 1 it is a sign that future galaxy profiling might be compromised by nearby sources unless effort is made to execute a model that also accounts for these sources.

\subsection{Colour Photometry}
\label{sec:methods_colour}

\begin{figure*}
	\includegraphics[width=5.8cm]{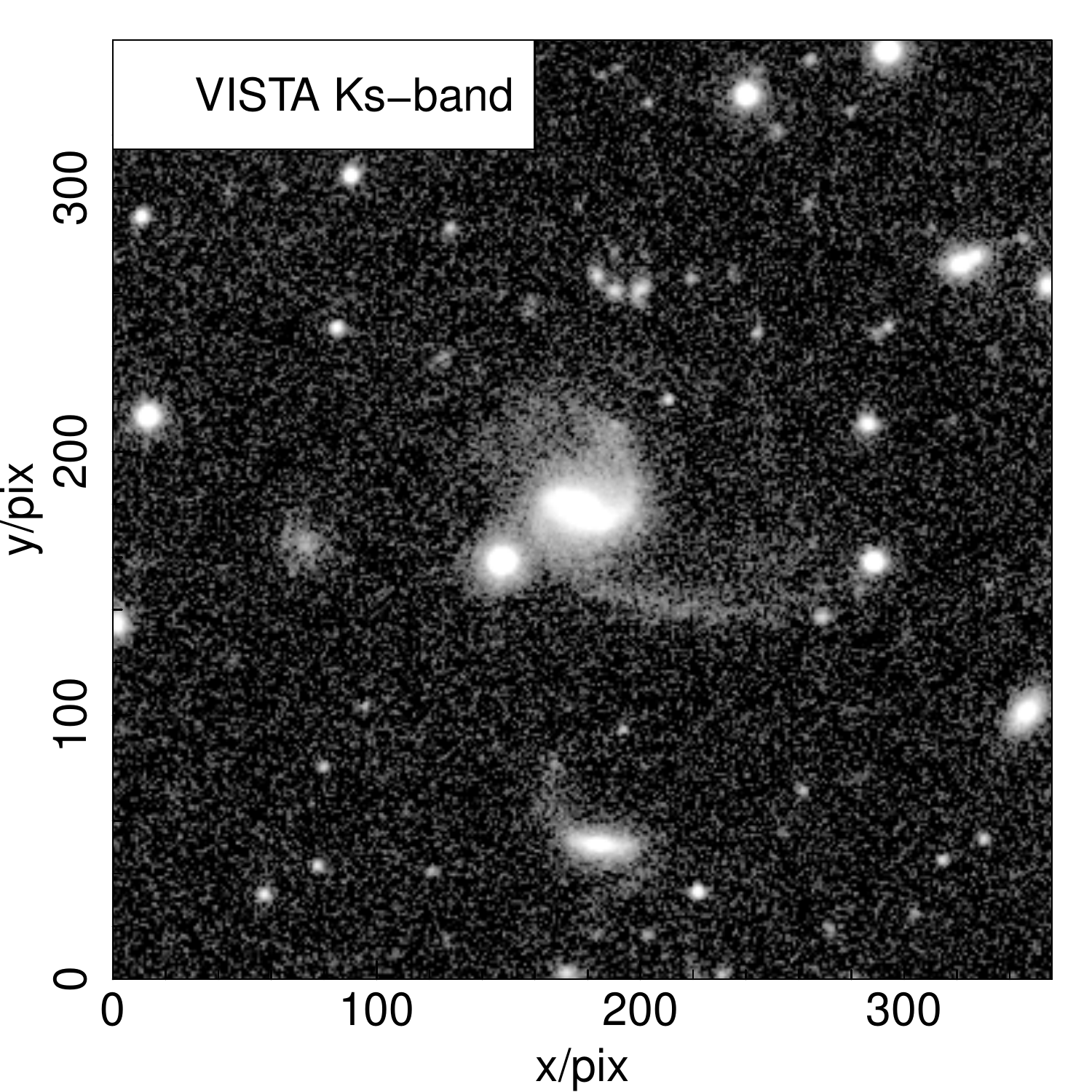}
	\includegraphics[width=5.8cm]{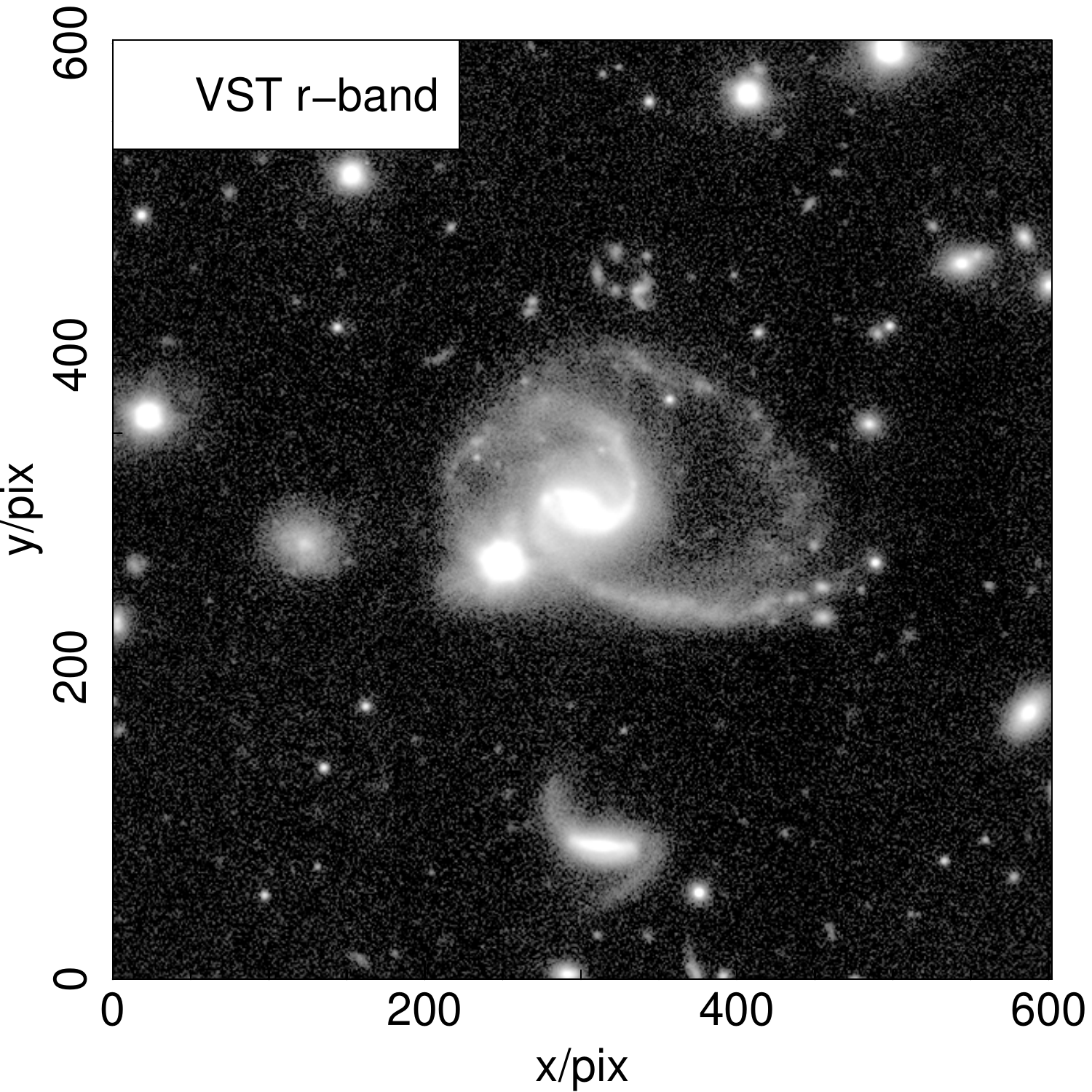}
	\includegraphics[width=5.8cm]{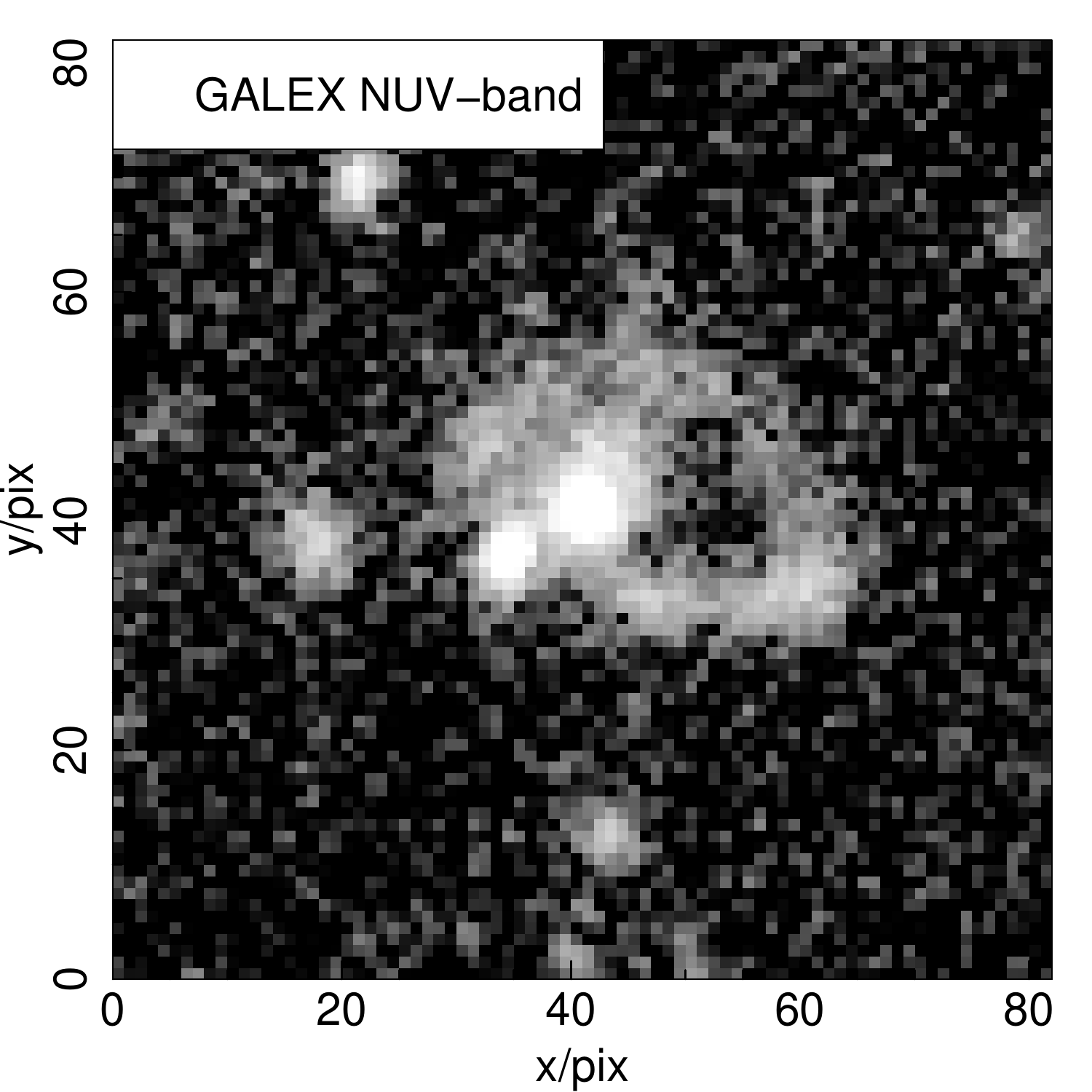}\\
	\caption{The \profound{} package comes with some highly WCS mismatched images of the same galaxy (GAMA galaxy ID G202627). The left panel shows VISTA Ks-band (pixel scale 0.339 asec/pix). The middle panel shows VST $r$-band (0.2 asec/pix). The right panel shows GALEX NUV-band (1.5 asec/pix). Each image has dimensions of 2'$\times$2'.}
	\label{fig:ImWarp}
\end{figure*}

Colour photometry is a catch-all term that usually refers to measuring fluxes in multiple bands using common apertures, where the differences in fluxes can be mapped onto traditional optical colours for visualization purposes. The higher level \profound{} function that provides the main interface to both source extraction and photometric analysis offers a few routes to extracting colour photometry. The top level interface can take a number of inputs that bypass internal routines to calculate them, e.g.\ segmentation maps, sky maps and object masks. Since the image provided does not have to be the same as the one used to create the segmentation map provided, it is easy to extract forced photometry by passing into \profound{} a pixel matched image that was observed using a different filter to the detection band, and turning off the option to dilate the segments.

\begin{figure}
	\includegraphics[width=\columnwidth]{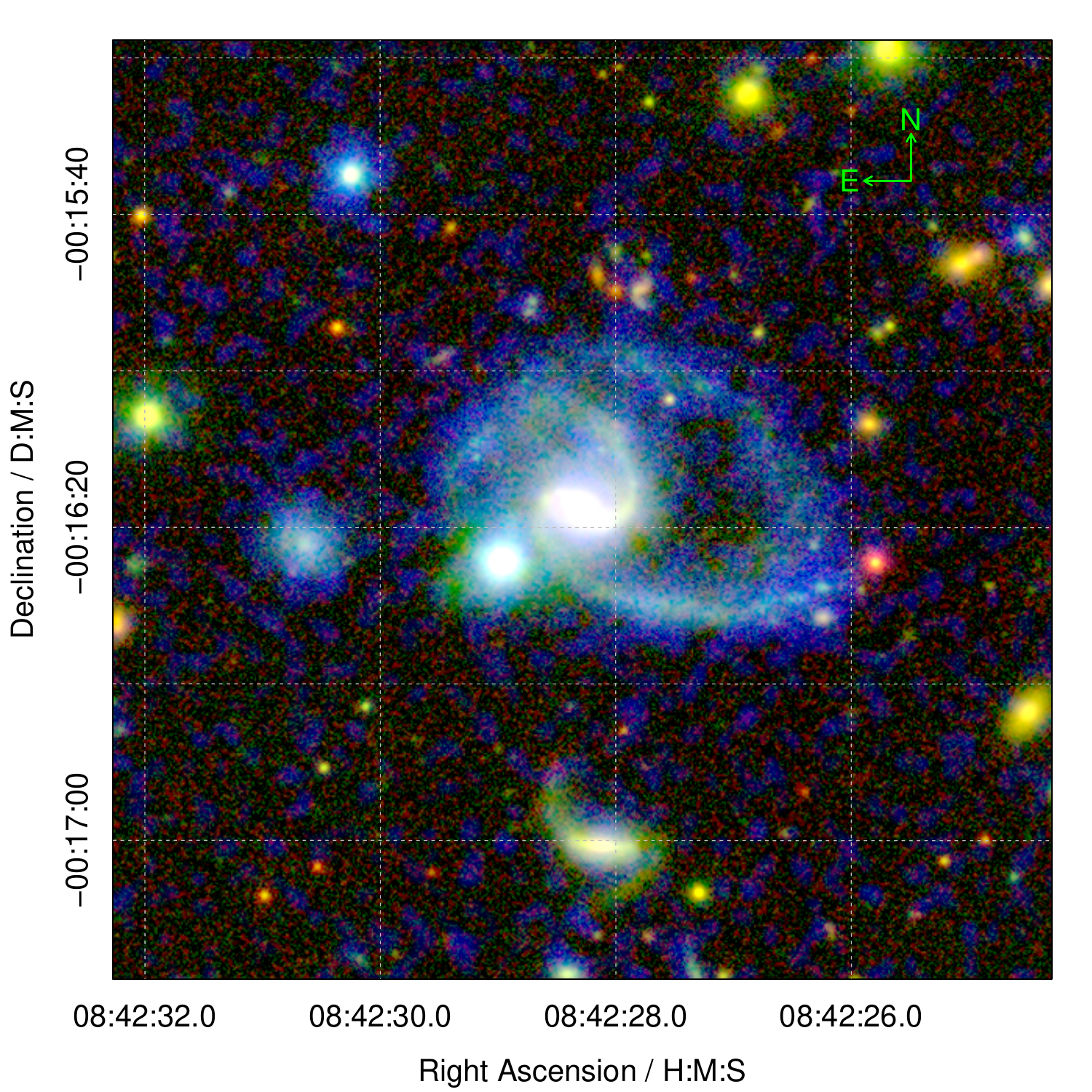}
    	\caption{RGB composite image of mismatching projections combing VISTA Ks-band, VST $r$-band and GALEX NUV-band (as shown in Figure \ref{fig:ImWarp}) for the red / green / blue channels respectively.}
    	\label{fig:KrNUV}
\end{figure}

A more advanced method, useful in the case where the point spread function (PSF) varies significantly between bands, is to allow the segments to dilate to best extract converged flux in the target band. This is referred to as soft colour photometry in \profound{}, and is a sensible method to extract total photometry across multiple bands with different depths and seeing conditions. In general the highest image quality and/or deepest band should be used as the detection image. Additionally, \profound{} includes routines to optimally stack images based on $S/N$ properties, in which case a stacked image can be used as the detection image (this was used for the \uv{} data analysis presented in Section \ref{sec:UV}).

One issue is that the above methods require the images to be pixel matched. The \profound{} package includes routines to remap images onto a common target Tan-Gnomonic world coordinate system (WCS), should the images not have a common projection. To do this \profound{} uses the image warping routines available in the {\sc Cimg} image analysis library, and accessible in {\sc R} via the {\sc imager} package.

An example of this being applied to mismatching VISTA Ks-band \citep[pixel scale 0.339 asec/pix][]{edge13}, Visual Survey Telescope $r$-band \citep[VST; 0.2 asec/pix][]{kuij15} and Galaxy Evolution Explorer NUV \citep[GALEX; 1.5 asec/pix][]{mart05} can be seen in Figure \ref{fig:ImWarp}. The remapping allows us to make a coordinate matched RGB colour image shown in Figure \ref{fig:KrNUV}, where the VISTA Ks-band and GALEX NUV-band data are remapped onto the WCS of the VST $r$-band data. The upsampling conserves flux, and by default uses bilinear interpolation (bicubic or nearest pixel, and forward or backward mapping, are also options).

\begin{figure*}
	\includegraphics[width=5.8cm]{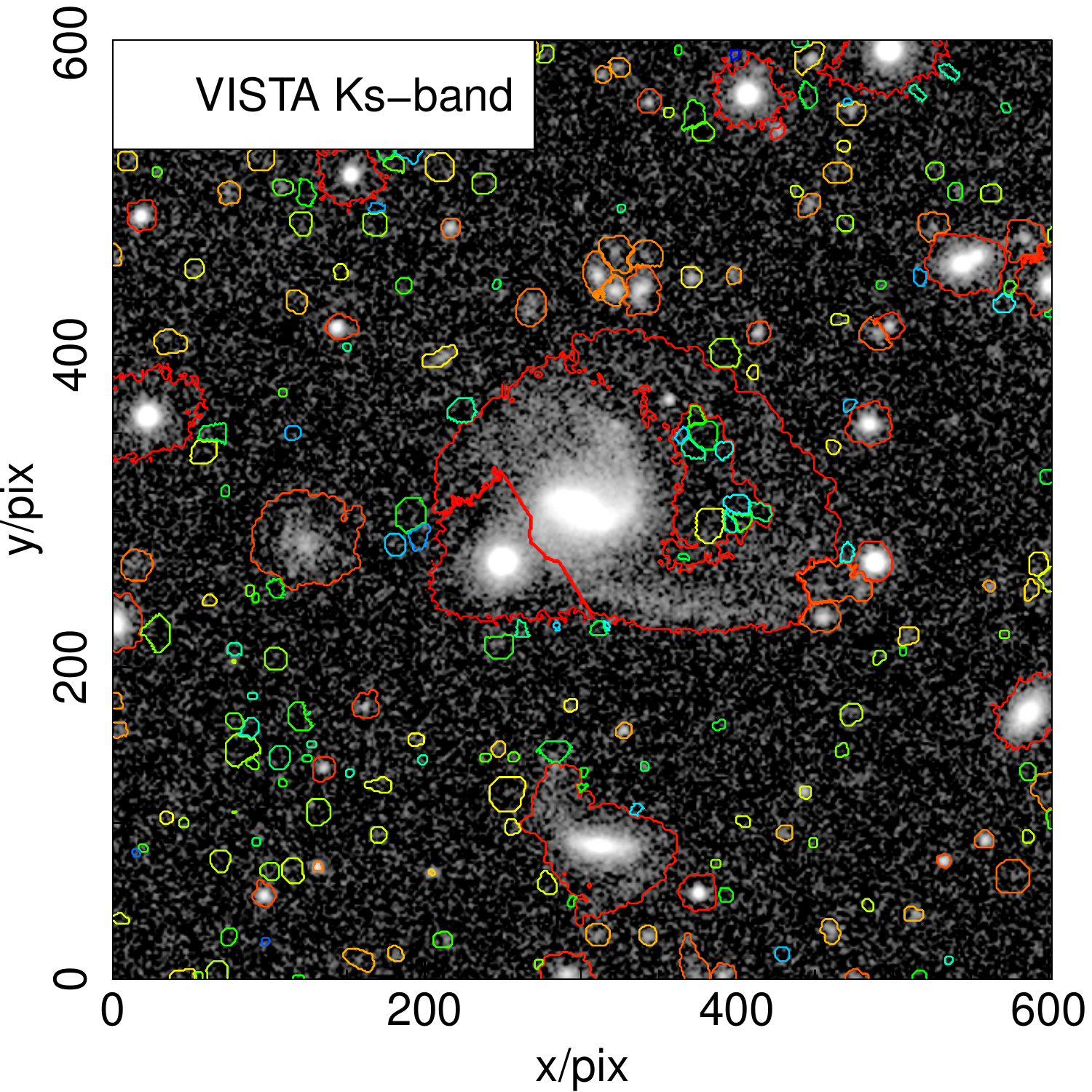}
	\includegraphics[width=5.8cm]{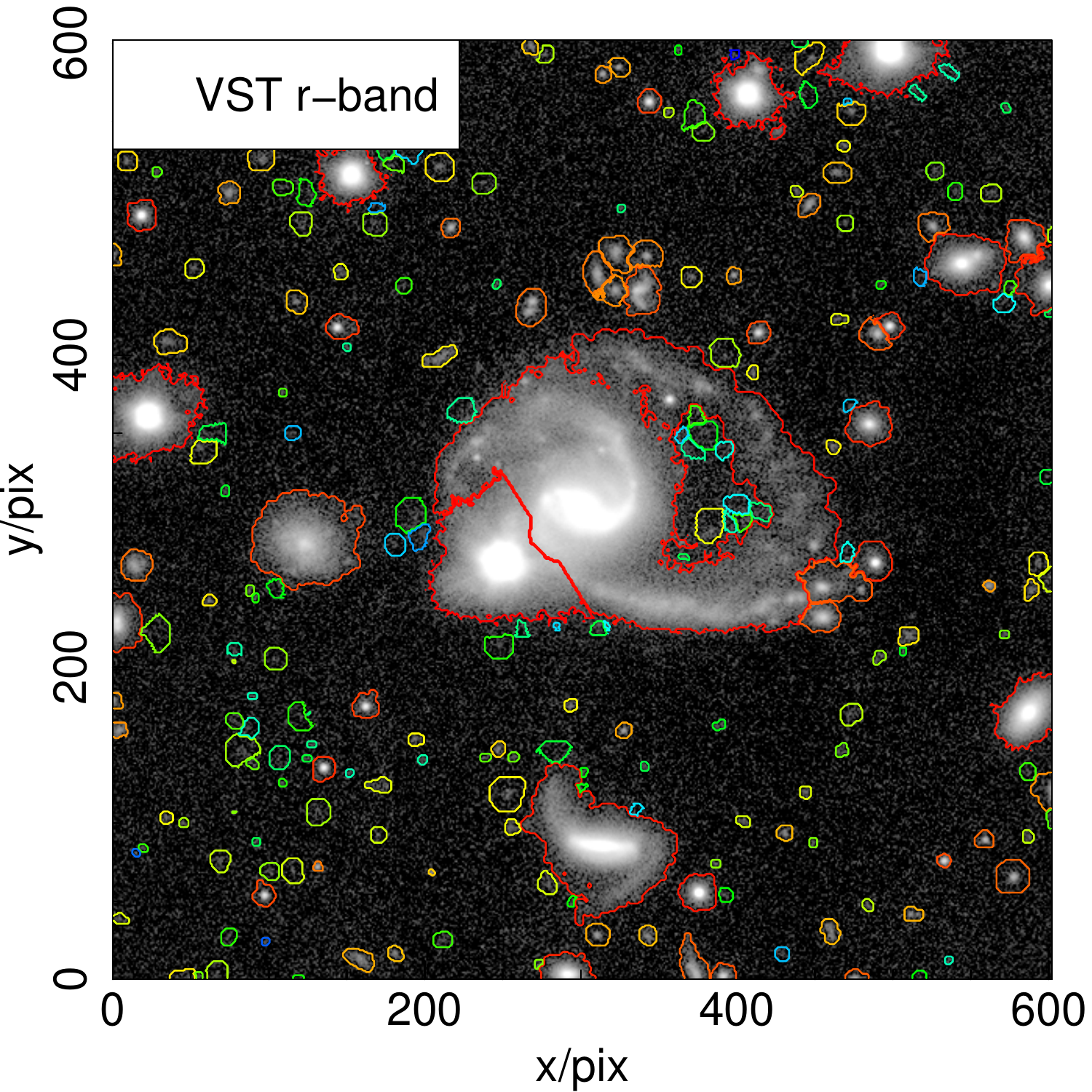}
	\includegraphics[width=5.8cm]{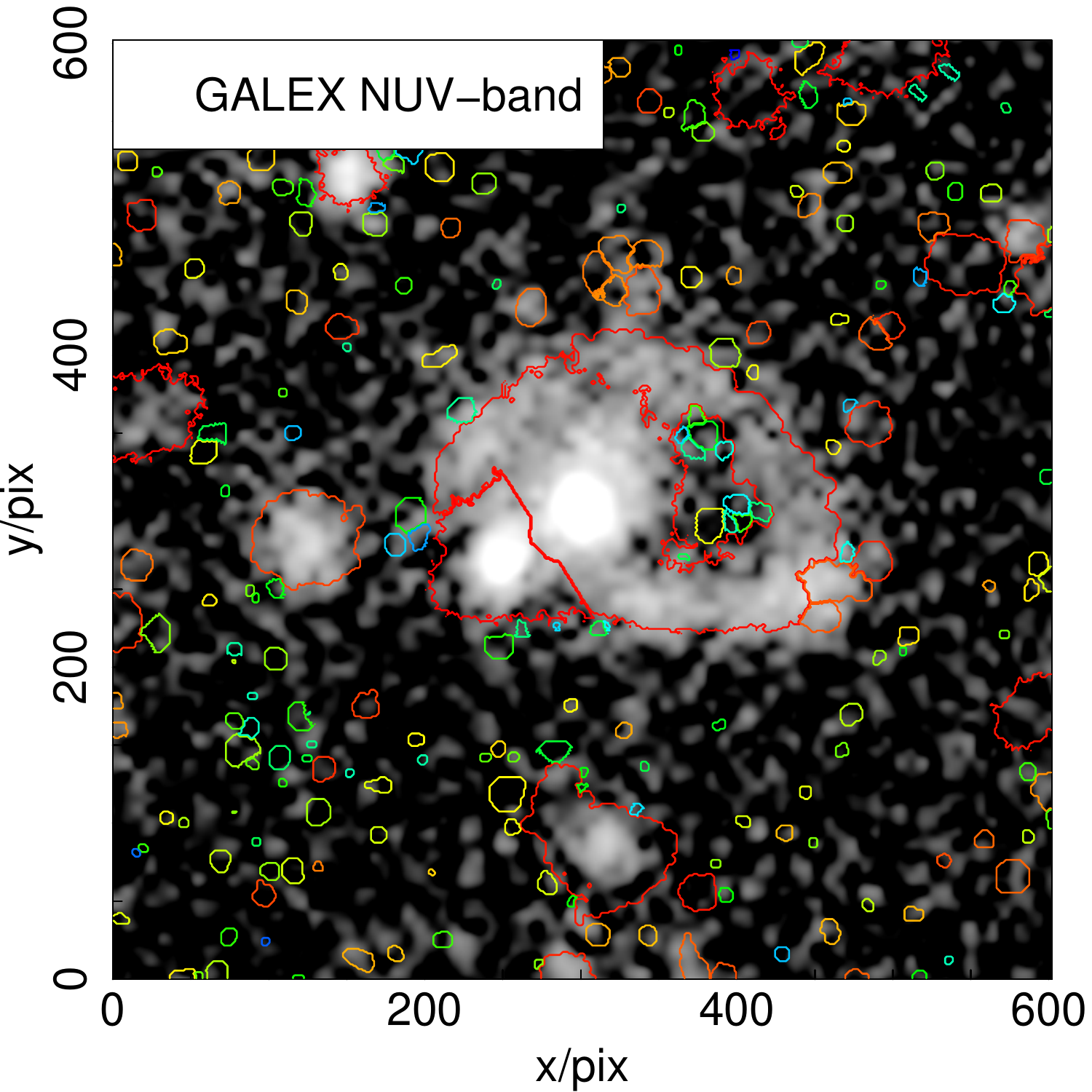}\\
	\caption{Schematic view of the \profound{} segments defined using a common WCS system. In this case we use the WCS scheme from the VST $r$-band image as seen in Figure \ref{fig:ImWarp}. The left panel shows the segments overlaid on the VISTA Ks-band. The middle panel shows the segments overlaid on the VST $r$-band they were defined with. The right panel shows the segments overlaid on the GALEX NUV-band. Each image has dimensions of 2'$\times$2'. It is notable that some of the GALEX NUV flux extends outside of the segment. \profound{} can capture this additional flux if it is allowed to further dilate the provided segment (this is the default mode).}
	\label{fig:ImWarppro}
\end{figure*}

Using these remapped images it is simple to extract matched segment photometry by applying segments extracted from a detection band (in this case the VST $r$-band) on the remapped target bands. Figure \ref{fig:ImWarppro} shows what this extraction might look like internally, where it is clear the majority of the GALEX NUV flux associated with the central spiral galaxy is enclosed by the VST $r$-band derived segments. Some of the fainter NUV features would be very hard to extract blindly, but should produce a reasonable signal when extracted in such a forced manner.

The above approaches are the best solution to extracting matched aperture total photometry. However, more accurate spectral energy distributions (SEDs) are often derived from using only the brighter inner parts of sources, since the outskirts usually include lower $S/N$ pixels which will act to increase the scatter between colours. Popular approaches to such colour photometry include fixed apertures (e.g.\ 2 asec circles placed on each source) or computing colours with a certain surface brightness level. The latter is achieved trivially in \profound{}, since one of the objects returned is the segmentation map before dilation, i.e.\ the segmentation that only includes pixels that are independently above some surface brightness threshold. An example of such a bright segmentation map is shown in Figure \ref{fig:segim_orig}, where the pixels identified are much brighter than the fully dilated segmentation map shown in the right panel of Figure \ref{fig:segim_orig}. By using this map and turning off the option to dilate the segments high surface brightness colours can be extracted, leading to less scatter in the colour photometry measured (as we see in detail later using \uv{} data).

Finally, a hybrid colour is possible, where the bright segment map is used as the starting point in each target image, but the segments are allowed to dilate independently in each target band in order to achieve converged flux. This is often similar in output to just applying the dilated segmentation map to each target band without allowing for independent dilation, however it can be a sensible option if the detection image has a much larger PSF than one or more of the target bands, where the dilated aperture might be much larger than is actually necessary, and includes a large quantity of sky pixels which will lower the fidelity of the photometry extracted.

\section{Combining \profound{} with \profit{}}
\label{sec:combine}

As discussed above, much of the design philosophy behind \profound{} was to provide good quality inputs for \profit{} galaxy profiling. This includes: careful sky subtraction; sigma map construction via estimating the local sky-RMS map; good quality segmentation maps for extended sources and accurate initial conditions for the profile parameters to use when fitting with the \profit{} engine.

Here we use these various elements of \profound{} to prepare the example VIKING data for a large multi-component fit\footnote{for full details see the `Complex Fit' vignette at http://rpubs.com/asgr/.}. Running \profound{} in default mode, but with the `boundstats' option turned on, creates all the outputs we need. In this example we aim to fit the central group of objects that have touching segments.

\begin{figure*}
	\profound{} estimates for the model components
	\includegraphics[width=15cm]{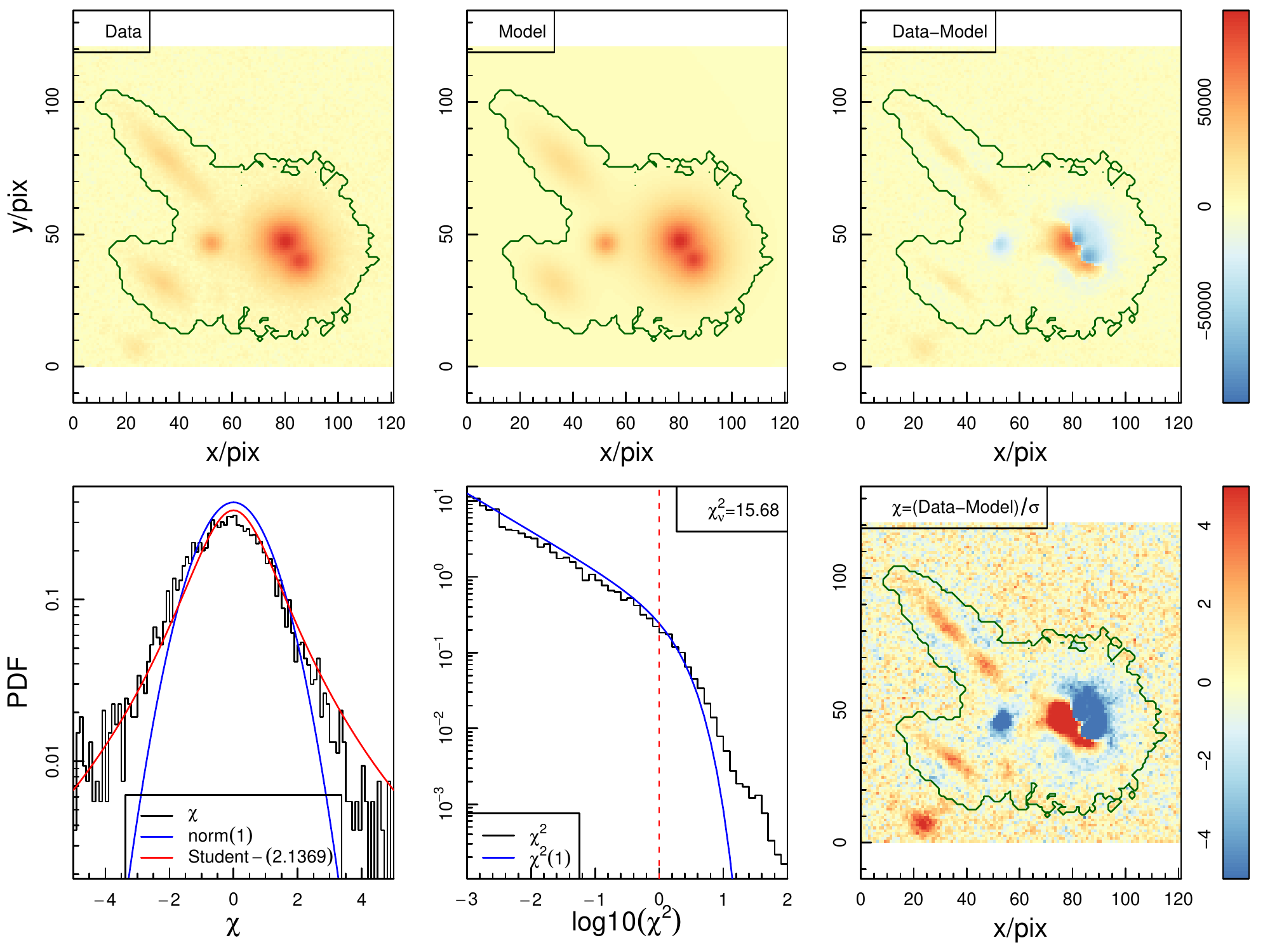}\\
	\profit{} optimised estimates for the model components
	\includegraphics[width=15cm]{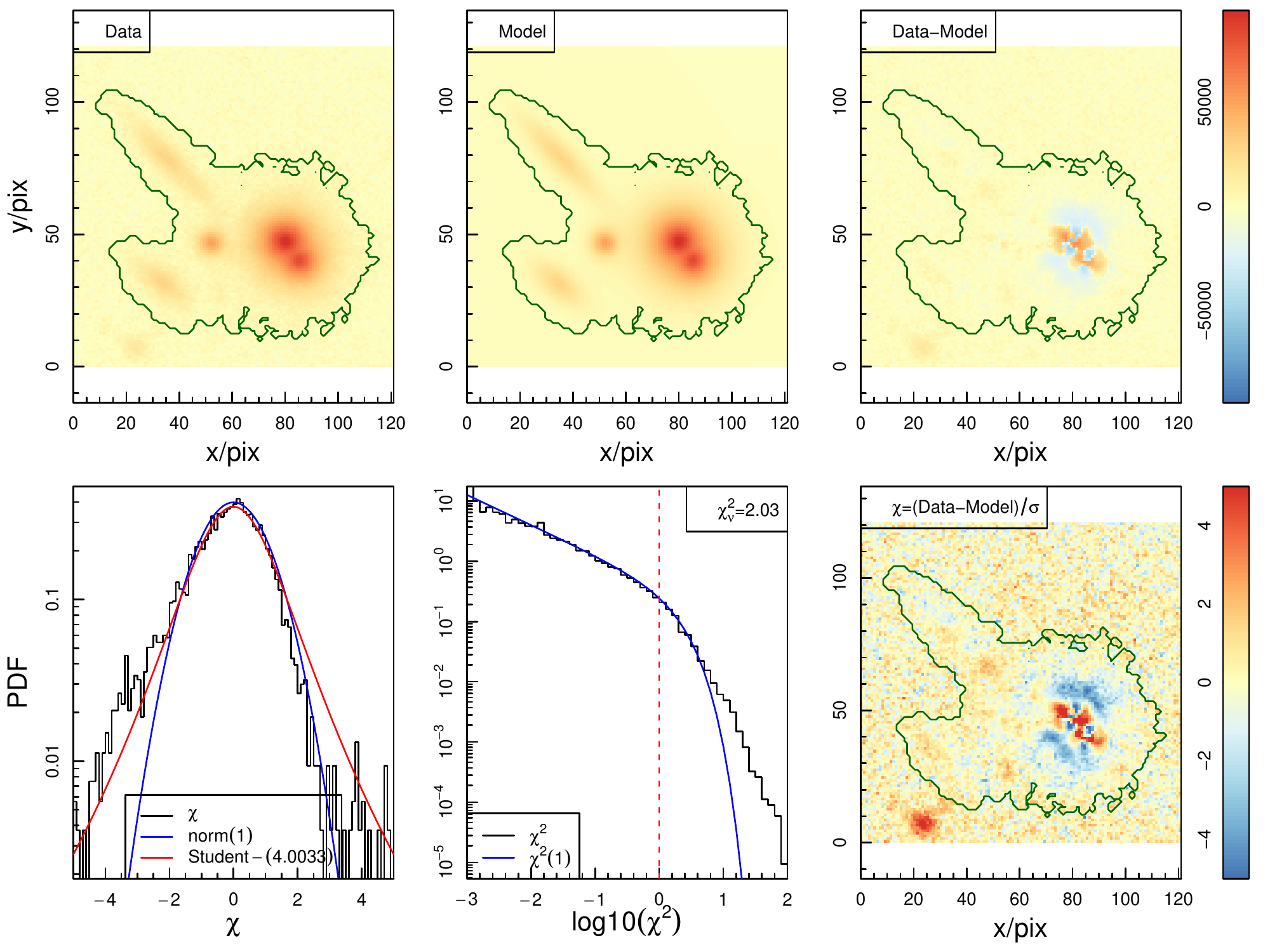}\\
    	\caption{Example of fitting a group of confused sources by combining the initial sky subtraction, segmentation map, sigma map and parameter estimates from \profound{} with the profiling and fitting engine of \profit{}.}
    	\label{fig:profound_profit}
\end{figure*}

Figure \ref{fig:profound_profit} shows the main results, where the top panels show the initial parameter estimates taken from \profound{}, and the bottom panels show the \profit{} BFGS optimised solution. This particular fit uses \sersic{} profiles for the two visually extended elliptical sources, and Moffat profiles for the remaining three objects which have PSF like characteristics. The Moffat PSF required for convolving the image and modelling the point-sources was estimated from fitting a number of isolated bright PSFs.

Overall we can achieve an excellent and rapidly converged simultaneous fit using this approach. The two extremely bright stars in the bottom-right of the fit region have some residual structure, but the relative flux residuals are generally small (a fraction of a percent). The other three sources are very well modelled, in particular the two extended elliptical sources that have profiles very close to pure exponential discs.

The input \profound{} and output \profit{} source fluxes all agree within 0.4 mag, and the differences are typically less than 0.1 mag. Even the two close bright stars are well estimated by \profound{}, with the differences being 0.07 mag fainter and 0.16 mag brighter. The total modelled flux in the fit region agrees to better than 1\% with the extracted \profound{} flux. The $R_e$ for the \sersic{} index is also well estimated, the \profound{} input increasing by $\sim$5\% for both of the clearly extended sources.

To achieve a rapidly converged fit for such complexes of objects the key requirements are that the objects are well segmented along flux saddle-points, and the initial estimates for the fluxes and sizes are within a factor of $\sim2$ of the correct solution. \profound{} can easily achieve these requirements if it is run in a sensible (usually near to default) manner. This suggests that using \profound{} combined with \profit{} as part of an automated pipeline is a reasonable goal for future large scale decomposition tasks.

\section{Simulations}
\label{sec:sims}

\subsection{\profit{} Simulations}

To check the performance of \profound{} we ran a number of tests using simulated data that was designed to approximately mimic the sky variations, sky RMS, PSF, magnitudes, sizes and profiles of a mixture of stars and galaxies in a typical VIKING survey frame using the image generation capabilities of \profit{} to make the simulated frames. {\change Table \ref{tab:param} details the various parameters and sampling ranges used when generating the simulated images. Code to replicate very similar types of simulations are also available online for user experimentation\footnote{see Simulated Images vignette at http://rpubs.com/asgr/}.} In order to ensure we have sources going much deeper than the noise threshold of the VIKING data, the selection of magnitudes, source sizes and shapes was approximated using the deeper \uv{} survey recovered using \profound{} (see Section \ref{sec:UV}). The PSF full width half max (FWHM) was chosen to be at the poorer extreme of values observed for VIKING: $\sim$1.7 asec, or 5 pixels at the 0.339 asec/pixel scale of VIKING images \citep{edge13}.

\begin{table}
	\centering
	\caption{The simulation setup parameters}
	\label{tab:param}
	\begin{tabular}{l | l} 
		\hline
		Simulation Parameter & Value\\
		\hline
		N simulations		&	100		\\
		Sky				&	0		\\
		Sky RMS			&	10		\\
		magnitude zero point &	30		\\
		x image pixels		& 	1000		\\
		y image pixels		& 	1000		\\
		PSF FWHM		&	5 pixels	\\
		\hline
		N stars			&	200		\\
		Magnitude range	&	15--23	\\
		Magnitude Power-law slope	&	1.5	\\
		\hline
		N galaxies			&	200		\\
		Magnitude range	&	15--23	\\
		Magnitude Power-law slope	&	2.0	\\
		$R_e$ Poisson $\lambda$	&	5	\\
		\sersic{} index range	&	1--4		\\
		Axial-ratio range	&	0.3--1	\\
		Boxiness range		& 	-0.3--0.3	\\
	\end{tabular}
\end{table}

\begin{figure*}
\centering
	\includegraphics[width=\columnwidth]{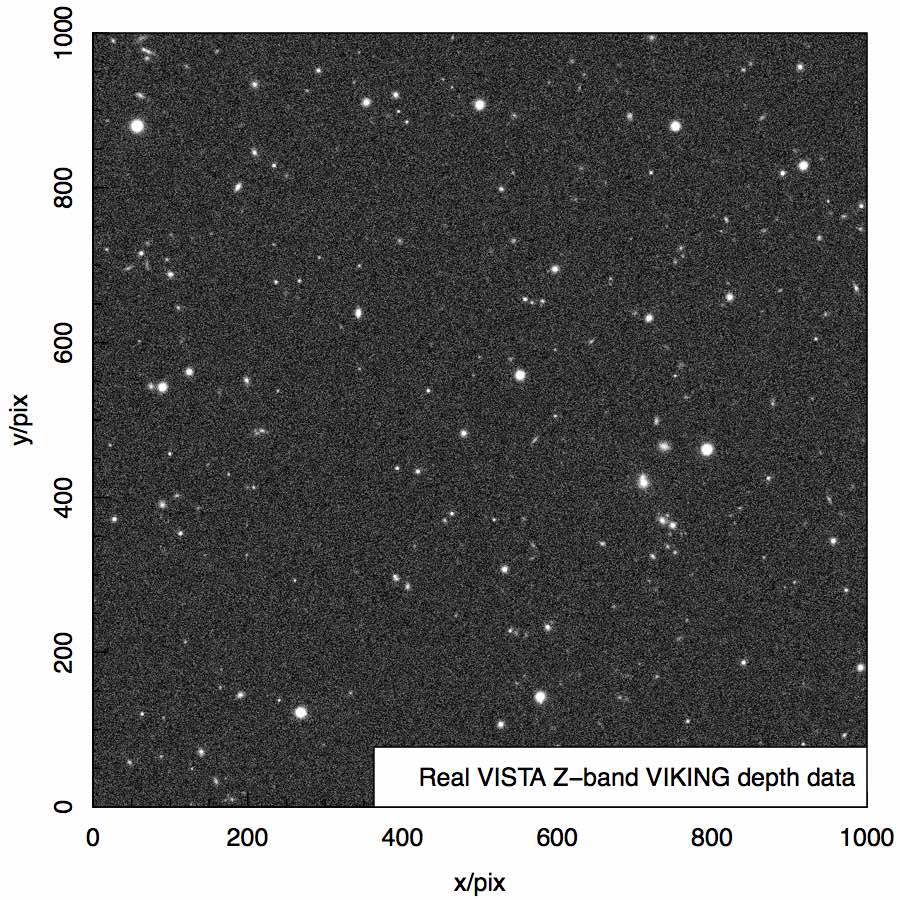}
	\includegraphics[width=\columnwidth]{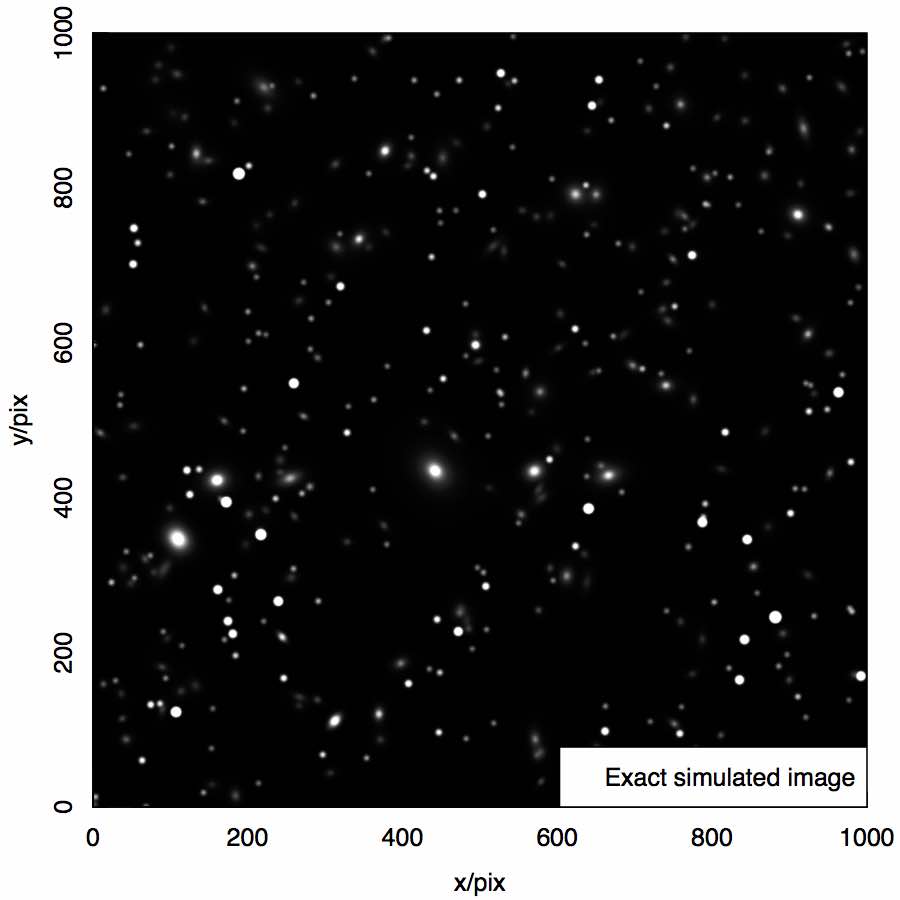}\\
	\includegraphics[width=\columnwidth]{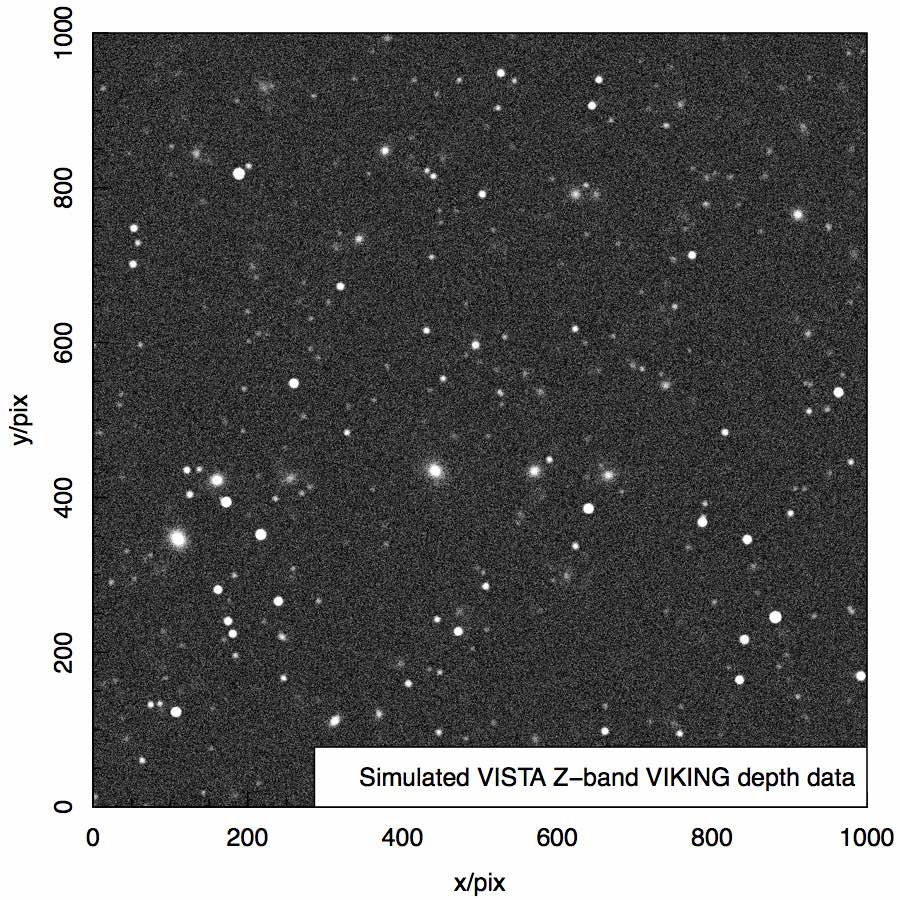}
	\includegraphics[width=\columnwidth]{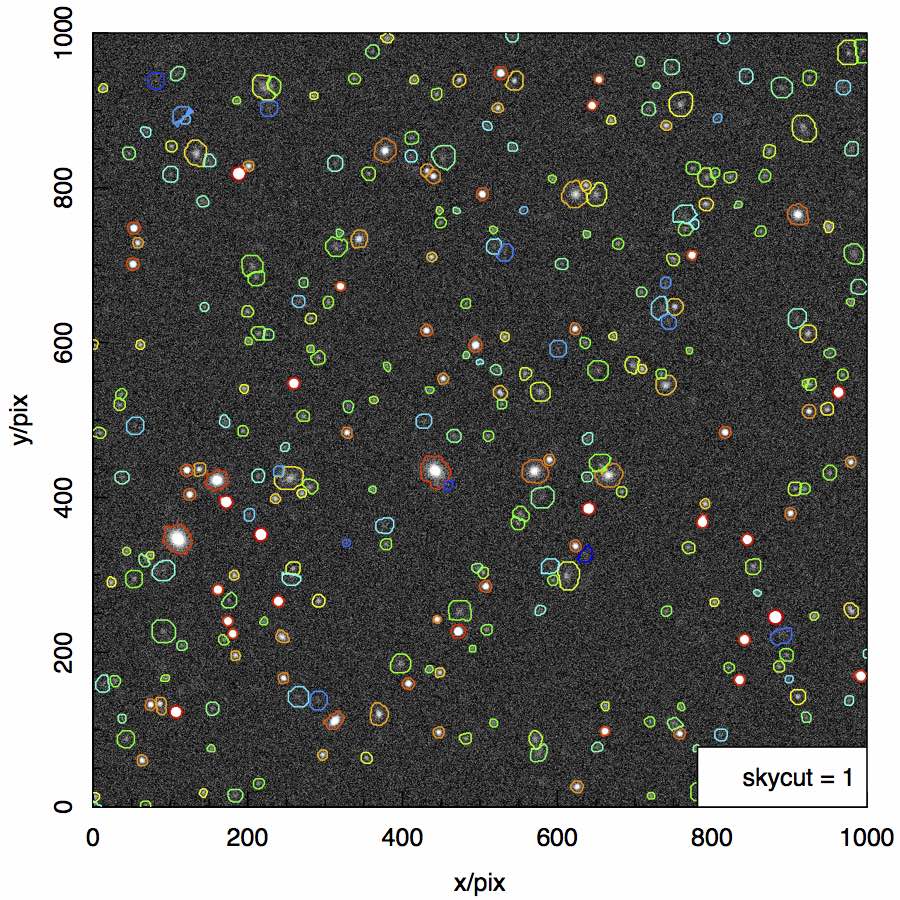}\\
	\caption{Top-left is a real VISTA Z-band VIKING depth frame. Top-right panel shows the pure model created by \profit{} with no noise added. Bottom-left panel adds realistic VIKING-like noise along with a variable sky background. The bottom-right panel shows the extracted \profound{} segments.}
	\label{fig:sims}
\end{figure*}

Figure \ref{fig:sims} is an example of the noise free model generated by \profit{} for the typical distribution of sources used for our simulations, the addition of noise and sky (estimated from \profound{} on real VIKING data) to create more realistic looking data, and the extraction of sources using \profound{}. In this example almost all of the stars were successfully recovered, and around half of the galaxies are detected, the rest being below the $1\sigma$ surface brightness threshold of the VIKING survey.

\profound{} was run with close to default settings, with the difference being the de-blend tolerance (how many sky RMS deviations to use to de-blend sources during the watershed stage) was set to 1. These settings mimic the qualitatively optimal settings we have determined for processing VIKING and \uv{} data with \profound{} (which in turn informed the majority of the default \profound{} settings), but the broad results are quite robust to small changes in these settings.

One hundred 1k$\times$1k frames were generated randomly with \profit{} model stars and galaxies and extracted with \profound{}, with 200 stars and 200 extended galaxies generated per frame. This produces a final catalogue of 40k stars and galaxies generated. This is used to compute the false-positive and true-positive rates for stars and galaxies, and also quantify the measurement biases in \profound{} compared to the intrinsic sources. The latter is important since one of the main design aims of \profound{} (along with better sky subtraction and flux converged segmentation maps) is to create reasonable initial conditions for \profit{} fits. These do not need to be perfect, but it helps to be reasonably close to the global maximum likelihood in order to speed up the fitting time. Based on our experience with the simulations presented in \citet{robo17}, a factor of two in flux and/or size is a reasonable starting point for efficient convergence (in detail this statement is clearly algorithm dependent, and \profit{} offers no particular restriction on the optimization routine, giving out-the-box access to over 100).

\begin{figure*}
\centering
	\includegraphics[width=\columnwidth]{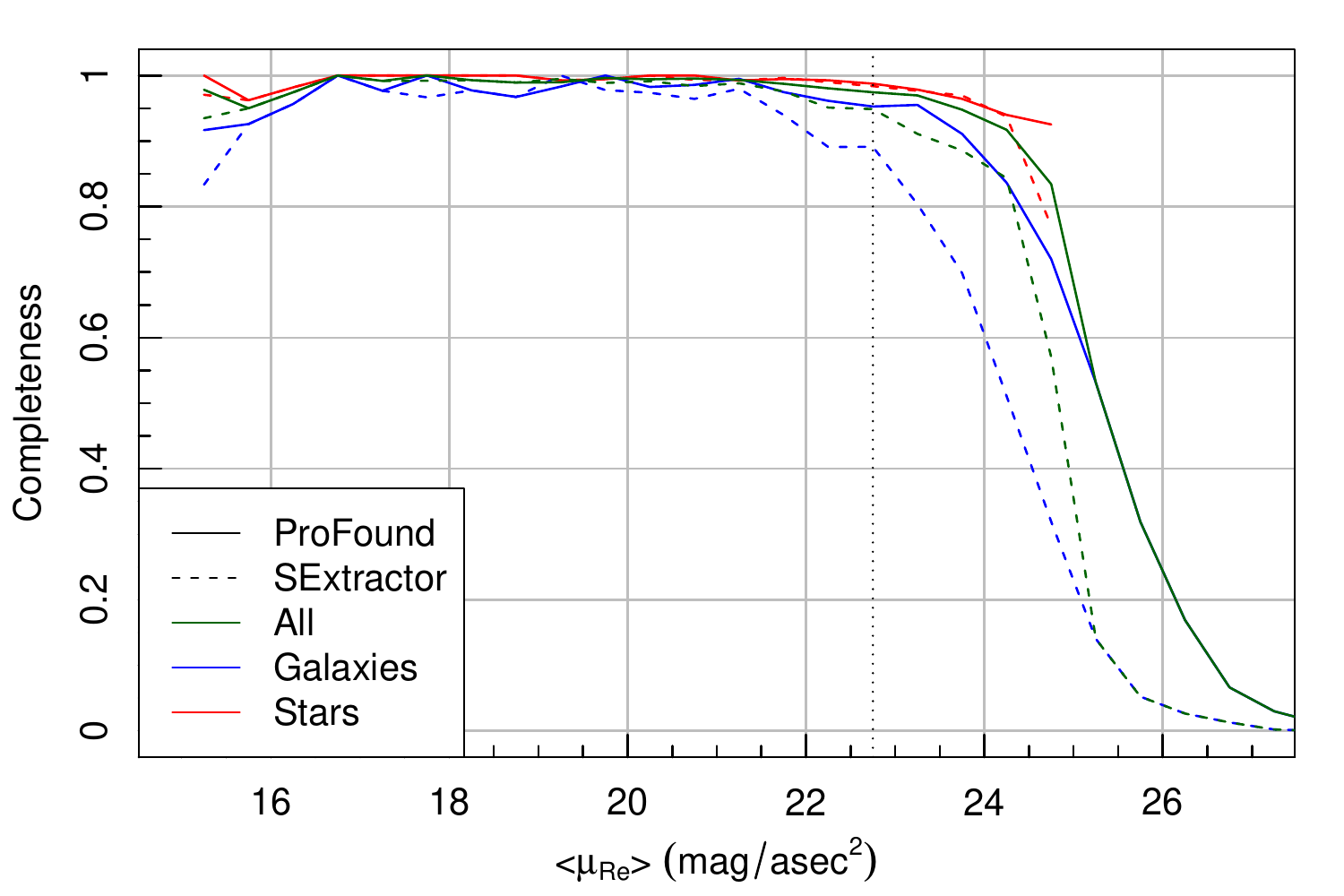}
	\includegraphics[width=\columnwidth]{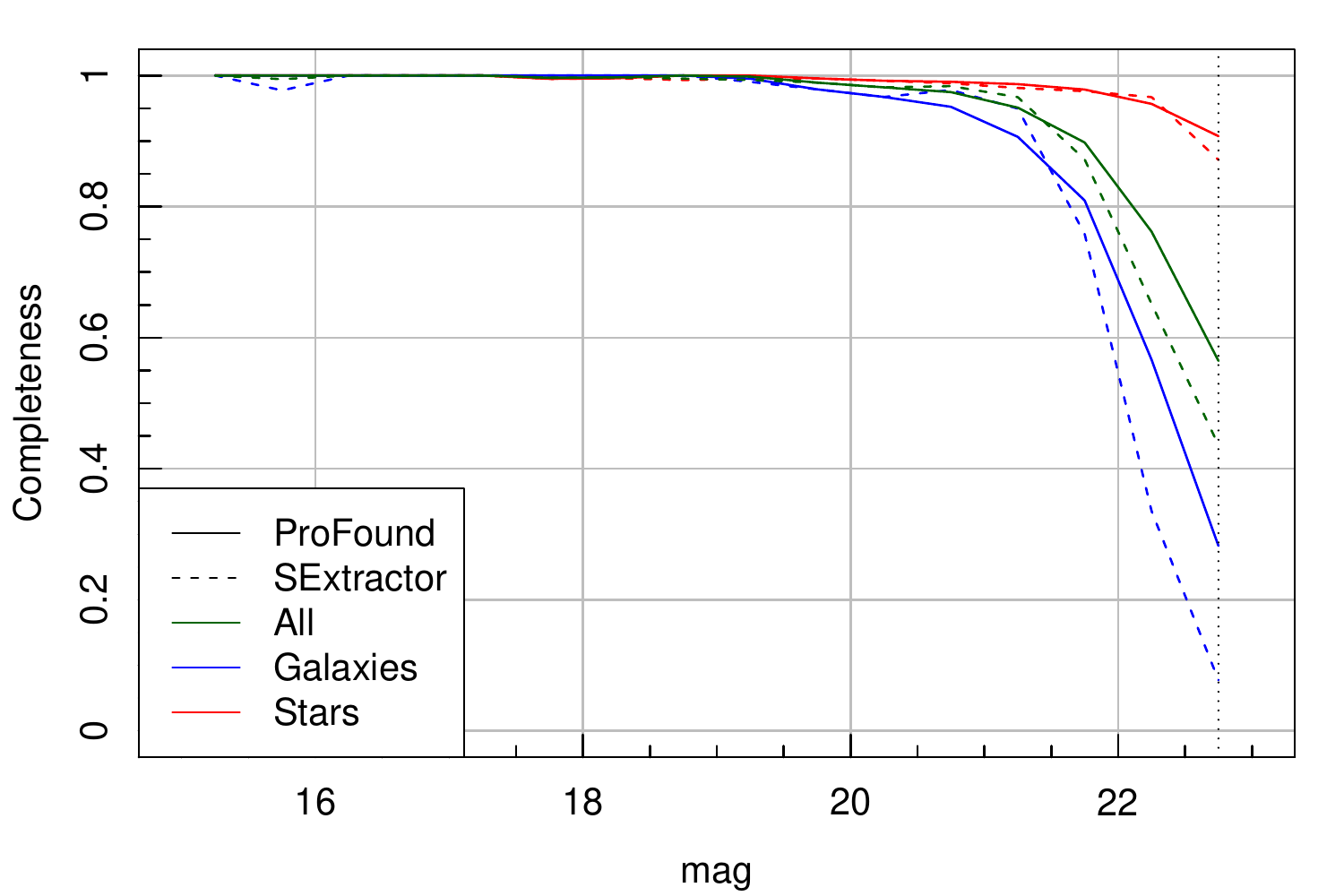}\\
	\includegraphics[width=\columnwidth]{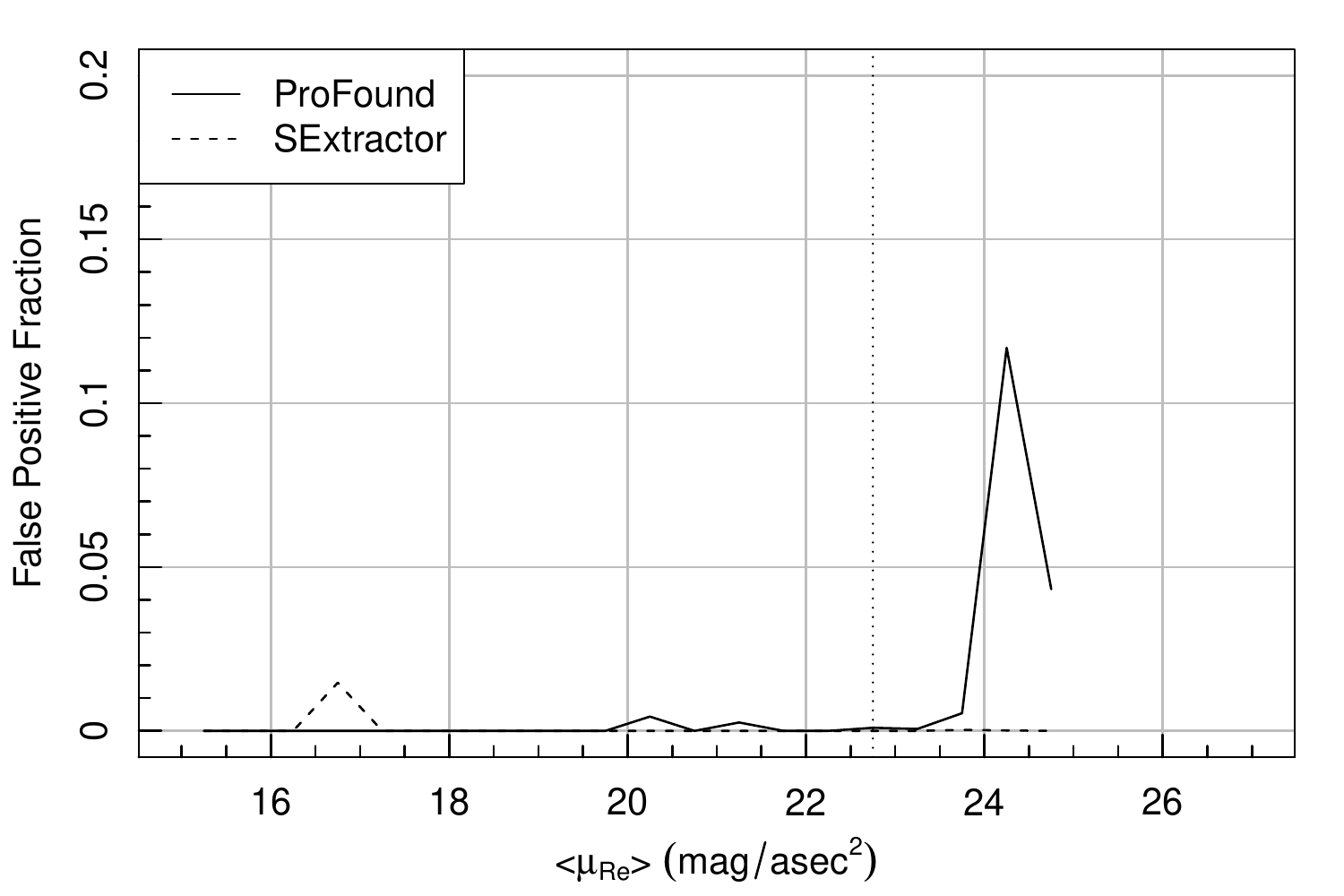}
	\includegraphics[width=\columnwidth]{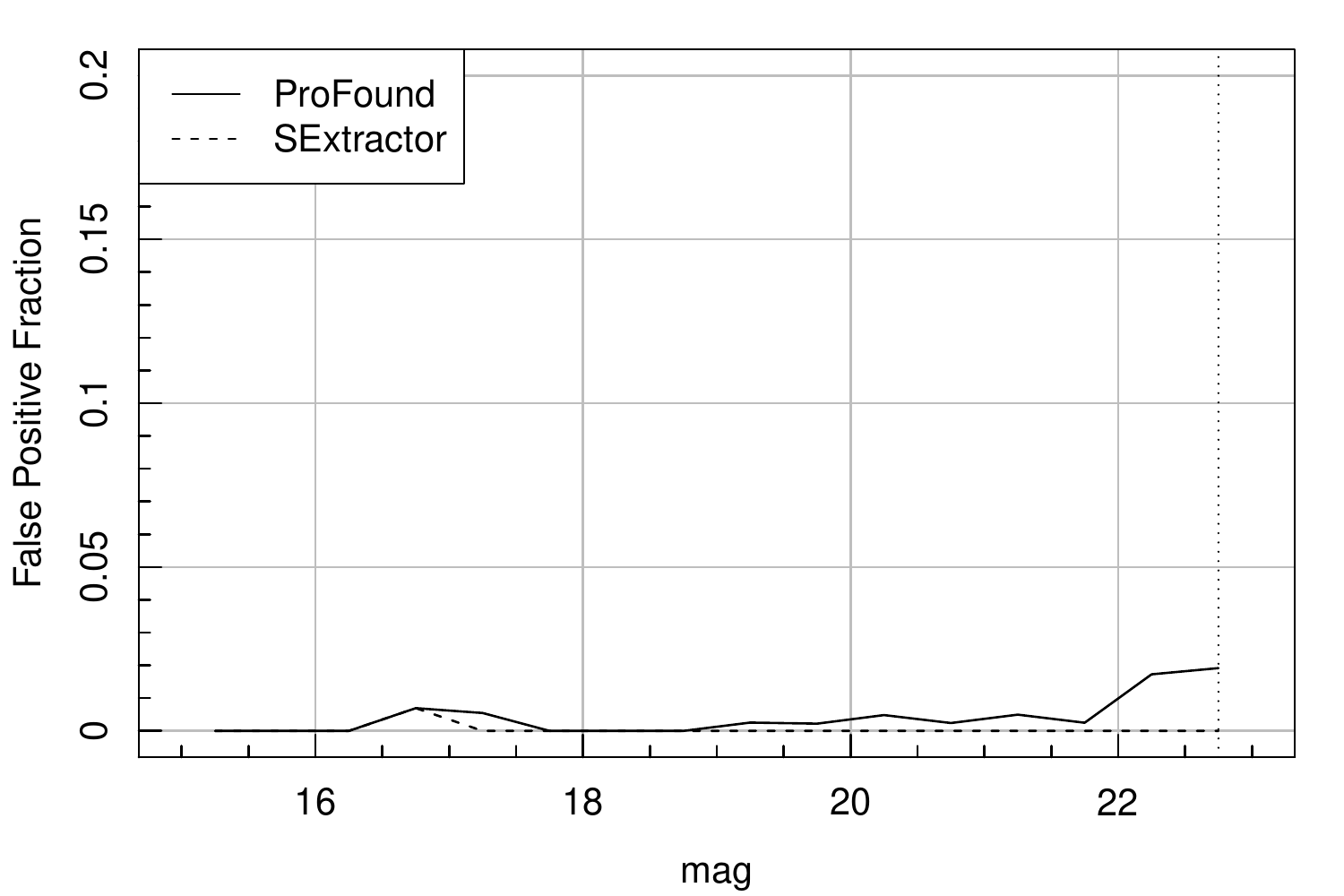}\\
	\caption{Top-panels show the detection completeness of stars and galaxies. Bottom-panels show the false-positive rate of spurious detections in \profound{}. Left-panels are as a function of mean surface brightness within $R_e$. Right-panels are as a function of total magnitude. The approximate surface brightness and magnitude limits used for simulating the data is shown as a vertical dashed line.}
	\label{fig:sims_compFP}
\end{figure*}

Figure \ref{fig:sims_compFP} shows the main completeness and spurious source extraction results for \profound{} and \sex{} both run with default extraction parameters. The main trends with surface brightness and magnitude are as should be expected, where fainter objects are harder to accurately extract. As source brightness gets close to the $1\sigma$ surface brightness limit of the data the completeness drops off very sharply and the false-positive rate increases. Note that in the default mode used (skycut=1, which uses the $1\sigma$ sky RMS level as the extraction limit) the false-positive rate is always far below the true-positive completeness. This is a consistent finding when running simulations at a large range of depths, and suggests \profound{} can be safely operated in this mode. The false-positive rate tends to get close to the true-positive rate for the faintest sources when skycut$\sim$0.8, after which point you almost certainly would not want to push the extraction much further since most of the additional sources will be spurious. It should be noted that this limit is not trivially an extraction limit, since internal pixel smoothing and local clustering information is used to flag pixels that are likely to belong to real sources. Depending on the quality of the data and the pixel-to-pixel covariance the reasonable limit of skycut extraction might need to be higher than the value used here, however it is unlikely you could push much lower.

{\change It is notable that the default extraction parameters of \sex{} are a bit more conservative, extracting fewer sources (and suffering the associated incompleteness) but with a lower false-positive rate. This comparison should not be considered exhaustive, since very different curves are possible by changing the parameter setup of both \profound{} and \sex{}, however it does give an indication of how things compare with little or no tuning for the data. Pushing deep into the sky noise is a complex topic that we do not aim to explore fully here, however efforts are under way to explore how best to operate \profound{} in order to extract extremely faint and extended low surface brightness sources.}

\begin{figure*}
\centering
	\includegraphics[width=\columnwidth]{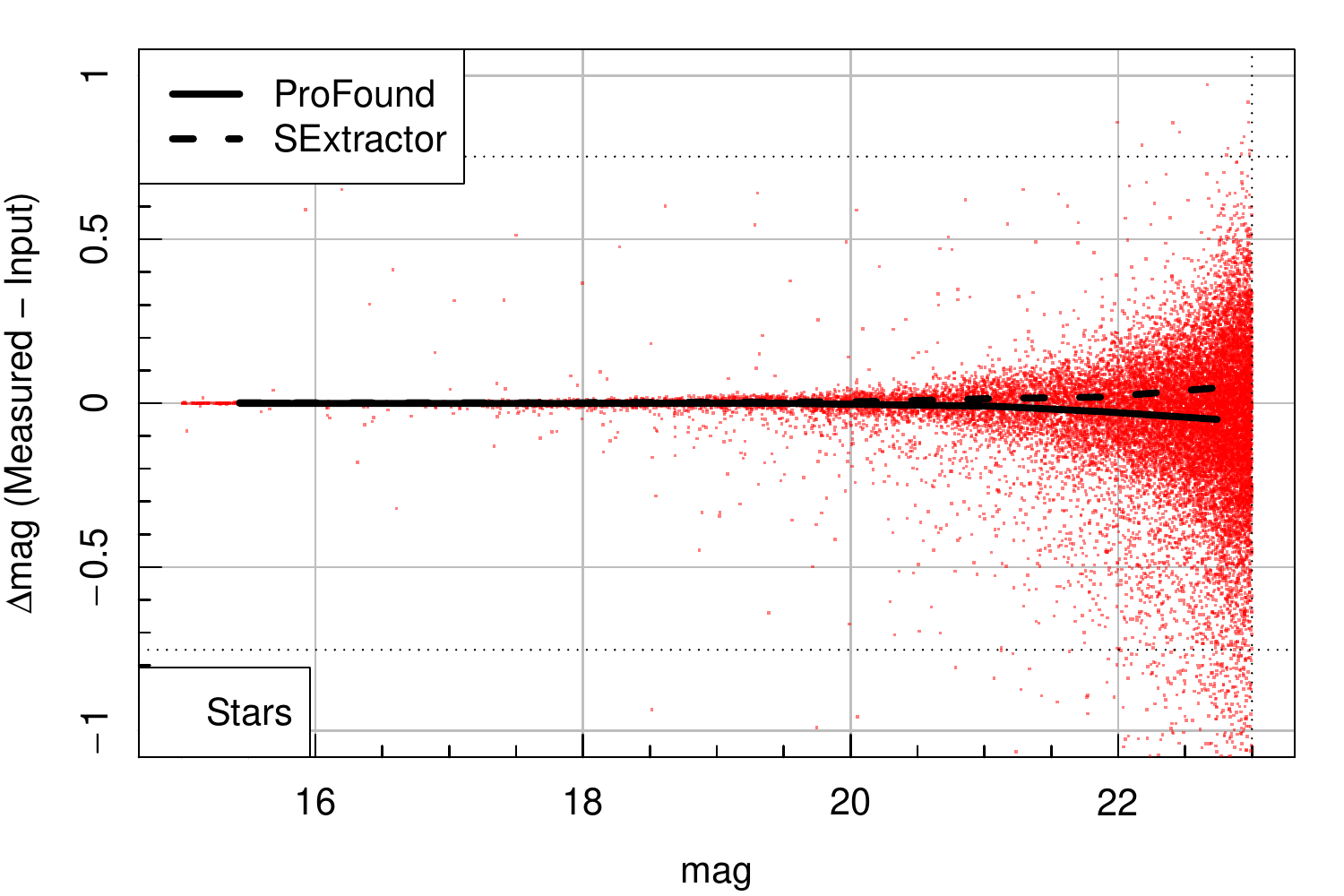}
	\includegraphics[width=\columnwidth]{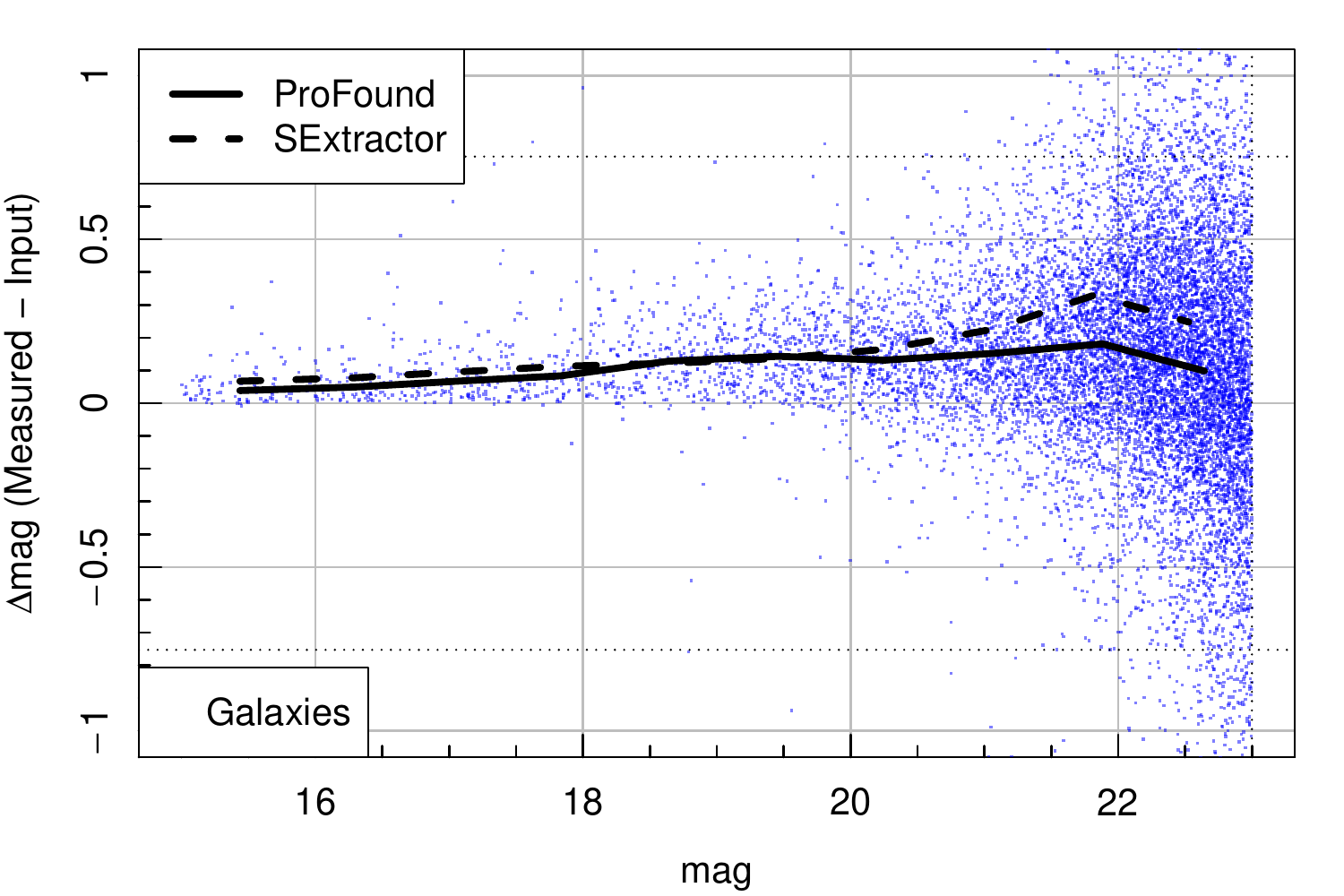}\\
	\includegraphics[width=\columnwidth]{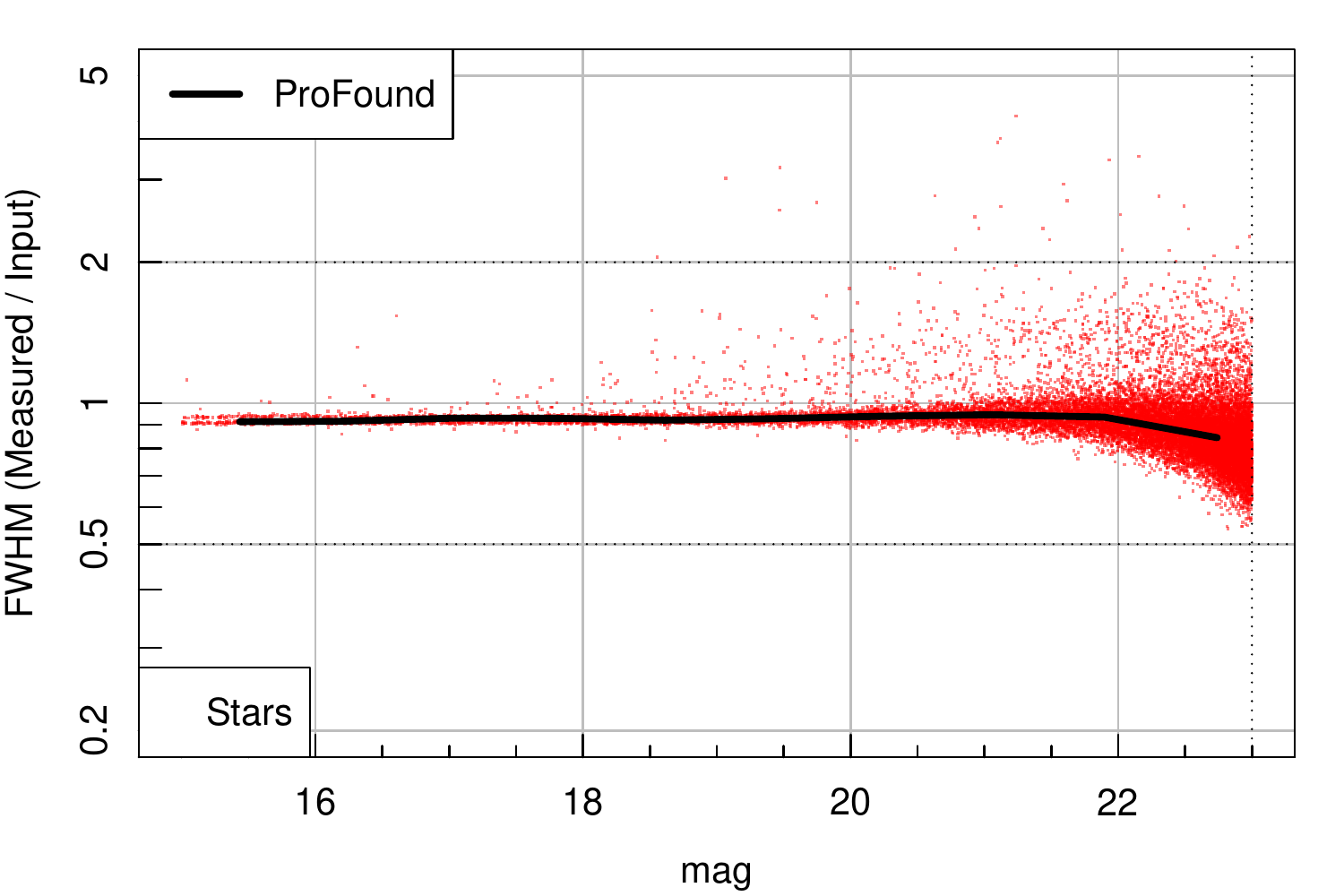}
	\includegraphics[width=\columnwidth]{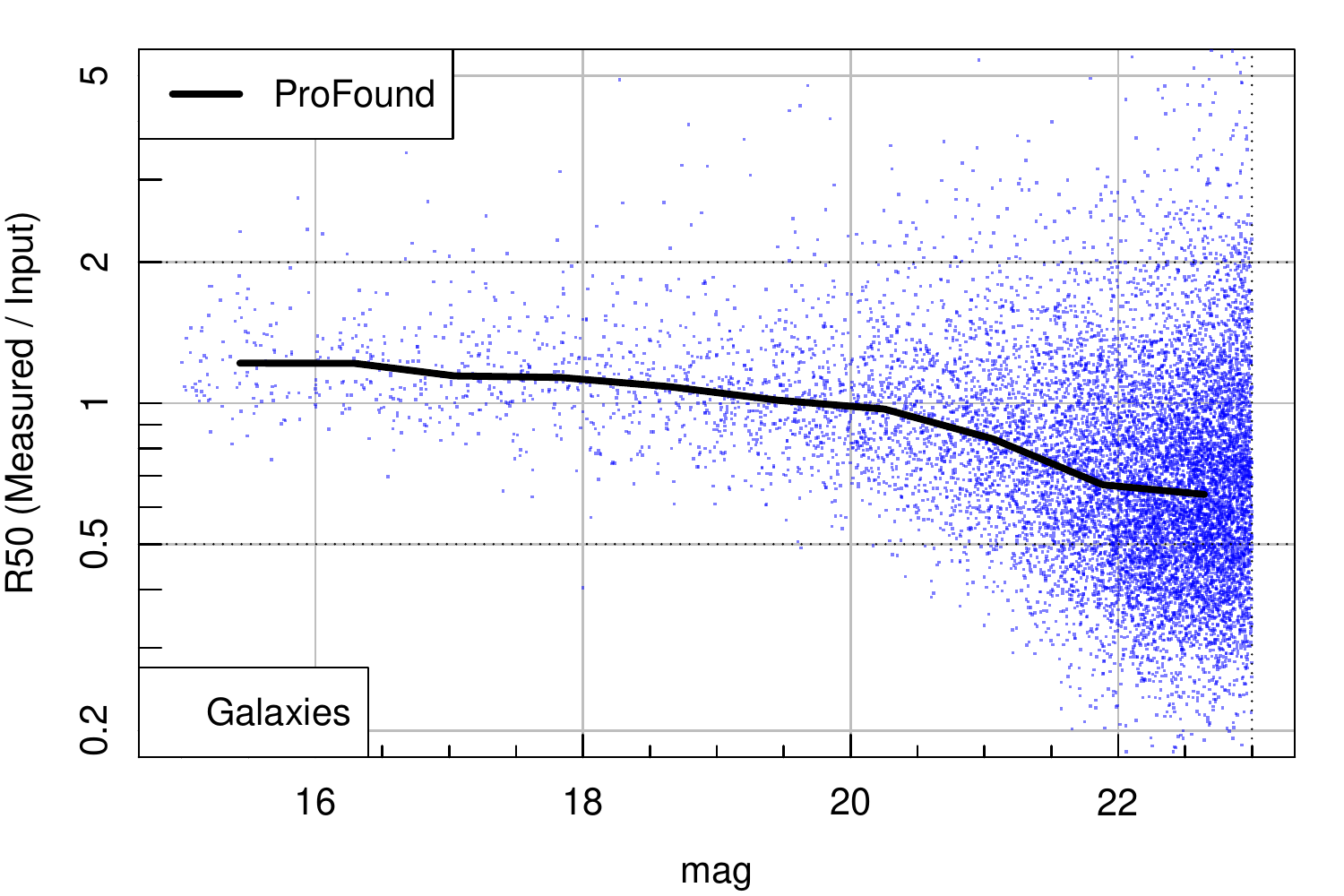}\\
	\caption{Main simulation results. The left panels show the star results in red. The right panels show the galaxy results in blue. The top panels show the difference between input and extracted magnitudes for both \profound{} (points and running median line) and \sex{} AUTO (running median line). The bottom panels show differences in the estimated objects sizes (full width half max [FWHM] for stars, $R_{50}$ for galaxies) but only for \profound{} (\sex{} does not directly compute these sizes). The horizontal dotted lines show a factor of two error in flux or size, where 99\% of stars and 70\% of galaxies are within this tolerance.}
	\label{fig:sims_magR50}
\end{figure*}

Figure \ref{fig:sims_magR50} shows the accuracy in the extraction of magnitudes and sizes for stars and galaxies in the simulated images. As seen in the left panels, in general stars can be extracted with better accuracy in terms of flux and size for a given magnitude. This is because the average surface brightness is typically much higher than for extended sources (i.e.\ galaxies). The extracted flux of stars show no systematic behaviour in the running median, even at the faint magnitude limit. The extracted sizes tend to show a small negative bias for brighter stars, which is due to pixel discreteness. At the faint magnitude limit, the sizes of stars becomes under-estimated because the dilation is halted by pixel flux noise.

The galaxy extractions show more scatter and also some systematic effects as a function of magnitude. Galaxies tend to have under-estimated fluxes at all magnitudes, but not surprisingly this effect is stronger for fainter objects. {\change \profound{} extracts fluxes for extended sources that are closer to the intrinsic values than \sex{} for all magnitudes, reflecting the utility of our dilated flux converged aperture approach.} As we saw in Figure \ref{fig:segim_iters}, extended sources tend to require more iterations before the flux is considered to be converged. Fainter sources tend to require more dilations to converge, which also means a greater fraction of their flux is introduced during dilation (in these simulations nearly 60\% of flux is introduced during dilation for sources that require six iterations). In all these cases it is noise in the image that halts the dilation (since any drop in flux immediately terminates the dilation process), hence it is increasingly difficult to extract the same fraction of intrinsic flux for the faintest sources. Instead the sky RMS starts to dominate over the object flux and the dilation halts. The behaviour of the extracted size is a bit more complicated, where very bright galaxies tend to have their sizes slightly overestimated. The is because large and brighter galaxies are more likely to be merged with nearby sources, further inflating their size. Towards the faint magnitude limit of the survey, the galaxies start to become under-estimated in terms of size, but the running median always stays within a factor of two of the input simulations.

In summary, \profound{} is able to extract objects at close to the noise limit without introducing a large number of false-positive detections, and the sources extracted have photometric properties that have relatively little bias with source flux and within the tolerance we would want for further profiling using \profit{}.

\subsection{LSST Simulations}
\label{sec:LSSTsims}

In order to more directly compare \profound{} against other source finding codes recently discussed in the literature, we ran it in default mode on some publicly available LSST simulation data \citep{conn10, zhen15}. This data was used as a testing data set in \citet{zhen15}, which presented a source finding approach that used a mixture of techniques to optimally recover sources blindly, with the focus on future large scale imaging surveys such as LSST. To allow a simple comparison we used default settings throughout except for the sky background filter size, which was set to 64$\times$64 in order to match the mode \sex{} had been run on using the same test data.

Running exactly the same images as analysed in \citet{zhen15} (Deep-32 and Deep-36, D32 and D36 from here) results in more detections than found in \citet{zhen15} (both the new extraction software presented and \sex{}). To ensure consistency we re-ran \sex{} with the parameters suggested, which resulted in an identical number of recovered sources to those presented in \citet{zhen15}. Since the method of defining a true-positive match was not clearly described, we utilised our own matching criterion based on proximity in spatial position and flux, these becoming the three dimensions we use to determine true matches. Measuring the sky match distance in arc-seconds and the flux in magnitudes, we define a match to be successful if the 3D match distance is within a radius of 2, i.e.\ if the intrinsic and recovered sources lie on top of each other, they are only matched in their fluxes are within two magnitudes.

Using this matching scheme, we find that of the 1548/1571 good sources (i.e.\ flag\_good=TRUE) recovered by \profound{} from the D32/D36 images, 1387/1413 were considered to be true matches (91.3\% and 88.3\% respectively). In direct comparison \sex{} recovered 1441/1386, of which 1249/1182 were considered to be true matches (86.5\% and 85.2\% respectively). The total number of sources recovered using \sex{} is identical to the analysis in \citet{zhen15}, however we find more true matches with our matching criteria (in that work 1189/1138 of the \sex{} sources are considered to be true matches). This suggests that our matching criterion is a bit more generous, but it at least allows for a fairly direct comparison between extraction methods. Just running with default parameters, \profound{} recovers more sources at a higher true-positive rate than \sex{}, and more sources at a similar true-positive rate to the code presented in \citet{zhan15}. A direct comparison with this code is harder given the differences in the matching criteria, but we certainly recover a lot more sources: they extract 1433/1375 sources with a claimed true-positive rate $\sim$91\%.

It is possible to increase the true-positive rate of \profound{} by changing a number of the available parameters. Equally, limiting the catalogue to the brightest 1433/1375 sources (to better match the extraction depth in \citet{zhen15}) increases the true-positive rate to 93.8\%/91.3\%. {\change The appropriate balance of extraction depth and fidelity will be science dependent, but it is possible to investigate this thoroughly through the use of receiver operating characteristic (ROC) curves (true-positive versus false-positive plots). This process can only be tackled through comparison to an absolute truth baseline, hence simulations should always be used when selecting appropriate parameters for a new data set.}

5Hence the more important finding is that \profound{} performs in a highly competitive manner, even when run with default parameters.

\section{Application to Ultra-VISTA Data}
\label{sec:UV}

The upcoming Deep Extragalactic VIsible Legacy Survey (DEVILS; Davies et~al. in prep.) on the Anglo Australian Telescope (AAT) will be conducting single-fibre redshift focussed spectroscopy for nearly 100\% of galaxies down to a total Y magnitude limit of 21.2. DEVILS will be targeting three main regions: two VISTA Deep Extragalactic Observations \citep[VIDEO;][]{jarv13} survey fields and the \uv{} field in the Cosmic Evolution Survey \citep[COSMOS;][]{scov07}. The source catalogues for \uv{} data release three (DR3) use a tailored run of \sex{} for total photometry and a mixture of forced aperture sizes for forced colour photometry \citep{mccr12}. For the purposes of DEVILS we need well converged total magnitudes and source apertures for likely galaxies. As discussed in \citet{wrig16} and \citet{andr17}, the apertures returned by \sex{} often fail in crowded regions in a manner that creates spurious apertures that loop around nearby bright sources. When visually inspecting the \sex{} total photometry catalogues made available for the \uv{} regions it became clear a large number of bright sources (above our Y 21.2 mag limit) have unusually large apertures and erroneously bright flux measurements. Rather than manually intervene and run the fixed apertures through \lambdar{} \citep[mimicking the work flow discussed in][]{wrig16} we ran \profound{} with close to default settings, creating a useable extra-galactic source catalogue in the process (Davies et~al. in prep will discuss the construction of the DEVILS input catalogues in detail).

\begin{figure*}
	\includegraphics[width=\columnwidth]{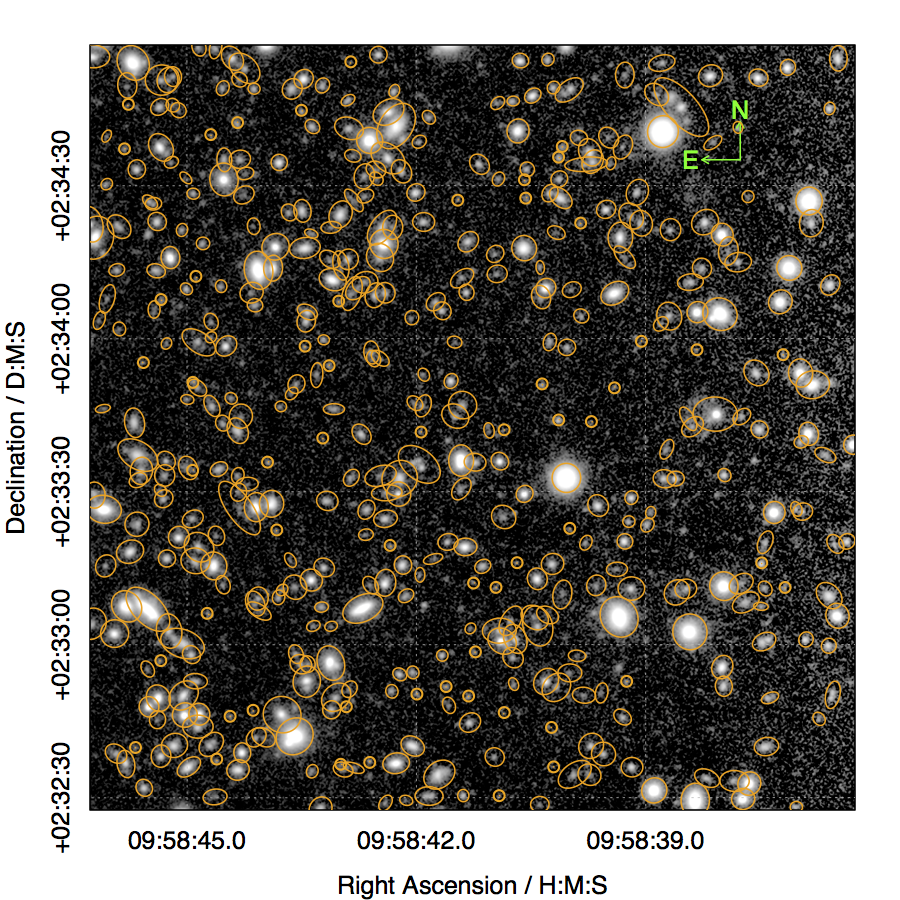}
	\includegraphics[width=\columnwidth]{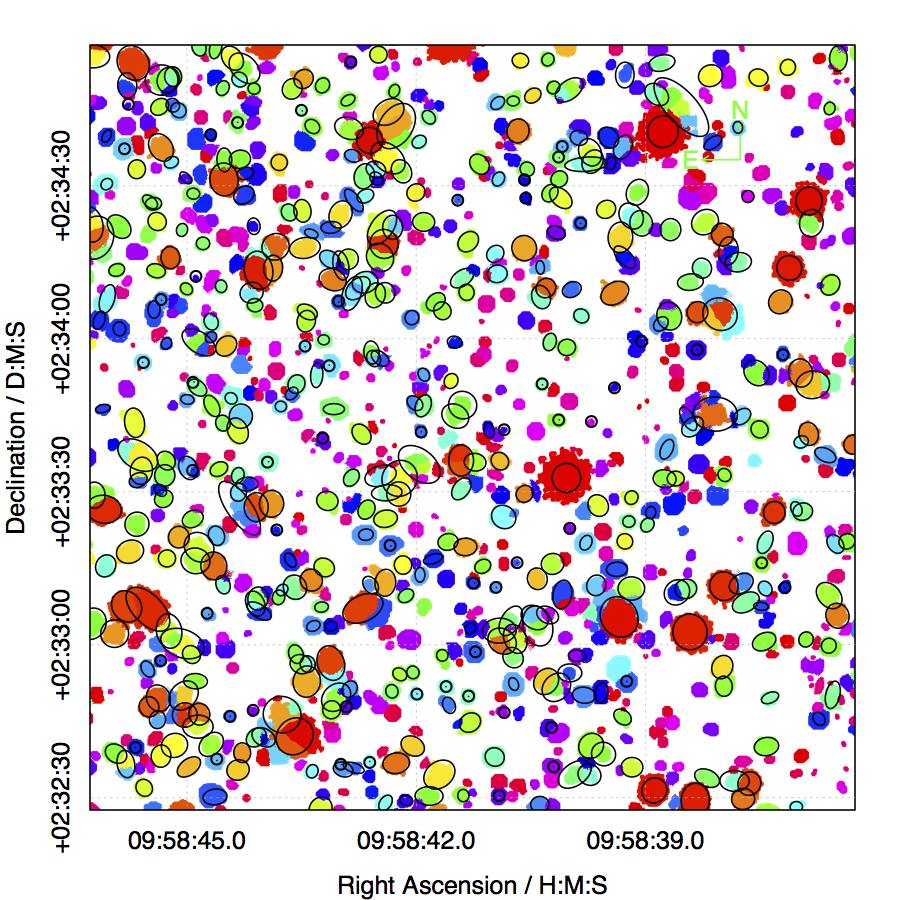}\\
    	\caption{The left panel shows an example stacked region taken from \uv{}, where the publicly released \sex{} catalogue AUTO apertures are overlaid as orange ellipses. The right panel shows the corresponding \profound{} segments in colour (redder means brighter sources and bluer means fainter sources), with the same \sex{} AUTO apertures overlaid in black. In isolated regions there is excellent qualitative agreement between the \profound{} segments and the \sex{} AUTO apertures.}
    	\label{fig:sexaps-v-prosegs}
\end{figure*}

Figure \ref{fig:sexaps-v-prosegs} shows a sub region of the \uv{} survey. The left panel shows an all-band (Y, J, H, Ks) stacked image overlaid with the AUTO (Kron like) apertures from the full \sex{} catalogue that had been run on the \uv{} Y-band data. The right panel shows the \profound{} pixel segments overlaid with the same apertures. It is clear that in isolated regions the AUTO apertures from \sex{} and the dilated flux converged segments from \profound{} agree rather well. However, in some of the more crowded regions it is clear the segmentation and de-blending is producing quite different extractions, e.g.\ the top-right region near the green compass shows an example of multiple sources being grouped together in the \sex{} de-blend. Also, due to the mode \profound{} was run in, the extraction depth is typically deeper with \profound{}, i.e. there are effectively no \sex{} objects missed by \profound{}, but there are some fainter (but visually real) sources extracted by \profound{} missing in the public \uv{} \sex{} based catalogue.

\begin{figure}
	\includegraphics[width=\columnwidth]{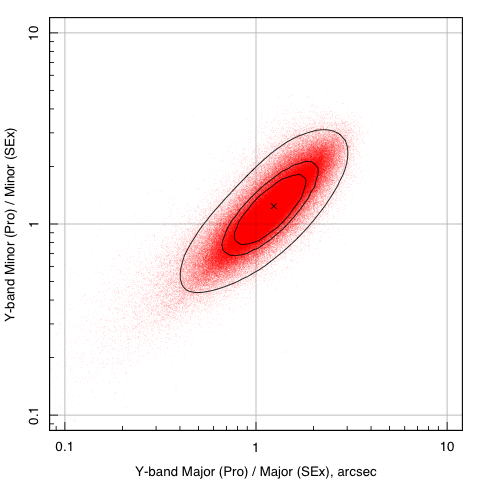}
    	\caption{Comparison between the publicly released \sex{} AUTO aperture major/minor axes and the \profound{} extracted major/minor axes. \profound{} does not directly use the ellipses it estimates for any photometry, but they can be used by other code, e.g.\ \lambdar{}. The three black contours show the high-density region containing 50\% / 68\% and 90\% of the data, and the cross shows the median point.}
    	\label{fig:major-minor}
\end{figure}

Whilst there is much similarity between the \sex{} apertures and the \profound{} segments, in general the \profound{} segments appear to be larger. This seems to be particularly true for very bright stars (where the \sex{} aperture does not encompass all of the halo light in general) and very faint objects. Figure \ref{fig:major-minor} shows the difference quantitatively, where we compare the \sex{} AUTO aperture values for the semi-major and semi-minor axes against the ellipse implied sizes for the same quantities in \profound{}. Whilst these apertures are produced by \profound{}, they are not directly used in the photometry (only the pixel segments are used when computing photometric properties). Instead they are typically used for diagnostic tests and to aid star galaxy separation (which we describe later in this paper).

In general the estimated ellipses agree quite well, but it is clear there is a lot of correlation between the major and minor axes comparison, i.e.\ \sex{} and \profound{} tend to agree on the shape of the ellipse, but there can be systematic differences in the total size of the objects. There is a small bias towards \profound{} finding slightly larger apertures in general (the cross is above the 1--1 lines in both x and y), but the majority of aperture sizes agree within a factor of two.

\begin{figure*}
	\includegraphics[width=\columnwidth]{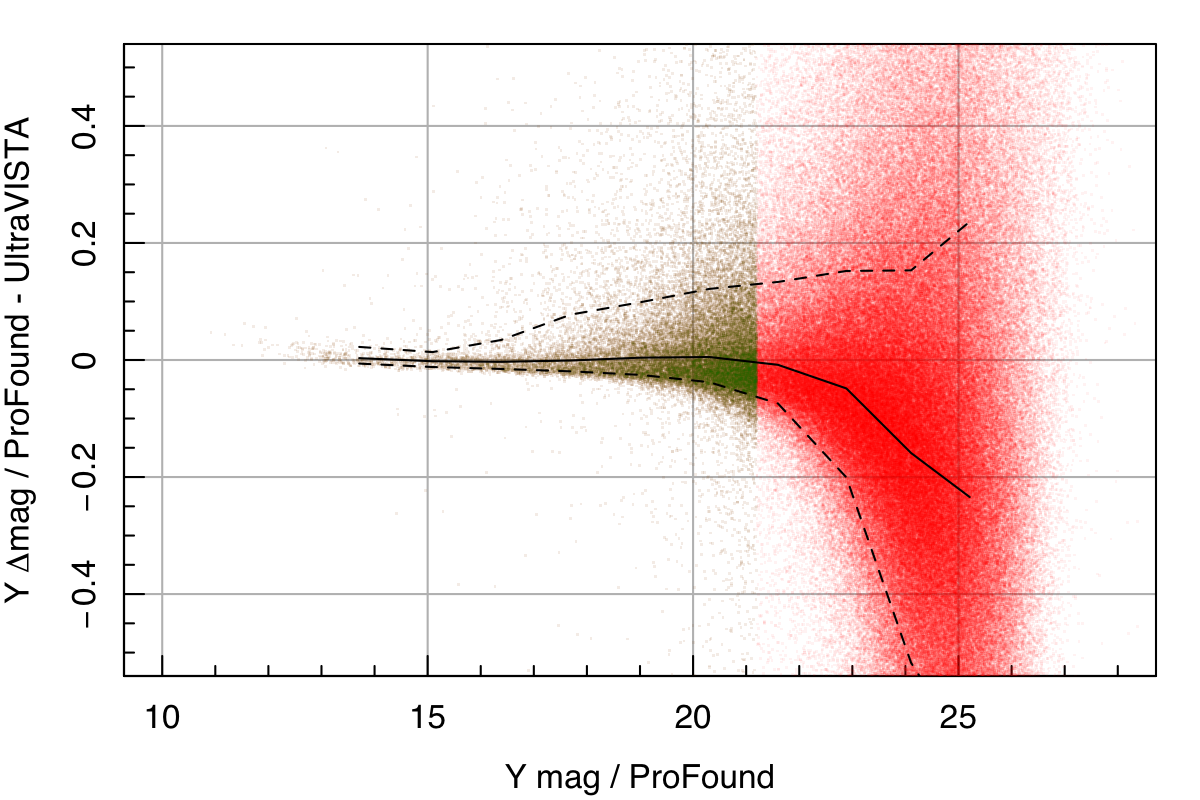}
	\includegraphics[width=\columnwidth]{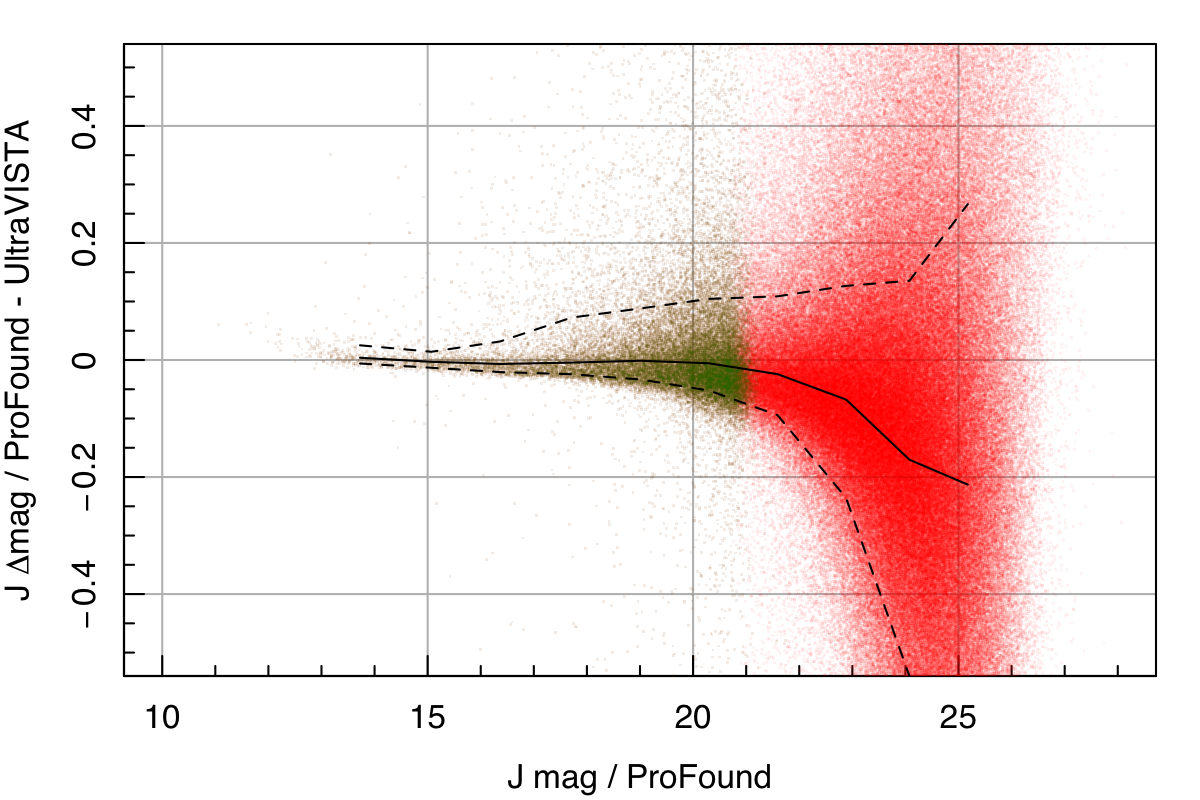}\\
	\includegraphics[width=\columnwidth]{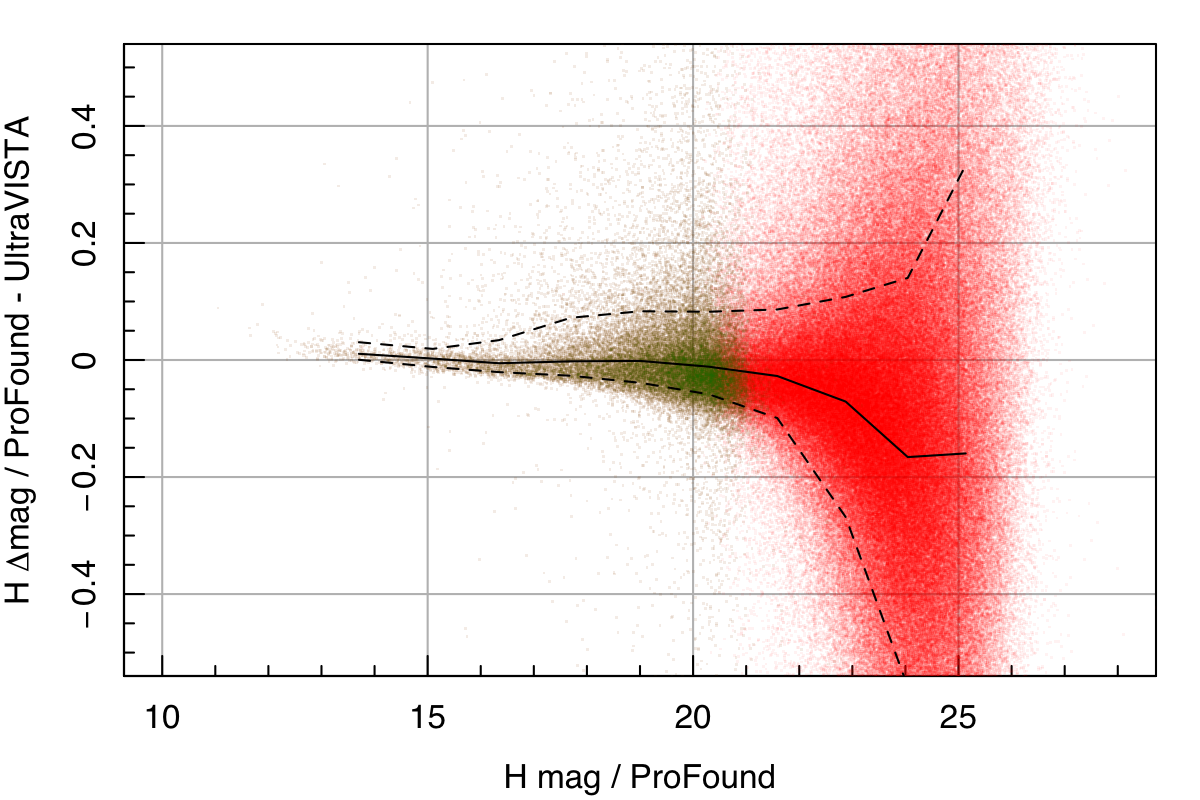}
	\includegraphics[width=\columnwidth]{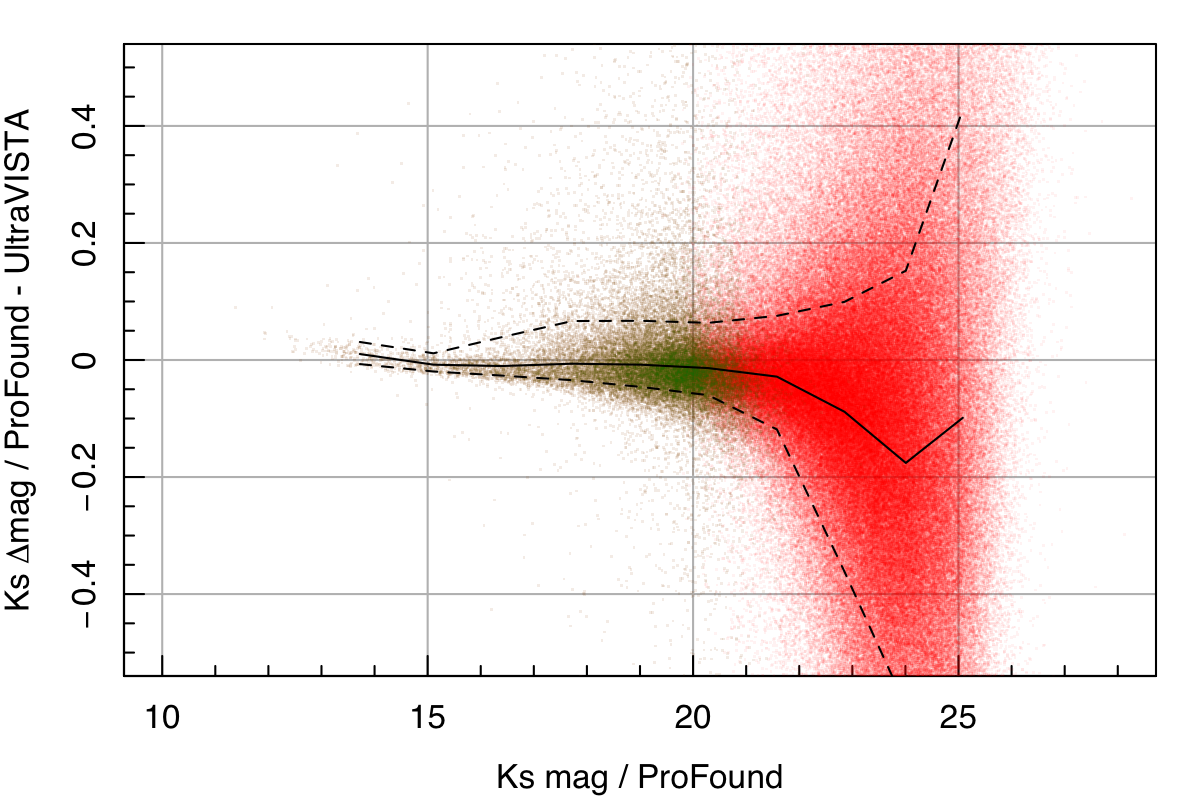}
    	\caption{Comparison of \profound{} dilated segment magnitudes versus \sex{} AUTO magnitudes for Y/J/H/Ks bands for 1`` matched sources. The solid black line shows the running median of the data, and the dashed lines show the range of 1$\sigma$ scatter around the median. The limit of the DEVILS sample (Y$<21.2$) is shown in dark green in each panel. For all bands we see that on average \profound{} returns more flux (i.e. the magnitude difference drops below 0 in these plots). The difference is typically less than 0.2 mag until $\sim23$mag, after which the amount of object scatter increases significantly. The typically brighter source extraction is the reason the faint limit has some diagonal structure.}
    	\label{fig:sex-pro-mag}
\end{figure*}

Figure \ref{fig:sex-pro-mag} shows how the extracted \profound{} magnitudes compare to the Kron-like AUTO magnitudes returned by \sex{}. For brighter objects with unambiguous segmentation solutions the \profound{} and \sex{} photometry agrees very well for all bands (well within 0.1 mag for the majority of sources). \sex{} does tend to have a number of brighter sources with close to double the flux of the \profound{} source. These cases tend to be where \profound{} has de-blended two very close by stars into two sources, whereas they are seen as being a single source in the published \uv{} catalogue. In these cases the de-blended solution in \profound{} appears to be preferable.

\begin{figure*}
	\includegraphics[width=18cm]{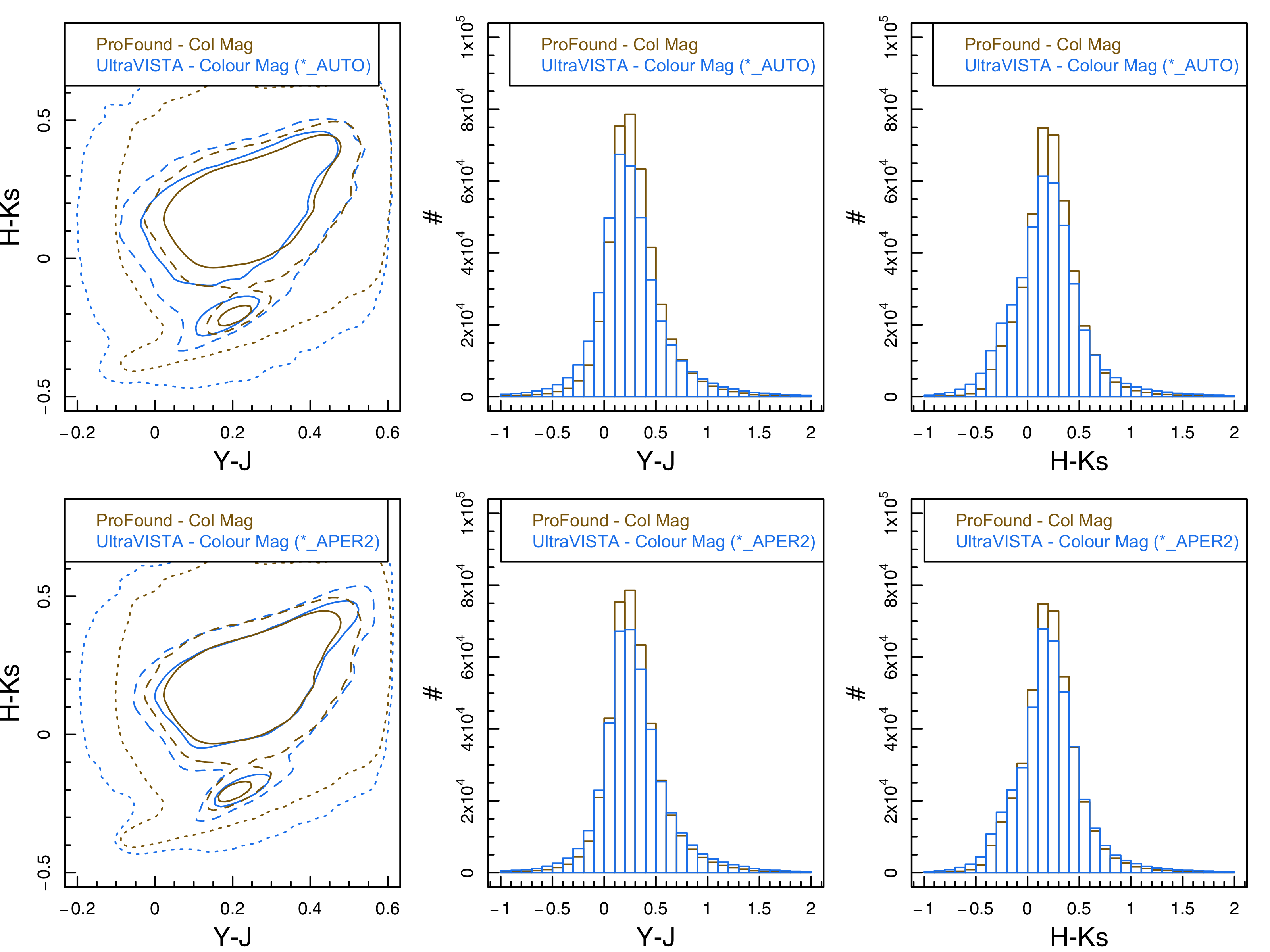}
    	\caption{Comparison of the NIR colour distributions between the released \sex{} \uv{} photometric catalogues and those generated by \profound{}. In these plots the sources have been explicitly matched against each other to ensure a fair comparison, though the same trends are evident if all detections at a comparable total magnitude cut are compared instead. The left panels show NIR colour-colour contour plots, where the contours contain the highest density 50\%/68\%/95\% of the data. The other panels show the bluest (Y-J, middle column) and reddest (H-Ks, right column) colour histograms for the comparison region. In the top and bottom rows the \profound{} photometry shown is the same, however the \sex{} \uv{} photometry is varied between the band optimised AUTO photometry (top row) and the fixed 2'' circular aperture photometry (bottom row).}
    	\label{fig:sex-pro-col}
\end{figure*}

Mimicking the analysis made in \citet{wrig16} and \citet{andr17}, we use the narrowness of the colour distributions to judge the quality of the matched aperture photometry. The logic is that any colour aperture error will generate an additional random component of noise, and hence an increase in the spread of the intrinsic colour distributions. I.e.\ there are few scenarios where any additional random error could decrease the colour distribution intrinsic scatter \citep[see][for a general discussion on intrinsic scatter and how best to model it]{robo15}. As discussed in \citet{andr17}, the 2`` aperture colours in \uv{} create very tight NIR colours, so these act as an excellent reference distribution for the matched segment \profound{} colours. Matched segment colours can be computed in a number of ways in \profound{}, but for this comparison we use the segmented pixels above a $2\sigma$ sky-RMS threshold.

Figure \ref{fig:sex-pro-col} shows the comparison of \profound{} and 2'' apertures taken from the published \sex{} derived \uv{} catalogue. Only two-way matching objects are shown for both distributions (i.e.\ the same ones as plotted in Figure \ref{fig:sex-pro-mag}). Since there are the same number of \profound{} and \uv{} objects in each colour histogram, higher peaks will generally highlight tighter colour distributions. The overall result is that \profound{} colours, run in matched segment mode, produce even tighter distributions than we find for the tightest \sex{} \uv{} colours. This is encouraging since the photometry of the main \uv{} survey was optimised for good quality colours for photo-z estimation.

A key issue for extra-galactic photometric surveys is correctly identifying stars and galaxies. This is especially important when creating input catalogues for follow-up surveys (e.g.\ DEVILS) since we do not want to waste time observing stars or miss real galaxy targets. \profound{} does not return a single parameter that flags stars and galaxies (like `Class-star' in \sex{}), but using a combination of size, surface brightness, ellipticity and concentration outputs it is possible to accurately identify likely stars in optical and NIR photometry. In the case of \uv{}, the NIR colours add highly constraining discriminative information.

\begin{figure*}
	\includegraphics[width=\columnwidth]{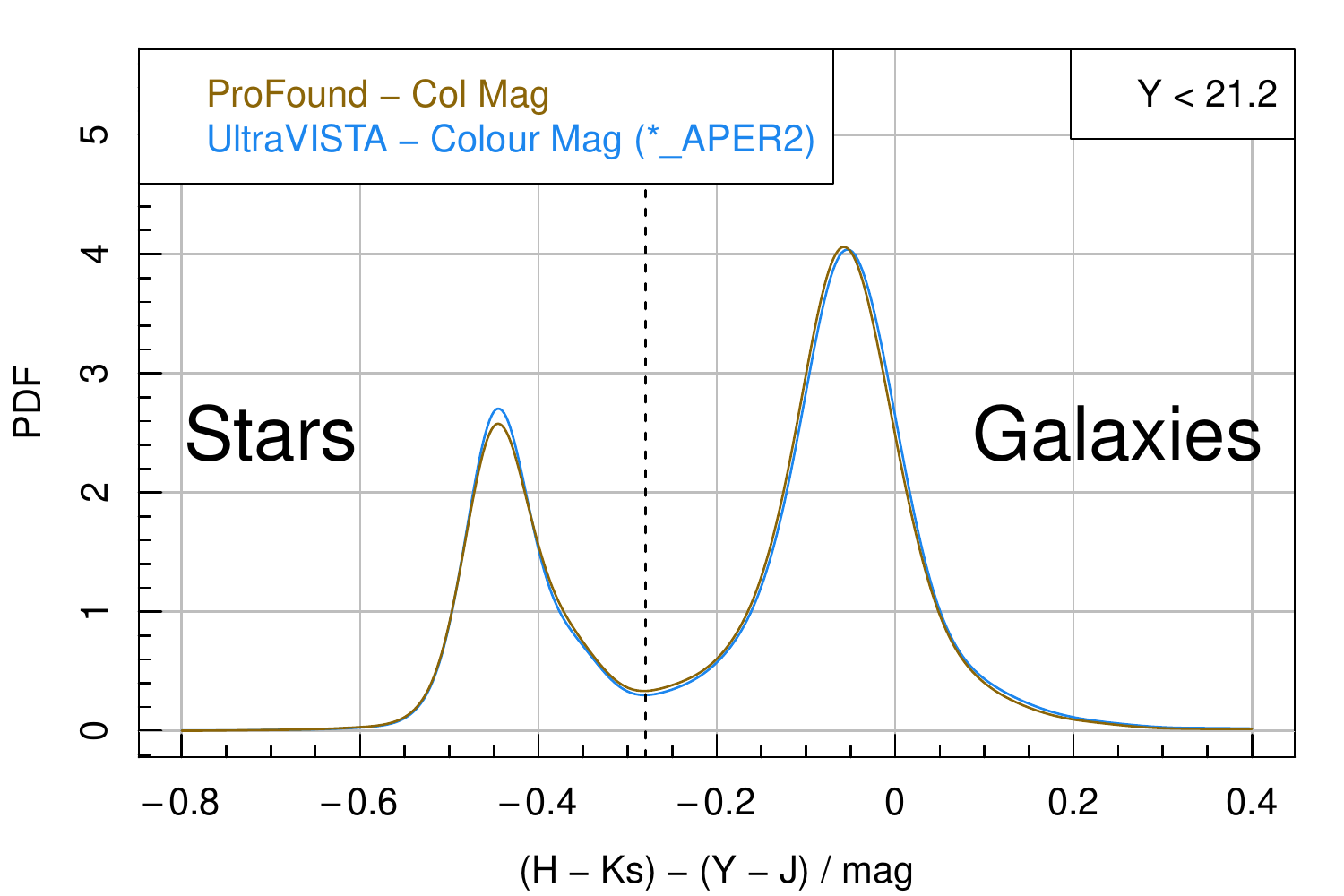}
	\includegraphics[width=\columnwidth]{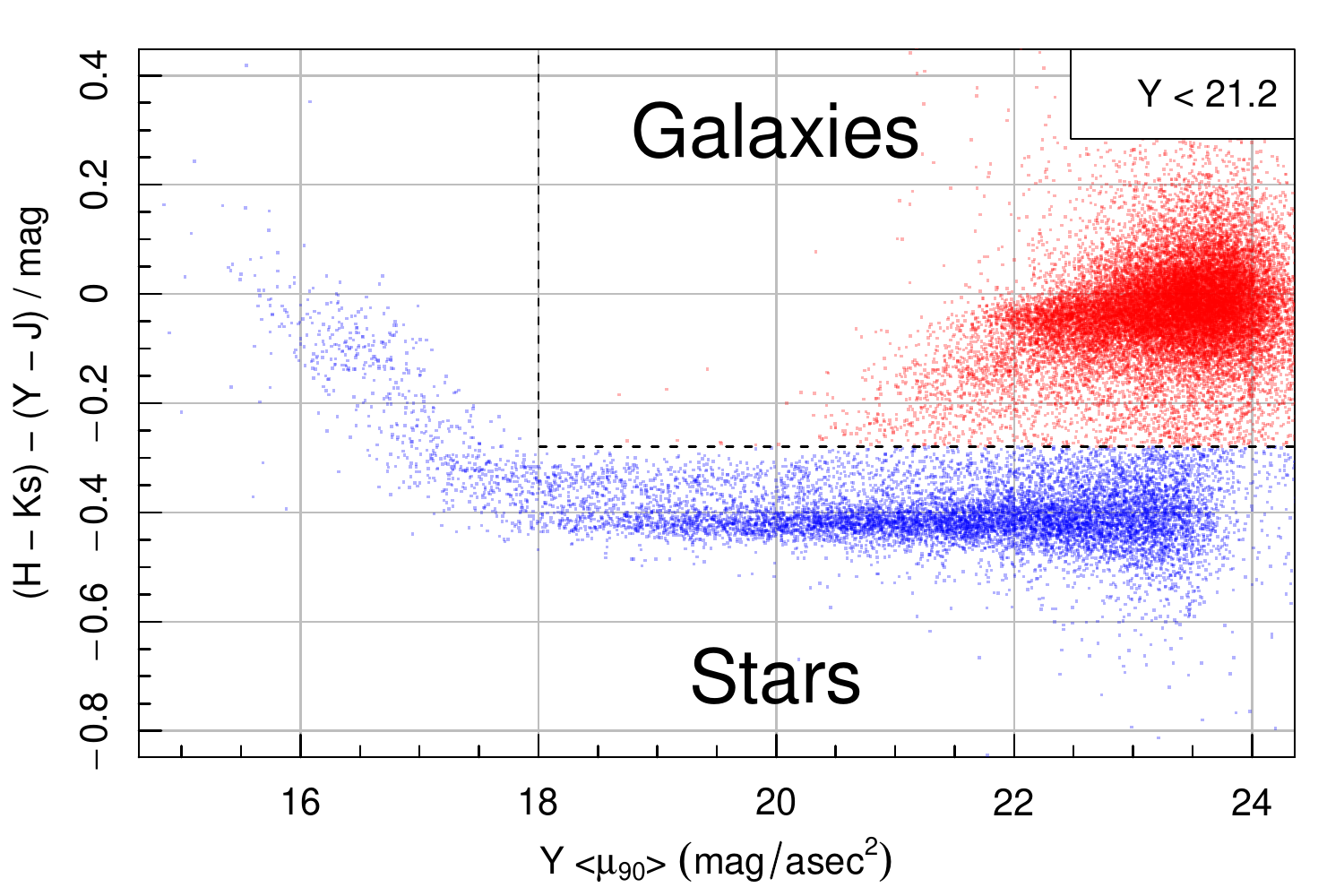}\\
	\includegraphics[width=\columnwidth]{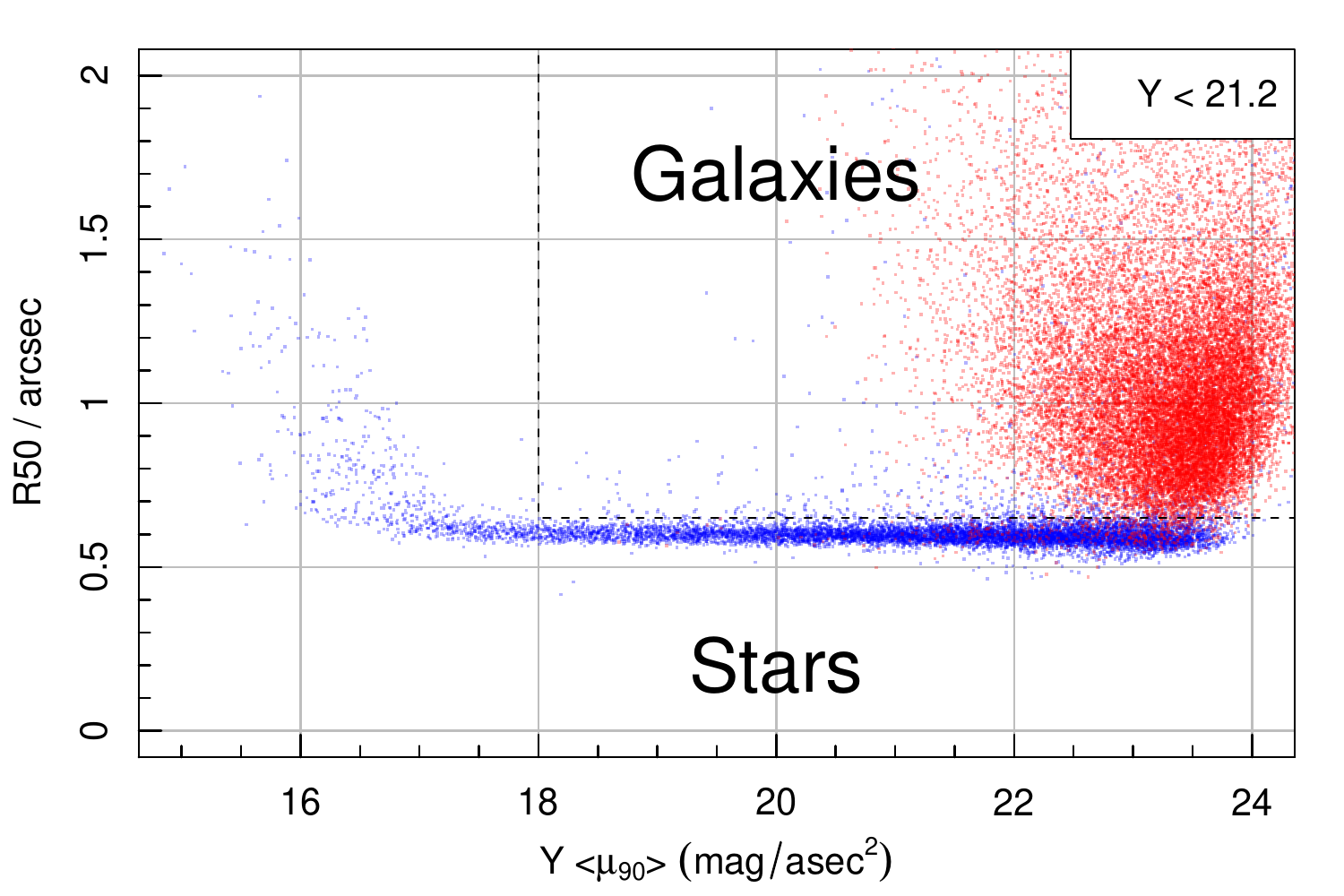}
	\includegraphics[width=\columnwidth]{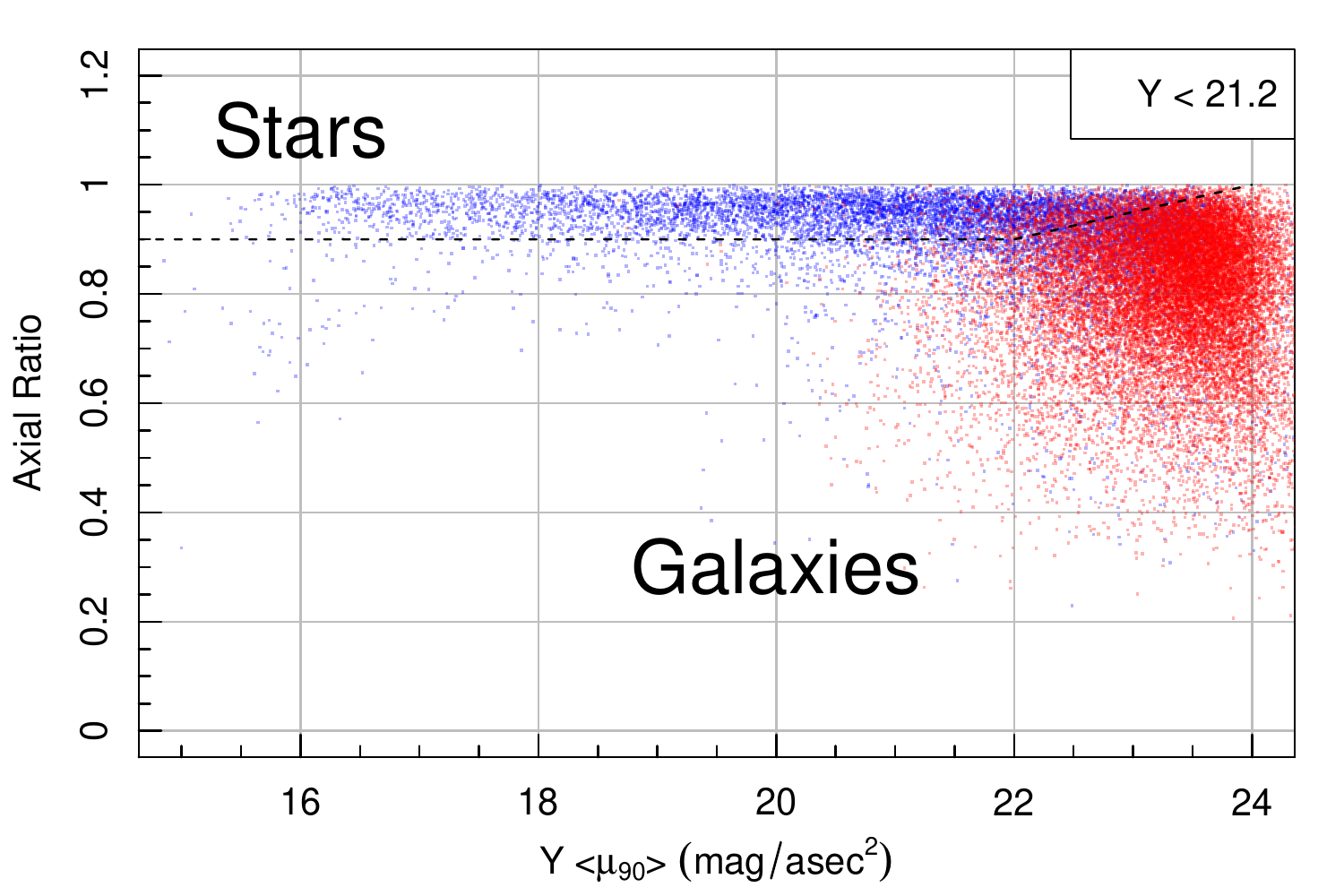}\\
    	\caption{Demonstration of the star galaxy separation possible using outputs from \profound{}. The top-left panel shows a simple multi-colour histogram ([H-Ks]-[Y-J]) where the use of the double colour term removes the main curvature seen in the left panels of Figure \ref{fig:sex-pro-col}. This is evident in the top-right panel where the same \profound{} multi-colour is shown on the y-axis against the mean surface brightness within the radius containing 90\% of the flux (the separation shown by the dashed line is applied to the subsequent panels by colouring nominal stars blue and galaxies red). The bottom-left panel shows the same surface brightness against R$_{50}$, and the bottom-right panel has axial ratio on the y-axis. The multi-colour clearly acts as a very effective star-galaxy separator on its own, with the additional projections acting mostly as consistency checks rather than improving the fidelity of the separation. Sources which might have been incorrectly assigned in the top-right panel are those that have a colour differing from the dominant population inside the dashed regions {\change (where the dashed regions are just indicative, and estimated by eye to approximately maximise the star galaxy discrimination)}.}
    	\label{fig:pro-stargal}
\end{figure*}

Figure \ref{fig:pro-stargal} shows a particularly successful projection of surface brightness and NIR colour. The NIR (H-Ks) - (Y-J) multi-colour creates an almost perfectly horizontal separation between stars and galaxies, but using the \profound{} surface brightness allows extremely bright stars with potentially erroneous colours (due to ghosting and saturation effects) do be easily selected. This selection is applied in a number of other geometric projections, where in each case the photometry has been extracted from the Y-band \uv{} data. It is clear that the NIR colours produce good separation agreement with the natural surface brightness versus size relationship, and also the surface brightness versus ellipticity relationship. These other two projections are not nearly as clear in isolation however, so in the case of \uv{} data the best possible star galaxy separation criteria must make use of accurate NIR colours. The details of the upcoming DEVILS survey star-galaxy separation using VISTA NIR colours and other geometric properties recovered with \profound{} will be discussed in detail in Davies et~al. in prep. In parallel to the processing of \uv{} data in order to create input catalogues for DEVILS, we are also using \profound{} to process VIKING survey data that covers the upcoming Wide Area Vista Extragalactic Survey \citep[WAVES;][circa 2021 start date]{driv16}\footnote{https://wavesurvey.org}. This future spectroscopic survey will cover $\sim$1300 square degrees and is a major project for the upgraded 4-metre Multi-Object Spectroscopic Telescope \citep[4MOST;][]{de-j14}. The preparation of the NIR VIKING data using \profound{} will be described in detail in an upcoming paper (Koushan et~al., in prep).

\section{Conclusions}

In this paper we have presented the new \profound{} source finding and extraction package. Its primary design goal is to provide all of the various inputs needed by the \profit{} galaxy modelling package, but it also serves as an effective stand-alone blind source finder. In these concluding remarks we will focus on the most novel aspects of the software compared to other open source astronomy alternatives.

\begin{itemize}

\item It uses a saddle point based source segmentation strategy that minimises the occurrence of source islands being created. A common artefact that essentially cannot be formed in \profound{} is an aperture for an extended source looping around a bright star. This has removed the necessity of manual aperture fixing for extended sources \citep[see][]{wrig16, andr17}.

\item It defines photometric properties using dilated apertures rather than ellipses. This maintains the major geometric features of the sources, whilst also guaranteeing good flux convergence (by default, though other properties can be used to define the convergence criterion).

\item It creates all of the basic parameter inputs for \profit{}, in particular the effective radius is computed for all sources and the ellipticity and geometric rotation are defined in the same sense as for \profit{} (removing simple conversion mistakes that are otherwise common).

\item It can operate in a number of modes for matched aperture (colour) photometry: e.g.\  simple per pixel segment matching, or with the capacity to let sources naturally dilate in each target band.

\item It returns a comprehensive set of meta data by default, in particular the segmentation, sky and sky RMS maps. It can also compute the grouped complexes of touching sources, making it easy to fit multiple potentially overlapping sources with \profit{}.

\item It offers a number of tools to aid the analysis of the source extraction and further prepare for running \profit{}, in particular it can create and sensibly track groups of touching/overlapping sources, and can extract surface brightness ellipses belonging to certain segments.

\item It is fully open source and LGPL-3 compliant and written using a mixture of available \R{} packages (available on the Comprehensive \R{} Archive Network; CRAN) and \R{} language, removing the need to compile the software for the majority of users.

\item As well as the main package being hosted and maintained on GitHub (github.com/asgr/ProFound) it offers a large number of long form tutorial vignettes to aid inexperienced users tackling complicated extraction and fitting problems (http://rpubs.com/asgr/).

\end{itemize}

{\change As with any piece of software, \profound{} has some limitations that we will be keen to address in the longer term. As mentioned previously, it is predominantly single-threaded, and does suffer memory limitation issues for very large images. The immediate aim is to improve the memory limitations, since this appears to present the biggest limitation in typical use cases (e.g.\ processing large surveys). Also, in comparison to other photometry packages, its approach to assign flux to blended sources is fairly rudimentary. However, it is now possible to provide \profound{} segmentation maps as inputs to \lambdar{}, which provides access to more sophisticated flux de-blending schemes.
A final comment is that by design the saddle-point segmentation approach, whilst avoiding the possibility of some classes of serious de-blend errors, lacks the flexibility to create segments deeply embedded within segments. It is possible to combine different segmentation maps to get round this limitation, but it does mean running the segmentation function more than once for highly extended sources if the initial map poses a problem for further profiling with \profit{}\footnote{see Segmentation Map vignette at http://rpubs.com/asgr/}. That said, there is often no easy way for even humans to distinguish between very compact clumpy substructure that belongs to a galaxy and foreground stars that do not. Even colour information (unless very extensive) does not always remove this inherent ambiguity.}

{\change In summary, \profound{} offers a number of novel approaches to tackling some common issues with image source detection and extraction. It is particularly well suited for preparing inputs for further processing with \profit{} and similar galaxy profiling software {\changetwo (e.g.\ Barsanti et~al. in prep, Casura et~al. in prep, Cook et~al. in prep, Hashemizadeh et~al. in prep)}. A number of ongoing projects are already using \profound{} for generic survey source extraction (Koushan et~al. in prep, Davies et~al. in prep). We are also actively investigating how best to use \profound{} for the low surface brightness extraction of extremely faint and extended sources in KiDS data (Turner et~al. in prep).}

In the future there is a clear opportunity to create a more tightly coupled user experience for the automated fitting of galaxy profiles (in the general spirit of \citet{bard12} and \citet{kelv12}). However, there is clear utility in using \profound{} simply as a source detection and extraction tool, so this explicit coupling has not been enforced at this stage. Prototypes of fully integrated packages that are inspired by some of the ideas and issues discussed in this work already exist, and will likely be released along with the corresponding scientific analysis over the next few years (e.g.\ Barsanti et~al. in prep, Casura et~al. in prep, Cook et~al. in prep, Hashemizadeh et~al. in prep).

\section*{Acknowledgements}

We would like to thank the referee, Emmanuel Bertin, for his helpful suggestions on how to present aspects of this work. Much of the work presented here was made possible by the free and open \R{} software environment \citep{rcor16}. All figures in this paper were made using the \R{} {\sc magicaxis} package \citep{robo16}. This publication has made use of data from the VIKING survey from VISTA at the ESO Paranal Observatory, programme ID 179.A-2004 \citep{edge13}. Data processing has been contributed by the VISTA Data Flow System at CASU, Cambridge and WFAU, Edinburgh. Based on data products from observations made with ESO Telescopes at the La Silla Paranal Observatory under ESO programme ID 179.A-2005 and on data products produced by TERAPIX and the Cambridge Astronomy Survey Unit on behalf of the UltraVISTA consortium. \citep{mccr12}. Parts of this research were conducted by the Australian Research Council Centre of Excellence for All-sky Astrophysics (CAASTRO), through project number CE110001020. 







%
%


\bsp	
\label{lastpage}
\end{document}